\documentclass[a4paper,12pt]{article}
\usepackage[dvipdfmx]{graphicx}
\usepackage{amsfonts,amsthm,amsmath,amssymb,graphicx,float,mathtools,ulem}
\numberwithin{equation}{section}
\usepackage{subfigure}
\usepackage{color}
\usepackage{bbm} 
\usepackage{tensor} 
\usepackage{verbatim} 
\usepackage[
	  pagebackref=false,
	  colorlinks=true,
      linkcolor=blue,
      urlcolor=blue,
      filecolor=black,
      citecolor=red,
      pdfstartview=FitV,
      pdftitle={},
        pdfauthor={},
        pdfsubject={},
        pdfkeywords={},
        pdfpagemode=None,
        bookmarksopen=true
      ]{hyperref}
\usepackage{ascmac}

\begin{document}

\newtheorem{definition}{Definition}[section]
\newcommand{\be}{\begin{equation}}
\newcommand{\ee}{\end{equation}}
\newcommand{\bea}{\begin{eqnarray}}
\newcommand{\eea}{\end{eqnarray}}
\newcommand{\LE}{\left[}
\newcommand{\R}{\right]}
\newcommand{\nn}{\nonumber}
\newcommand{\Tr}{\text{Tr}}
\newcommand{\N}{\mathcal{N}}
\newcommand{\G}{\Gamma}
\newcommand{\vf}{\varphi}
\newcommand{\LL}{\mathcal{L}}
\newcommand{\Op}{\mathcal{O}}
\newcommand{\HH}{\mathcal{H}}
\newcommand{\arctanh}{\text{arctanh}}
\newcommand{\up}{\uparrow}
\newcommand{\down}{\downarrow}
\newcommand{\ket}[1]{\left| #1 \right>}
\newcommand{\bra}[1]{\left< #1 \right|}
\newcommand{\ketbra}[1]{\left|#1\right>\left<#1\right|}
\newcommand{\rd}{\partial}
\newcommand{\de}{\partial}
\newcommand{\ba}{\begin{eqnarray}}
\newcommand{\ea}{\end{eqnarray}}
\newcommand{\db}{\bar{\partial}}
\newcommand{\we}{\wedge}
\newcommand{\ca}{\mathcal}
\newcommand{\lr}{\leftrightarrow}
\newcommand{\f}{\frac}
\newcommand{\s}{\sqrt}
\newcommand{\vp}{\varphi}
\newcommand{\hvp}{\hat{\varphi}}
\newcommand{\tvp}{\tilde{\varphi}}
\newcommand{\tp}{\tilde{\phi}}
\newcommand{\ti}{\tilde}
\newcommand{\ap}{\alpha}
\newcommand{\pr}{\propto}
\newcommand{\mb}{\mathbf}
\newcommand{\ddd}{\cdot\cdot\cdot}
\newcommand{\no}{\nonumber \\}
\newcommand{\la}{\langle}
\newcommand{\lb}{\rangle}
\newcommand{\ep}{\epsilon}
 \def\we{\wedge}
 \def\lr{\leftrightarrow}
 \def\f {\frac}
 \def\ti{\tilde}
 \def\ap{\alpha}
 \def\pr{\propto}
 \def\mb{\mathbf}
 \def\ddd{\cdot\cdot\cdot}
 \def\no{\nonumber \\}
 \def\la{\langle}
 \def\lb{\rangle}
 \def\ep{\epsilon}
\newcommand{\mcl}{\mathcal}
 \def\g{\gamma}
\def\Tr{\text{tr}}

\begin{titlepage}
\thispagestyle{empty}

\begin{flushright}

\end{flushright}
\bigskip
\begin{center}
 \noindent{\large \textbf{Entanglement Dynamics of the Non-Unitary Holographic Channel}}\\

\vspace{1cm}
\renewcommand\thefootnote{\mbox{$\fnsymbol{footnote}$}}
Kanato Goto\footnote{kanato.goto@yukawa.kyoto-u.ac.jp}${}^{1,2,3}$, Masahiro Nozaki\footnote{mnozaki@ucas.ac.cn}${}^{4,3}$, Kotaro Tamaoka\footnote{tamaoka.kotaro@nihon-u.ac.jp}${}^{5}$ and Mao Tian Tan\footnote{maotian.tan@apctp.org}${}^{6,7}$\\

\vspace{1cm}

${}^{1}${\small \sl Department of Physics, Princeton University, Princeton, New Jersey, 08540, USA}\\
${}^{2}${\small \sl Center for Gravitational Physics and Quantum Information (CGPQI),\\ Yukawa Institute for Theoretical Physics,
Kyoto University, Kyoto 606-8501, Japan}\\
${}^{3}${\small \sl RIKEN Interdisciplinary Theoretical and Mathematical Sciences (iTHEMS), \\Wako, Saitama 351-0198, Japan}\\
${}^{4}${\small \sl Kavli Institute for Theoretical Sciences, University of Chinese Academy of Sciences,
Beijing 100190, China}\\

${}^{5}${\small \sl Department of Physics, College of Humanities and Sciences, Nihon University, \\Sakura-josui, Tokyo 156-8550, Japan}\\
${}^{6}${\small \sl Center for Quantum Phenomena, Department of Physics, New York University, 726 Broadway, New York, New York 10003, USA}\\
${}^{7}${\small \sl Asia Pacific Center for Theoretical Physics, Pohang, Gyeongbuk, 37673, Korea}\\

\if(
\author{Kanato Goto}\email[]{kanato.goto@riken.jp}
\affiliation{\it RIKEN Interdisciplinary Theoretical and Mathematical Sciences (iTHEMS), Wako, Saitama 351-0198, Japan}
\author{Masahiro Nozaki}\email[]{manozaki@ucas.ac.cn }
\affiliation{\it Kavli Institute for Theoretical Sciences and CAS Center for Excellence in Topological Quantum Computation, University of Chinese Academy of Sciences, Beijing, 100190, China}
\affiliation{\it RIKEN Interdisciplinary Theoretical and Mathematical Sciences (iTHEMS), Wako, Saitama 351-0198, Japan}
\author{Kotaro Tamaoka}\email[]{tamaoka.kotaro@nihon-u.ac.jp}

\affiliation{\it Department of Physics, College of Humanities and Sciences, Nihon University, Sakura-josui, Tokyo 156-8550, Japan}
)\fi

\setcounter{footnote}{0}
\renewcommand\thefootnote{\mbox{\arabic{footnote}}}
\vskip 2em
\end{center}
\begin{abstract}
We study the dynamical properties of a strongly scrambling quantum circuit involving a projective measurement on a finite-sized region by studying the operator entanglement entropy and mutual information (OEE and BOMI) of the dual operator state that corresponds to this quantum circuit. The time-dependence of the OEE exhibits a new dynamical behavior of operator entanglement, namely an additional fractional coefficient that accompanies the linear time growth of the OEE. For a holographic system, this is equivalent to an additional fractional coefficient that modifies the linear growth rate of the wormhole volume.
The time-dependence of the BOMI shows that the projective measurement may destroy the non-local correlations in this dual state. We also propose a gravity dual as well as a line-tension picture, which is an effective model, that describe this strongly scrambling quantum circuit.
\end{abstract}
\end{titlepage} 
\tableofcontents

\section{Introduction and Summary}
In recent years, quantum information-theoretic approaches have been used in many research areas ranging from string theory to condensed matter theory. For example, the entanglement entropy of a reduced density operator associated with a subsystem along with its spectrum
have been used to classify topological phases \cite{2005PhLA..337...22H,2006PhRvL..96k0404K,2006PhRvL..96k0405L,2008PhRvL.101a0504L,2010PhRvB..81f4439P}. In the AdS/CFT correspondence where a gravitational theory is equivalent to a non-gravitational theory, quantum information-theoretic methods and quantities have been used to construct the geometry of the gravitational theory from quantum states \cite{2006PhRvL..96r1602R,Ryu:2006ef,Almheiri:2014lwa,Pastawski:2015qua}.

In general, state-preparation and quantum-computing processes may be either non-unitary or non-equilibrium, or both.
These state-preparation processes are algorithms \cite{2014arXiv1411.4028F,2020PNAS..11725396P} and quench protocols \cite{2020PhRvR...2c3347R,2018PhRvL.120u0604A,2019PhRvB..99j4308M} that allow a system to be prepared in a state of interest. Operations in quantum computing can be either unitary or non-unitary \cite{1997PhDT.......232G,2014NJPh...16a3009V}. One way to construct a quantum computer is to build a circuit that is composed of layers of quantum gates. Within each layer, the quantum gates act on a spatially ordered array of qubits. Successive layers of these quantum gates implement a sequence of quantum computing operations.
If each step of the computation is viewed as a single time-step of the time evolution, such a quantum circuit implements an out-of-equilibrium dynamical process which may potentially be non-unitary.
Despite the substantial literature on these out-of-equilibrium processes, discussions pertaining to non-unitary operations are few and far between. This is especially true when it comes to studies involving conformal field theories. Some pioneering studies on projective measurements in two-dimensional conformal field theories ($2$d CFTs) can be found in \cite{Numasawa:2016emc,2022arXiv220912903A,2022arXiv221003083M}.


In this paper, we seek to fill in this gap by studying non-unitary time evolution in two-dimensional conformal field theories by using operator entanglement entropy (OEE), which is the von Neumann entropy for an operator state, since information about the dynamics is encoded in the entanglement structure of the operator state. We also compute the bipartite operator mutual information (BOMI) which is a linear combination of OEEs that quantifies the amount of correlation between two systems. In particular, we study the dynamical properties of a quantum circuit with a single projective measurement on a spatial region of finite size.
For the quantum circuit without the projective measurements, effective models have been proposed to describe the dynamics of operator entanglement.
In non-scrambling quantum mechanical systems such as a $2$d free fermion theory, the entanglement dynamics is known to be described by the relativistic propagation of localized objects called quasiparticles \cite{2018arXiv181200013N}.
In systems where strong information scrambling is present, such as holographic CFTs which are CFTs that possess a gravitational dual, it is described by the growth in time of a codimension-2 object \cite{2018PhRvX...8b1014N,2018arXiv180300089J,2018arXiv180300089J,2018arXiv181200013N,Kudler-Flam:2020yml,Goto:2021gve}. 
For example, in $2$d holographic CFTs, it is a one-dimensional object, and the effective model is called the line tension picture.
The projective measurement may change the entanglement structure associated with the region where the operation is performed.
Therefore, the projective measurement may also change the dynamics of the operator entanglement.
In the papers \cite{2019PhRvB..99v4307C,2019PhRvX...9c1009S,2019PhRvB.100m4306L,2018PhRvB..98t5136L}, the authors studied the dynamics of operator entanglement when local projective measurements are carried out at random spatial locations in each layer of a quantum circuit.
A phase transition with respect to entanglement occurs by modulating the rate at which projective measurements are performed: If the rate is low enough, the leading order term of the entanglement entropy is proportional to the volume of the subsystem (volume law), and if the rate is high enough, it is proportional to the area of the boundary of the subsystem (area law).
For further discussions of non-equilibrium processes involving random projective measurements and quantum error correcting codes, see \cite{2020PhRvL.125c0505C,2020PhRvX..10d1020G,2020PhRvB.101j4301B,2021PhRvB.103q4309F,2021PhRvB.103j4306L}.
\subsubsection*{The goal of this paper }
In this paper, we study how a projective measurement affects the entanglement dynamics of $2$d holographic CFTs. We also propose an effective model describing strong scrambling dynamics in the presence of a projective measurement.

\subsection*{Summary}
We have discovered a new dynamical behavior of OEE which we present in this paper (see Section \ref{Section:OEE} for details.). This new dynamical behavior is a linear time growth of OEE that is modified by additional fractional coefficients. During certain time intervals in the evolution, this proportionality coefficient results in an OEE growth that is half of the well-known linear growth of OEE in holographic systems as explained by the growth of a  wormhole in the gravitational dual \cite{2013JHEP...05..014H}.
We also propose a line tension model, which is an effective model describing the dynamics of quantum circuits, and a gravity dual of this quantum circuit.

\subsubsection*{Line tension picture}

The line tension picture is an effective hydrodynamic description of the entanglement production in a chaotic system. In a random unitary circuit, the entanglement production at a certain spot $x$ in the unitary circuit from time $t=0$ to $t=t_1$ is computed by the integral of the ``line-tension" along the curve that connects $x$ at $t=0$ and $t=t_1$. The appropriate form of the line-tension for the unitary circuit in a chaotic system is computed in \cite{2018arXiv180300089J}. We generalize the line-tension picture to include a single projective measurement that is performed at a certain time $t_p$ during the unitary time evolution from $t=0$ to $t=t_1$. The application of a projective measurement at a particular instant in time on a region of a unitary circuit leads to a removal of links in that region at that time.
This ``crack'' in spacetime is causally propagated by the unitary time evolution, creating a light-cone-like hole in the unitary circuit. Since there are no links to generate entanglement, the value of line-tension in this hole is zero. In this paper, we show that the entanglement calculated using this geometry of the unitary circuit agrees with that calculated by a holographic CFT in the coarse-grained limit.
\subsubsection*{The gravity dual of a finite-size projective measurement}
Such non-unitary processes are also interesting from a holographic point of view; in the language of the AdS/CFT correspondence, we introduce a projective measurement in the thermofield double state dual to the eternal black hole in AdS. An earlier work~\cite{Numasawa:2016emc} also studied the single-sided black hole at the instant of the projective measurement. The effect of a projective measurement can be understood holographically as the insertion of an end-of-the-world (EOW) brane by the AdS/BCFT correspondence \cite{Takayanagi:2011zk,Fujita:2011fp}. Namely, the erasure of entanglement by projection is equivalent to the erasure of a portion of spacetime in the gravity dual. We extend their analysis to the two-sided setting and consider the time evolution after a projective measurement in both cases. In particular, we focus on the time-evolution of a single-sided setup. 
The time evolution of the entanglement entropy after the measurement tells us how the effect of the projective measurement weakens for a given subsystem. From our results, we confirm that the light-cone structure of the EOW brane in the line-tension picture is consistent with the gravity description. Interestingly, the finite size effect of the EOW brane causes a quantitative change in the linear growth rate of the entanglement entropy.
\\

In addition to this new dynamical behavior of OEE, we find that finite-size projective measurements may cause the BOMI to vanish earlier than usual. This suggests that the finite-size projection reduces the non-local correlations present in the dual operator state of the time evolution operator.

\subsubsection*{Organization of this paper}
In Section \ref{Section:Preliminary}, we introduce the dual operator state of the time evolution operator with the finite-size projective measurement and then define the entanglement entropy and mutual information for this dual state.
We will also describe how to compute these quantities in the path-integral formalism.
In Section \ref{Section:OEE_and_BOMI}, we present the time-dependence of OEEs and BOMIs in $2$d holographic CFTs. In Section \ref{Section:line_tension_picture}, we propose the line tension model describing the operator state with the finite-size projective measurement. In Section \ref{Section:Gravitational_description}, we propose a gravity dual of the finite-size projective measurement.

\section{Preliminaries\label{Section:Preliminary}}
We consider the dynamical properties of a unitary time evolution with a Projective measurement of finite size in $2$d holographic CFTs. 
Let us suppose that the system is in a dynamical state where the density operator is defined as
\be\label{DensityMatrix1}
\rho(t_1,t_2)= \mathcal{N}^2e^{-(\epsilon_1+it_1)H} \ket{\mu} \bra{\mu}_{V} e^{-(\epsilon_2+it_2)H} \ket{\Psi_0}\bra{\Psi_0} e^{-(\epsilon_2-it_2)H}\ket{\mu} \bra{\mu}_V e^{-(\epsilon_1-it_1)H},
\ee
where $\ket{\Psi_0}$ is the initial state, $\mathcal{N}$ is a normalization constant such that $\Tr \rho(t_1,t_2)=1$ and $\epsilon_{i=1,2}$ are regularization parameters that guarantee that the OEE and BOMI are finite. In this non-equilibrium process, we start from the initial state $\ket{\Psi_0}$ and evolve it with the uniform Hamiltonian $H$ from $t=0$ to $t=t_2$. 
At $t=t_2$, we perform a projective measurement $P_V=\ket{\mu} \bra{\mu}_V$ on a spatial region $V$ of the system and then further evolve it in time from $t=t_1$ to $t=t_1+t_2$ with the uniform Hamiltonian. 
The region $V$ is defined as
\be
V=\left\{x\big{|} P_2 \le x \le P_1\right\}.
\ee
Immediately after the projective measurement, the reduced density matrix in the spatial region $V$ is projected onto the state $\ket{\mu}\bra{\mu}$.
\subsection{Operator state of a unitary time evolution operator with a projection}
Consider a time evolution operator with a projective measurement of finite size
\be \label{teo_w_pm}
K_{\mu}=e^{-(\epsilon_1+it_1)H} \ket{\mu} \bra{\mu}_V e^{-(\epsilon_2+it_2)H}.
\ee
By using the channel-state map \cite{2016JHEP...02..004H,nielsen_chuang_2010}, the dual state of the operator (\ref{teo_w_pm}) can be defined as
\be \label{kmu}
\ket{K_{\mu}}=\mathcal{N} \sum_{a,b}\bra{a}e^{-(\epsilon_1+it_1)H}\ket{\mu} \bra{\mu}_{V} e^{-(\epsilon_2+it_2)H}\ket{b} \ket{a}_1 \ket{b^*}_2,
\ee 
where the normalization constant $\mathcal{N}$ satisfies the condition
\be
\mathcal{N}^2 =\f{1}{\Tr\left[\ket{\mu}\bra{\mu}_V e^{-(\epsilon_1+\epsilon_2+i(t_1-t_2))H}\ket{\mu}\bra{\mu}_V e^{-(\epsilon_1+\epsilon_2-i(t_1-t_2))H}\right]}.
\ee
The state $\ket{b}$ is an eigenstate of the Hamiltonian $H$ and $\ket{\cdot^*}$ denotes the CPT conjugate of $\ket{\cdot}$.
The information on the projective measurement and the time evolution operator is encoded in the entanglement structure of this dual state which is also known as an operator state.
The operator state is defined on a doubled Hilbert space $\mathcal{H}=\mathcal{H}_1\otimes \mathcal{H}_2$.
The density matrix of the operator state $\ket{K_{\mu}}$ is given by 
\be\label{DensityMatrix2}
\begin{split}
\rho=&\ket{K_{\mu}} \bra{K_{\mu}} \\
=& \mathcal{N}^2 \sum_{a,b,c,d }\bra{a}e^{-(\epsilon_1+it_1)H}\ket{\mu} \bra{\mu}_{V} e^{-(\epsilon_2+it_2)H}\ket{b}\bra{d}e^{-(\epsilon_2-it_2)H}\ket{\mu}  \\
&\times\bra{\mu}_{V} e^{-(\epsilon_1-it_1)H}\ket{c} \ket{a}\bra{c}_1\otimes \ket{b^*}\bra{d^*}_2.
\end{split}
\ee
\subsection{Operator entanglement entropy (OEE)}
In this section, we will explain the method for computing the operator entanglement entropy (OEE) for single and double intervals. Divide the Hilbert into $A$ and its spatial complement to it, $\tilde{A}$. The interval $A$ is taken to be a single interval in a single Hilbert space which could be either $\mathcal{H}_1$ or $\mathcal{H}_2$. Let $X_1$ and $X_2$ denote endpoints of $A$ and assume that $X_1>X_2>0$. Define the reduced density matrix $\rho_{A}$ as $\Tr_{\overline{A}}\rho$ which is equal to
\be
\begin{split}
    \rho_{A} = \begin{cases}
    \mathcal{N}^2 \Tr_{\tilde{A}} \left[e^{-(\epsilon_1+it_1)H}\ket{\mu}\bra{\mu}_Ve^{-2\epsilon_2 H}\ket{\mu}\bra{\mu}_Ve^{-(\epsilon_1-it_1)H} \right] & \text{For~~} A \subset \mathcal{H}_1\\
    \mathcal{N}^2 \sum_{bd} \bra{d}e^{-(\epsilon_2-it_2)H}\ket{\mu}\bra{\mu}_Ve^{-2\epsilon_1 H}\ket{\mu}\bra{\mu}_Ve^{-(\epsilon_2+it_2)H} \ket{b} \Tr_{\tilde{A}}\left(\ket{b^*}\bra{d^*}\right) & \text{For~~} A \subset \mathcal{H}_2\\
    \end{cases},
\end{split}
\ee
where $\tilde{A}$ denotes the spatial complement to $A$ on $\mathcal{H}_{i=1,2}$, respectively. To compute the OEE for double intervals, let $A$ and $B$ denote sub-regions on $\mathcal{H}_{i=1}$ and $\mathcal{H}_{i=2}$ respectively.
Define $A\cup B$ as the union of $A$ and $B$ and the reduced density matrix associated with $A\cup B$ as
\be
\begin{split}
\rho_{A\cup B} &=\mathcal{N}^2 \sum_{a,b,c,d }\bra{a}e^{-(\epsilon_1+it_1)H}\ket{\mu} \bra{\mu}_{V} e^{-(\epsilon_2+it_2)H}\ket{b}\bra{d}e^{-(\epsilon_2-it_2)H}\ket{\mu} \bra{\mu}_{V} e^{-(\epsilon_1-it_1)H}\ket{c} \\
&~~~~~~\times\Tr_{\tilde{A}}\left(\ket{a}\bra{c}_1\right)\otimes \Tr_{\tilde{B}}\left(\ket{b^*}\bra{d^*}_2\right),
\end{split}
\ee
where $\tilde{B}$ and $\tilde{A}$ are the complements of $B$ and $A$ on $\mathcal{H}_{i=2}$ and $\mathcal{H}_{i=1}$ respectively.
The endpoints of $A$ are $X_1$ and $X_2$ and those of $B$ are $Y_1$ and $Y_2$. 
Assume that $X_1>X_2>0$ and $Y_1>Y_2>0$ for simplicity.
The OEEs for the single and double intervals are defined as
\be
\begin{split}
S_{A} =\lim_{n\rightarrow 1}S^{(n)}_{A}=-\Tr_A\rho_A \log{\rho_A},~~~~ S_{A \cup B}=\lim_{n\rightarrow 1}S^{(n)}_{A\cup B} =-\Tr_{A\cup B} \rho_{A\cup B}\log{\rho_{A\cup B}},   
\end{split}
\ee
where $S^{(n)}_{\alpha=A,A\cup B}$ is $n$-th R$\acute{e}$nyi operator entanglement entropy (ROEE) that is defined as 
\be
S^{(n)}_{\alpha}=\f{1}{1-n}\log{\left[\Tr_{\alpha}\left(\rho_{\alpha}\right)^n\right]}.
\ee
\subsubsection{Twist operator formalism}

\begin{figure}[htbp]
 \begin{minipage}{0.5\hsize}
  \begin{center}
   \includegraphics[width=70mm]{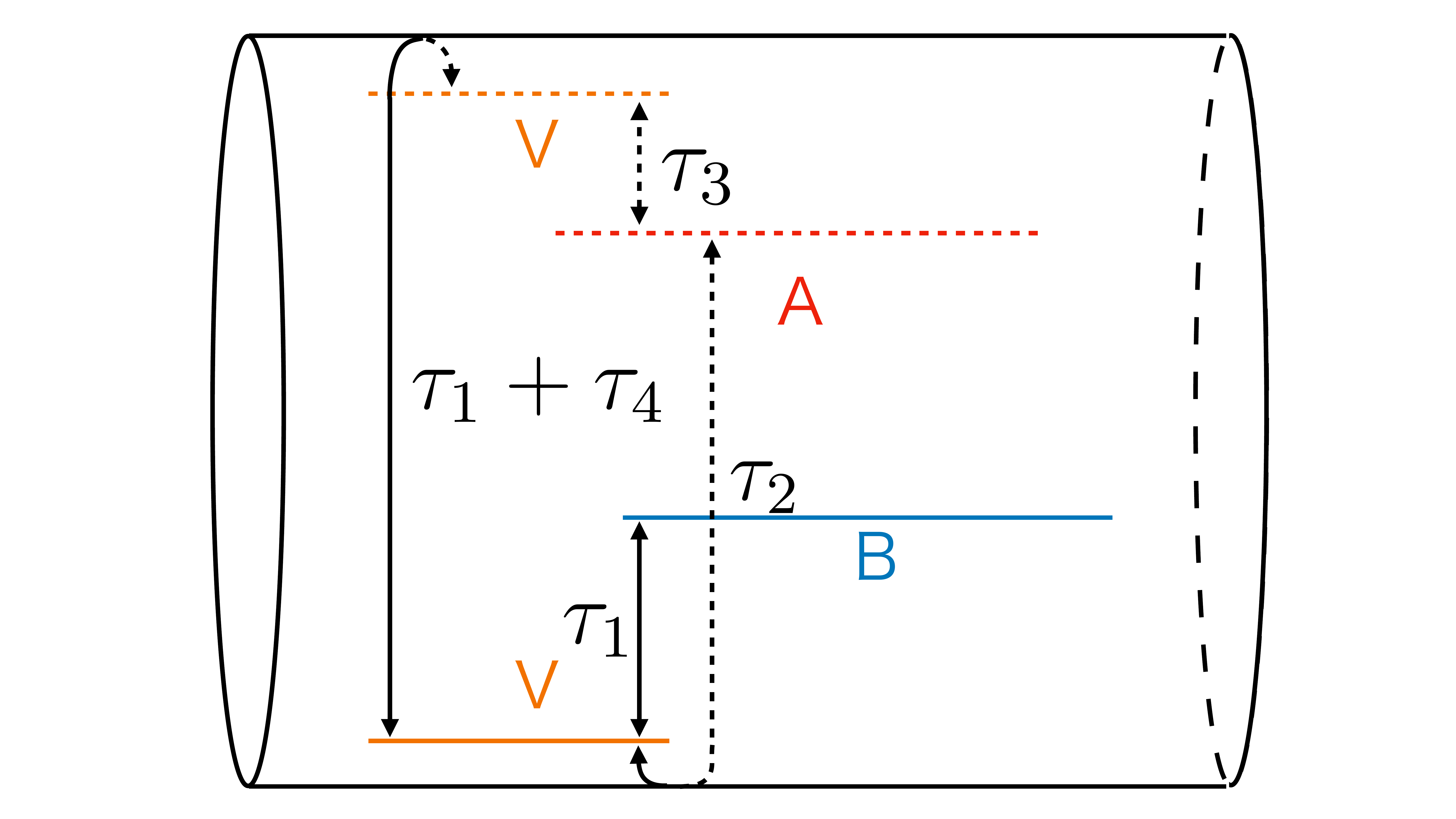}
   
   [a] Cylinder 
   
  \end{center}
 \end{minipage}
 \begin{minipage}{0.5\hsize}
  \begin{center}
   \includegraphics[width=70mm]{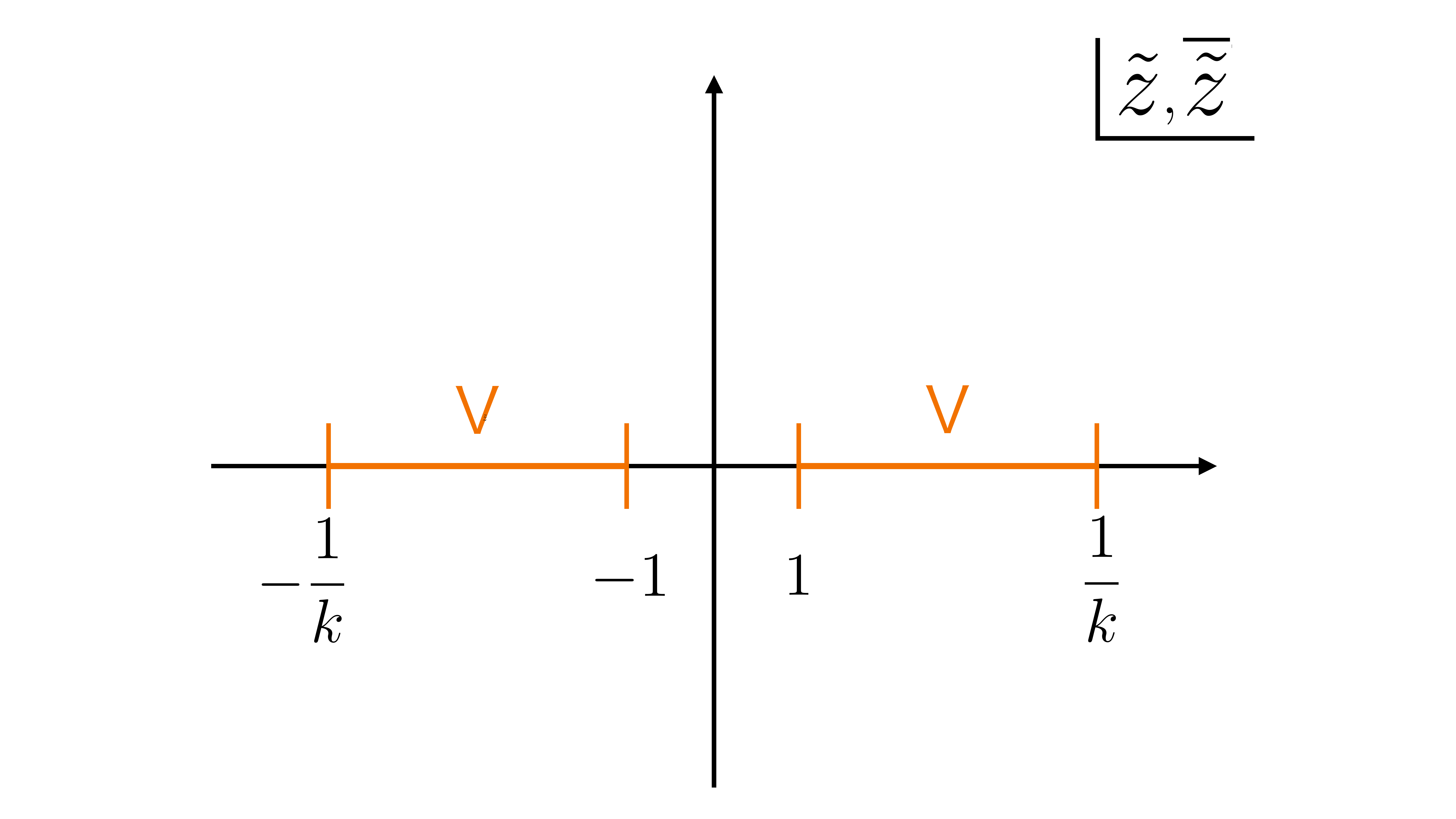}

  [b] Plane 

  \end{center}
 \end{minipage}
 \caption{ {\bf Left panel:} A cartoon of the thermal cylinder with two cuts. In this figure, the red and blue lines correspond to the subsystems $A$ and $B$ respectively. The orange lines illustrate the cuts corresponding to the projective measurement. {\bf Right panel:} A cartoon of the projective measurement on the plane. In this panel, we only show the location of the projective measurement. The parameter $k$ is defined by (\ref{parameter_k})}
 \label{Fig:loc_on_th_cy}
\end{figure}

\begin{figure}[htbp]
  \begin{tabular}{cc}
 \begin{minipage}{0.40\hsize}
  \begin{center}
   \includegraphics[width=50mm]{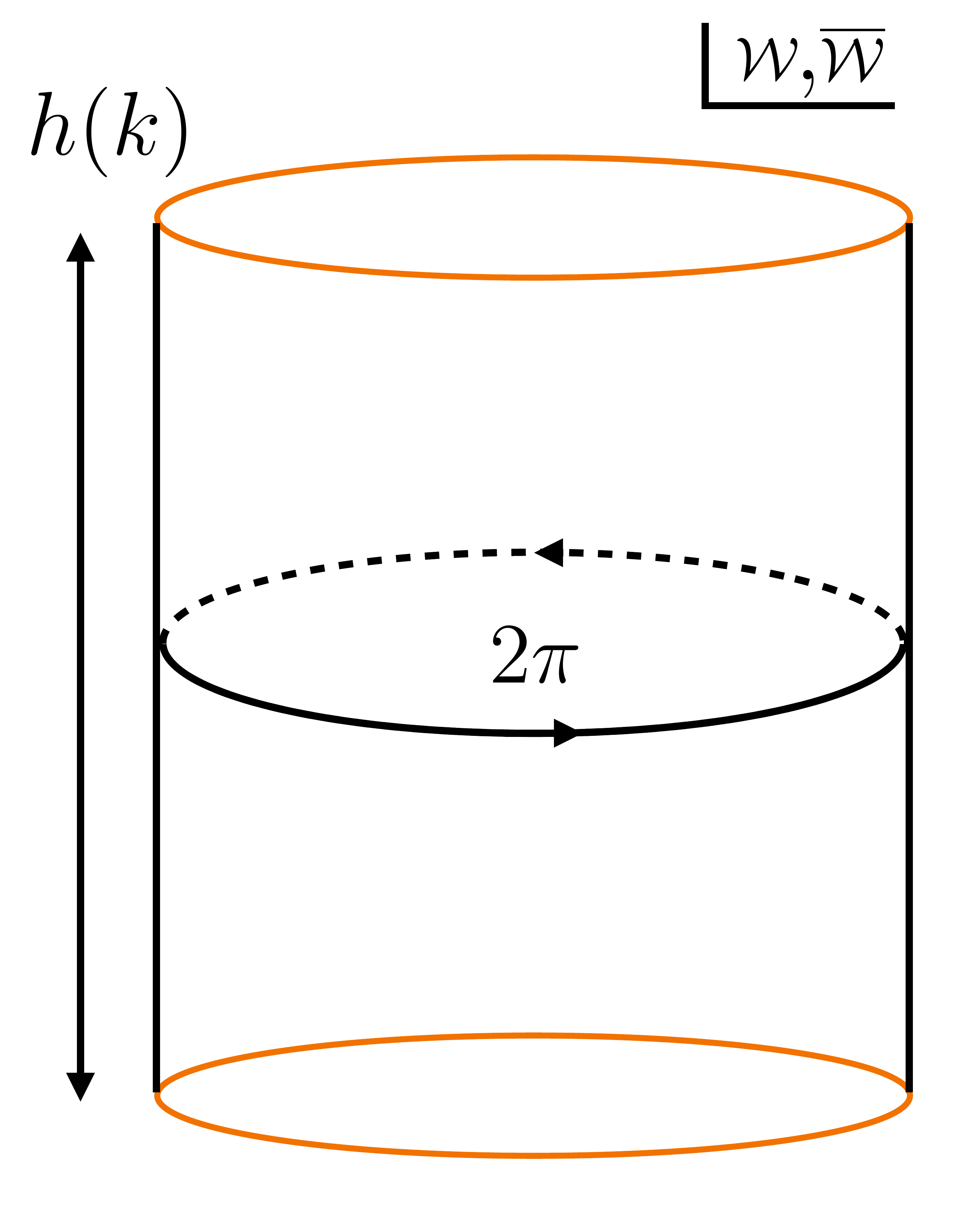}
   
   [a] Cylinder 
   
  \end{center}
 \end{minipage}

  \begin{minipage}{0.20\hsize}
  \begin{center}
   \includegraphics[width=30mm]{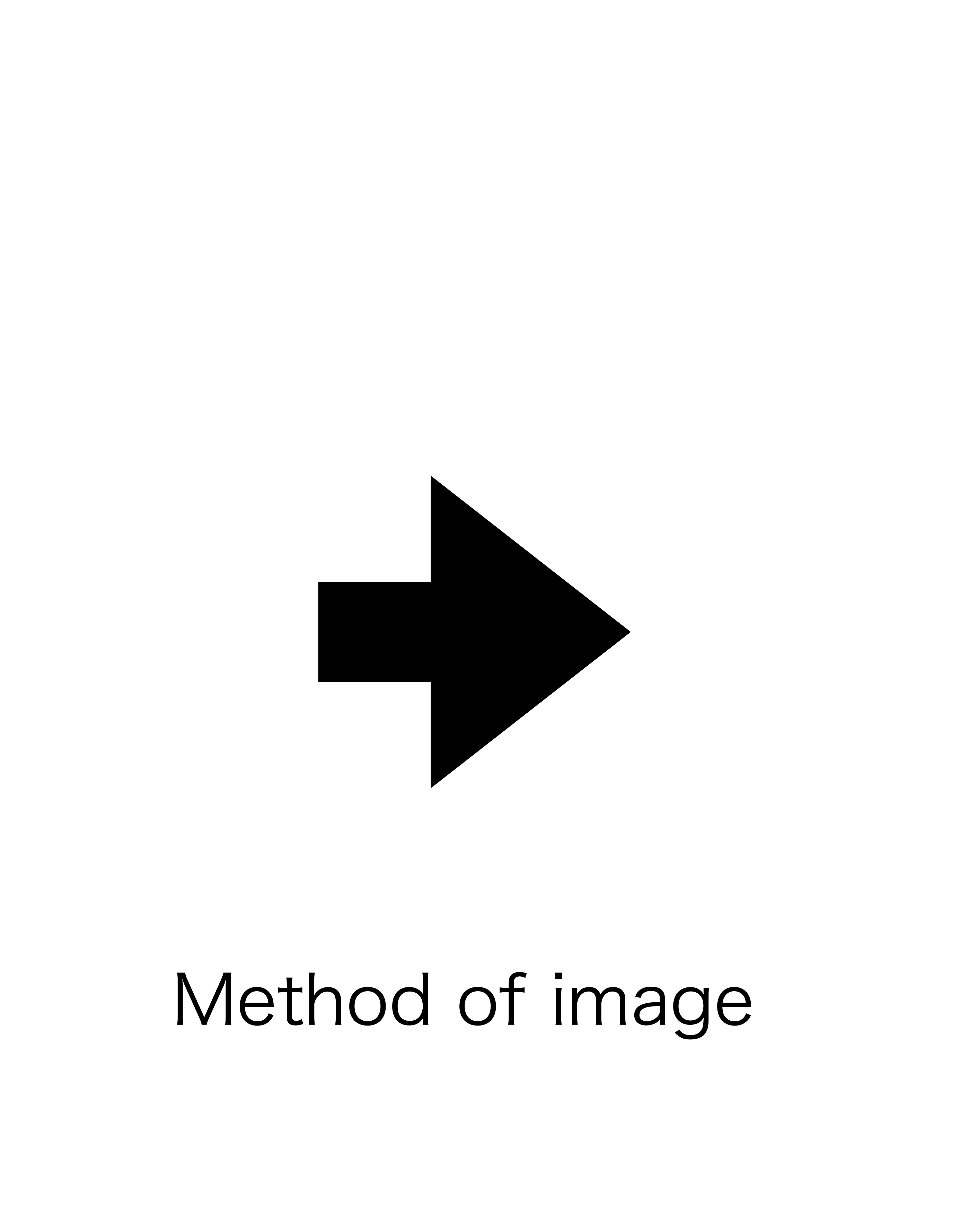}
  \end{center}
 \end{minipage}
 \begin{minipage}{0.40\hsize}
  \begin{center}
   \includegraphics[width=50mm]{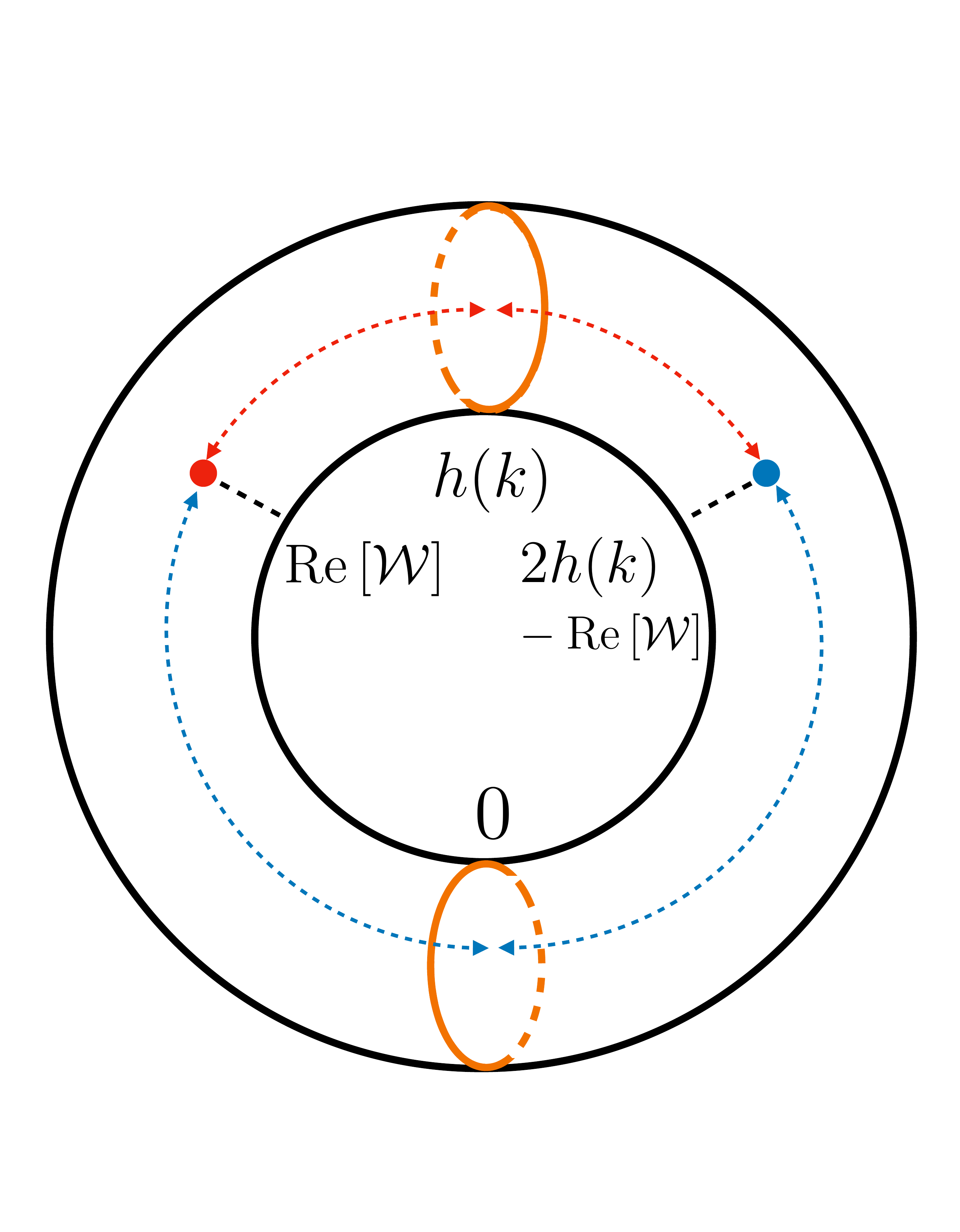}

  [b] Torus

  \end{center}
 \end{minipage}
\end{tabular}
 \caption{A cartoon of the cylinder and torus in the $\mathcal{W}$ and $\overline{\mathcal{W}}$ coordinate system. In Panel [a], the cylinder in $\mathcal{W}$ and $\overline{\mathcal{W}}$ coordinate system is shown. The regions where we perform the measurement is mapped onto the top and bottom of the cylinder which are shown here in orange. Panel [b] shows the real part of the location of the original operator (red point) and its image (blue point).} 
 \label{Fig:cylinder_and_torus}
\end{figure}

To compute the OEEs for single and double intervals in the twist operator formalism \cite{2004JSMTE..06..002C,Calabrese:2009qy}, define the Euclidean density matrix as
\be
\rho_{E}=\mathcal{N}^2_E\sum_{a,b,c,d}\bra{a}e^{-\tau_1 H}\ket{\mu}\bra{\mu}_Ve^{-\tau_2 H} \ket{b}\bra{d}e^{-\tau_3 H}\ket{\mu}\bra{\mu}_Ve^{-\tau_4 H}\ket{c} \ket{a}\bra{c}_1 \otimes \ket{b^*}\bra{d^*}_2,
\ee
where the normalization constant $\mathcal{N}_E$ obeys the condition
\be
\mathcal{N}^{-2}_E=\Tr \left(e^{-(\tau_1+\tau_4)H}\ket{\mu}\bra{\mu}_Ve^{-(\tau_2+\tau_3)H}\ket{\mu}\bra{\mu}_V\right).
\ee
In terms of the twist and anti-twist operators, the $n$-th ROEEs are given by
\be
\begin{split}
S^{(n)}_{A} &= \f{1}{1-n}\log{\left[\left \langle \sigma_n (X_1, \tau_{1}+\tau_3+\tau_4)\overline{\sigma}_n (X_2,\tau_{1}+\tau_3+\tau_4)  \right \rangle_{\sum_{i=1,2,3,4}\tau_i, \mu} \right]},\\
S^{(n)}_{B} &= \f{1}{1-n}\log{\left[\left \langle  \overline{\sigma}_n (Y_1, \tau_1) \sigma_n (Y_2, \tau_1)  \right \rangle_{\sum_{i=1,2,3,4}\tau_i, \mu} \right]},\\
S^{(n)}_{A\cup B} &= \f{1}{1-n}\log{\left[\left \langle \sigma_n (X_1, \tau_{1}+\tau_3+\tau_4)\overline{\sigma}_n (X_2,\tau_{1}+\tau_3+\tau_4) \overline{\sigma}_n (Y_1, \tau_1) \sigma_n (Y_2, \tau_1)  \right \rangle_{\sum_{i=1,2,3,4}\tau_i, \mu} \right]},
\end{split}
\ee
where the two and four point functions are defined on the thermal cylinder (see Fig. \ref{Fig:loc_on_th_cy}). 
Here, the complex coordinates are 
\be
\begin{split}
    &w_{i=1,2}=X_{i=1,2}+i(\tau_1+\tau_3+\tau_4),~ \overline{w}_{i=1,2}=X_{i=1,2}-i(\tau_1+\tau_3+\tau_4),\\
    &w_{3}=Y_1+i\tau_1,~\overline{w}_3=Y_1-i\tau_1,~w_{4}=Y_2+i\tau_1,~\overline{w}_4=Y_2-i\tau_1,\end{split}
\ee
where we assumed that $\tau_1+\tau_4=\tau_2+\tau_3=2\epsilon$.
The locations of the projective measurement along the Euclidean time direction are $\tau=0$ and $\tau=\tau_1+\tau_4$. The subsystems $B$ and $A$ are situated at $\tau=\tau_1$ and $\tau_1+\tau_3+\tau_4$ respectively.
Map the cylinder to the plane by using the exponential maps
\be
\tilde{z}=e^{\f{\pi(w-P_2)}{2\epsilon}}, \overline{\tilde{z}}=e^{\f{\pi(\overline{w}-P_2)}{2\epsilon}}
\ee
The projective measurement at $\tau =0$ is mapped to $e^{\f{\pi P_2}{2\epsilon}}\le\Re[\tilde{z}], \Re[\overline{\tilde{z}}]\le e^{\f{\pi P_1}{2\epsilon}}$ while the measurement at $\tau =\tau_1+\tau_4$ is mapped to  $-e^{\f{\pi P_1}{2\epsilon}}\le\Re[\tilde{z}], \Re[\overline{\tilde{z}}]\le -e^{\f{\pi P_2}{2\epsilon}}$ as in panel [b] of Fig \ref{Fig:loc_on_th_cy}. 
Thus, in this coordinate system, the locations of intervals are symmetric about $\Re[\tilde{z}]=0$ and $\Re[\overline{\tilde{z}}]=0$. We perform one last conformal map from the plane to a finite cylinder \cite{2015PhRvB..92g5108R,2015arXiv150307771R,2016JSMTE..06.3109R,Kusuki_2020} as in Fig. \ref{Fig:cylinder_and_torus}:
\be \label{cm_p_ft}
\begin{split}
&\mathcal{W} (\tilde{z}) =\f{h(k)}{2}+\f{h(k)\cdot sn^{-1}(\tilde{z},k^2)}{2K(k^2)}, ~\overline{\mathcal{W} }(\overline{\tilde{z}}) =\f{h(k)}{2}+\f{h(k)\cdot sn^{-1}(\overline{\tilde{z}},k^2)}{2K(k^2)}, \\
&h(k) = \f{2\pi K(k^2)}{K(1-k^2)}, sn^{-1}(\tilde{z},k^2) =\int^{\tilde{z}}_0\f{dx}{\sqrt{(1-x^2)(1-k^2x^2)}}, K(t^2) =\int^{1}_0\f{dx}{\sqrt{(1-x^2)(1-t^2x^2)}},
\end{split}
\ee 
where the parameter $k$ is defined as
\be \label{parameter_k}
\f{1}{k}=e^{\f{\pi (P_1-P_2)}{2\epsilon}}>1.
\ee 
Therefore, $k$ is smaller than $1$. 
The height of the finite cylinder is $h(k)$ and its periodicity is $\Im[\mathcal{W}]\sim \Im[\mathcal{W}]+2\pi$ and $\Im[\overline{\mathcal{W}}]\sim \Im[\overline{\mathcal{W}}]+2\pi$.
The locations of the projective measurement in $\mathcal{W}$ and $\overline{\mathcal{W}}$ coordinates are $\Re\left[\mathcal{W}\right]=\Re\left[\overline{\mathcal{W}}\right]=0, h(k)$. 
Finally, we arrive at the expressions for the $n$-th ROEEs for the single and double intervals which are given by
\be
\begin{split}
S^{(n)}_{A} &= \f{1}{1-n}\log{\left[\left \langle \sigma_n (w_1, \overline{w}_1)\overline{\sigma}_n (w_2, \overline{w}_2)  \right \rangle_{2(\epsilon_1+\epsilon_2), \mu} \right]}\\
&=\f{h_{n}}{1-n}\log{\left[\left(\Pi_{i=1,2}\f{d\mathcal{W}(w_i)}{dw_i}\f{d\overline{\mathcal{W}}(\overline{w}_i)}{d\overline{w}_i}\right)\right]}+\f{1}{1-n}\log{\left[\left \langle \sigma_n (\mathcal{W}_1, \overline{\mathcal{W}}_1)\overline{\sigma}_n (\mathcal{W}_2, \overline{\mathcal{W}}_2)  \right \rangle_{\text{Cylinder}} \right]}, \\
S^{(n)}_{B} &= \f{1}{1-n}\log{\left[\left \langle \overline{\sigma}_n (w_3, \overline{w}_3)\sigma_n (w_4, \overline{w}_4)  \right \rangle_{2(\epsilon_1+\epsilon_2), \mu} \right]}\\
&=\f{h_{n}}{1-n}\log{\left[\left(\Pi_{i=3,4}\f{d\mathcal{W}(w_i)}{dw_i}\f{d\overline{\mathcal{W}}(\overline{w}_i)}{d\overline{w}_i}\right)\right]}+\f{1}{1-n}\log{\left[\left \langle \overline{\sigma}_n (\mathcal{W}_3, \overline{\mathcal{W}}_3)\sigma_n (\mathcal{W}_4, \overline{\mathcal{W}}_4)  \right \rangle_{\text{Cylinder}} \right]}, \\
S^{(n)}_{A \cup B} &= \f{1}{1-n}\log{\left[\left \langle \sigma_n (w_1, \overline{w}_1)\overline{\sigma}_n (w_2, \overline{w}_2) \overline{\sigma}_n (w_3, \overline{w}_3)\sigma_n (w_4, \overline{w}_4) \right \rangle_{2(\epsilon_1+\epsilon_2), \mu} \right]}\\
&=\f{h_{n}}{1-n}\log{\left[\left(\Pi_{i=1,4}\f{d\mathcal{W}(w_i)}{dw_i}\f{d\overline{\mathcal{W}}(\overline{w}_i)}{d\overline{w}_i}\right)\right]}\\
&+ \f{1}{1-n}\log{\left[\left \langle \sigma_n (\mathcal{W}_1, \overline{\mathcal{W}}_1)\overline{\sigma}_n (\mathcal{W}_2, \overline{\mathcal{W}}_2) \overline{\sigma}_n (\mathcal{W}_3, \overline{\mathcal{W}}_3)\sigma_n (\mathcal{W}_4, \overline{\mathcal{W}}_4) \right \rangle_{\text{Cylinder}}\right]},
\end{split}
\ee
where $h_n =\f{c}{24n}(n^2-1)$ is the conformal dimensions of the twist operators. 
The symbol $\left\langle \cdot \right \rangle_{\text{cylinder}}$ denotes the non-universal piece of the OEE which depends on the particular $2$d CFT considered.

\subsubsection*{Non-universal piece of (OEE) in $2$d holographic CFT}
In the $2$d holographic CFT, we will compute the non-universal piece using the method of images. The image points of the original twist operators are located at $\Re[\mathcal{W}]=\Re[\overline{\mathcal{W}}]=0,h(k)$ on the torus which is periodic under the identification $\Im[\mathcal{W}] \sim \Im[\mathcal{W}] + 2\pi$ and $\Re[\mathcal{W}] \sim \Re[\mathcal{W}] + 2h(k)$. The location of the image corresponding to $\mathcal{O}(\mathcal{W}, \overline{\mathcal{W}})$ is 
\be \label{bc1}
\begin{split}
    \tilde{\mathcal{W}}=2nh(k)-\overline{\mathcal{W}}, \overline{\tilde{\mathcal{W}}}=2nh(k)-\mathcal{W},
\end{split}
\ee
where $n$ is an integer.
This location (\ref{bc1}) preserves the conformal boundary condition.
In the original cylindrical coordinates $w$ and $\overline{w}$, the location of the images are given by
\be
\tilde{w}=\overline{w}+i(4m+2)\epsilon, \overline{\tilde{w}}=w+i(4m+2)\epsilon,
\ee
where $m$ is an integer.\\

\if[0]
*********************************************\textcolor{red}{MN: I will remove this part later}*********************************************

\textcolor{red}{The location of image corresponding to $\mathcal{O}(\mathcal{W}, \overline{\mathcal{W}})$ is given by
\be \label{bc1}
\begin{split}
    \tilde{\mathcal{W}}=i\Im[\mathcal{W}]+2h(k)-\Re[\mathcal{W}]=2h(k)-\overline{\mathcal{W}}, \overline{\tilde{\mathcal{W}}}=-i\Im[\mathcal{W}]+2h(k)-\Re[\mathcal{W}]=2h(k)-\mathcal{W}.
\end{split}
\ee
The location of image corresponding to other boundary condition is 
\be \label{bc2}
\tilde{\mathcal{W}} =h(k)+\mathcal{W},~~\overline{\tilde{\mathcal{W}}} =h(k)+\overline{\mathcal{W}}.
\ee
Let $\tilde{w}$ and $\overline{w}$ denote the location of the images in $w$ and $\overline{w}$.
The location of the image corresponding to (\ref{bc1}) is 
\be
\tilde{w}=\overline{w}, \overline{\tilde{w}}=\overline{w}.
\ee
This corresponds to the boundary condition where the mirror exists at $\text{Im}[w]=0$.
The location of the image corresponding to (\ref{bc2}) is 
\be
\tilde{w}=w\pm 2i\epsilon, \overline{\tilde{w}}=\overline{w}\pm 2i\epsilon.
\ee
This corresponds to the boundary condition which is the periodic boundary condition with the period, $w \simeq w+2i\epsilon$ and $\overline{w} \simeq \overline{w}+2i\epsilon$.$\rightarrow$ {\bf We should fix this part.}}

*********************************************\textcolor{red}{MN: I will remove this part later}*********************************************\\
\fi
In terms of twist and anti-twist operators, the non-universal pieces of $S_A$, $S_B$, and $S_{A\cup B}$ are respectively given by
\be
\begin{split}
    &\f{1}{1-n}\log{\left[\left \langle \sigma_n (\mathcal{W}_1, \overline{\mathcal{W}}_1)\overline{\sigma}_n (\mathcal{W}_2, \overline{\mathcal{W}}_2)  \right \rangle_{\text{Cylinder}} \right]}\\
    &~~~~~~~~~~~~~~~~~~~~~~~~~~~=\f{1}{2(1-n)}\log{\left[\left \langle \sigma_n (\mathcal{W}_1, \overline{\mathcal{W}}_1)\overline{\sigma}_n (\mathcal{W}_2, \overline{\mathcal{W}}_2)   \overline{\sigma}_n (\tilde{\mathcal{W}}_1, \overline{\tilde{\mathcal{W}}}_1)\sigma_n (\tilde{\mathcal{W}}_2, \overline{\tilde{\mathcal{W}}}_2)\right \rangle_{\text{Torus}} \right]},\\
    &\f{1}{1-n}\log{\left[\left \langle \overline{\sigma}_n (\mathcal{W}_3, \overline{\mathcal{W}}_3)\sigma_n (\mathcal{W}_4, \overline{\mathcal{W}}_4)  \right \rangle_{\text{Cylinder}} \right]}\\
    &~~~~~~~~~~~~~~~~~~~~~~~~~~~=\f{1}{1-n}\log{\left[\left \langle \overline{\sigma}_n (\mathcal{W}_3, \overline{\mathcal{W}}_3)\sigma_n (\mathcal{W}_4, \overline{\mathcal{W}}_4)  \sigma_n (\tilde{\mathcal{W}}_3, \overline{\tilde{\mathcal{W}}}_3)\overline{\sigma}_n (\tilde{\mathcal{W}}_4, \overline{\tilde{\mathcal{W}}}_4)\right \rangle_{\text{Torus}} \right]},\\
    &\f{1}{1-n}\log{\left[\left \langle \sigma_n (\mathcal{W}_1, \overline{\mathcal{W}}_1)\overline{\sigma}_n (\mathcal{W}_2, \overline{\mathcal{W}}_2) \overline{\sigma}_n (\mathcal{W}_3, \overline{\mathcal{W}}_3)\sigma_n (\mathcal{W}_4, \overline{\mathcal{W}}_4) \right \rangle_{\text{Cylinder}}\right]}\\
     &=\f{1}{1-n}\log\Bigg{[}\big  \langle \sigma_n (\mathcal{W}_1, \overline{\mathcal{W}}_1)\overline{\sigma}_n (\mathcal{W}_2, \overline{\mathcal{W}}_2) \overline{\sigma}_n (\mathcal{W}_3, \overline{\mathcal{W}}_3)\sigma_n (\mathcal{W}_4, \overline{\mathcal{W}}_4) \\
     &~~~~~~~~~~~~~~~~~~~~~~~~~\times\overline{\sigma}_n (\tilde{\mathcal{W}}_1, \overline{\tilde{\mathcal{W}}}_1)\sigma_n (\tilde{\mathcal{W}}_2, \overline{\tilde{\mathcal{W}}}_2) \sigma_n (\tilde{\mathcal{W}}_3, \overline{\tilde{\mathcal{W}}}_3)\overline{\sigma}_n (\tilde{\mathcal{W}}_4, \overline{\tilde{\mathcal{W}}}_4)\big \rangle_{\text{Torus}}\Bigg]\\
\end{split}.
\ee
In the von Neumann limit, $n\rightarrow 1$ and the OEEs are given by the summation of geodesic lengths, $\mathcal{L}_i$, subject to the homology condition:
\be \label{non-uni-hol}
S_{\mathcal{V}=A,B,A\cup B} = \f{1}{2}\times\f{1}{6c} \sum_{i}\mathcal{L}_i.
\ee
Let us make a comment on the gravity dual these geodesic lengths in (\ref{non-uni-hol}) are calculated. Define the moduli parameters of the torus as 
\be
\tau=\f{ih(k)}{\pi},~ \overline{\tau}= \f{-ih(k)}{\pi},
\ee
where these parameters are purely imaginary. 
As in \cite{Witten:1998zw}, the gravity dual of the torus depends on the moduli parameters.
In the high temperature region ($h(k)<\pi$), the gravity dual to the thermal state on the circle is a BTZ black hole, while in the low temperature region ($h(k)>\pi$), the gravitational dual is a thermal AdS$_3$. 
Since the parameter $h(k)$ depends on the size of the projective measurement as in (\ref{parameter_k}), a phase transition of the gravity dual is induced by varying the size of the region where we perform the projective measurement, so the geodesic length depends on the size of the region where the measurement is implemented.

\subsubsection*{Gravity dual for a small measurement region, $1\gg\f{P_1-P_2}{\epsilon}>0$}
Let us consider the gravity dual in the small measurement region where $1\gg\f{P_1-P_2}{\epsilon}>0$. 
In this regime, the moduli parameters are given by
\be
\tau(k) \approx \f{2i}{\pi}\log{\left(\f{16\epsilon}{\pi (P_1-P_2)}\right)}, \overline{\tau}(k) \approx \f{-2i}{\pi}\log{\left(\f{16\epsilon}{\pi (P_1-P_2)}\right)}, h(k) \approx 2\log{\left(\f{16\epsilon}{\pi (P_1-P_2)}\right)} \gg 1.
\ee
Therefore, the gravity dual is given by the thermal AdS$_3$. 
\if[0]
Then, the correlation function of twist and anti twist operators is given by the area of minimal surface in the thermal AdS$_3$ geometry\footnote{I should fix this explanation.}:
\be
\begin{split}
\left \langle \sigma_n (\mathcal{W}, \overline{\mathcal{W}})\bar{\sigma}_n (\mathcal{W}', \overline{\mathcal{W}}')\right \rangle \approx 
\sum_{n=-\infty}^{\infty}\left(\f{\epsilon^2}{4}\right)^{2h_{n}}\f{1}{\left|\sinh{\left(\f{\mathcal{W}-\mathcal{W}'+2nh(k)}{2}\right)}\right|^{4h_{n}}}
\end{split}
\ee
\fi
\subsubsection*{The gravity dual in the large measurement limit, $\f{P_1-P_2}{\epsilon}\gg 1$}
In the large measurement region where  $\f{P_1-P_2}{\epsilon}\gg 1$, the moduli parameters are approximated by
\be
\tau \approx \f{2 i \epsilon}{P_1-P_2}, ~\bar{\tau}\approx \f{-2 i \epsilon}{P_1-P_2}, h(k) \approx \f{2\pi \epsilon}{P_1-P_2} \ll 1.
\ee
Therefore, the gravity dual in this region is the BTZ black hole.

\if[0]
Then, the correlation function of twist and anti twist operators is given by the area of minimal surface in BTZ black hole geometry\footnote{I should fix this explanation.}:
\be \label{atfunction}
\begin{split}
\left \langle \sigma_n (\mathcal{W}, \overline{\mathcal{W}})\bar{\sigma}_n (\mathcal{W}', \overline{\mathcal{W}}')\right \rangle \approx 
\sum_{m=-\infty}^{\infty}\left(\f{a h(k)}{4\pi}\right)^{4h_n}\f{1}{\left| \sin{\left[ \f{\pi(\mathcal{W}-\mathcal{W}'+i 2m\pi )}{2h(k)}\right]}\right|^{4h_n}}
\end{split}
\ee

\fi

\subsection{Location of the twist 
operators after analytic continuation}
Let us now turn to a detailed discussion of the time-dependence of the twist and anti-twist operators on the torus after an analytic continuation to Lorentzian signature.
Define the spatial and Euclidean time coordinates of the operator as
\be
\begin{split}
&X^{\text{effective}}_{x}=\f{\mathcal{W}_x-\overline{\mathcal{W}}_x}{2i},~\tilde{X}^{\text{effective}}_{x}=\f{\tilde{\mathcal{W}}_x-\overline{\tilde{\mathcal{W}}}_x}{2i}=-X^{\text{effective}}_x,\\
&\tau^{\text{effective}}_x=\f{\mathcal{W}_x+\overline{\mathcal{W}}_x}{2},\tilde{\tau}^{\text{effective}}_x=2nh-\f{\tilde{\mathcal{W}}_x+\overline{\tilde{\mathcal{W}}}_x}{2},
\end{split}
\ee
and let $\tau^{\text{Real}}_{x}$ and $\tau^{\text{Imaginary}}_{x}$ be the real and imaginary parts of $\tau_x$ respectively.
Then, analytically continue the Euclidean time intervals to
\be \label{ac}
\begin{split}
    \tau_1=\epsilon+it_1,~ \tau_2=\epsilon+it_2,~ \tau_3=\epsilon-it_2,~ \tau_4=\epsilon-it_1,
\end{split}
\ee
where we set $\epsilon_1=\epsilon_2=\epsilon$, for simplicity.  
After performing this analytic continuation, $sn^{-1}(\tilde{z},k^2)$ becomes a purely imaginary function, so $X^{\text{effective}}_x$ and $\tilde{X}^{\text{effective}}_x$ are now real functions.
On the other hand, the analytically-continued variables, $\tau^{\text{effective}}_x$ and $\overline{\tau}^{\text{effective}}_x$, are complex functions. Let $\tau^{\text{Real}}_x$ and $\tilde{\tau}^{\text{Real}}_x$ be their corresponding real parts and $\tau^{\text{Imaginary}}_x$ and $\tilde{\tau}^{\text{Imaginary}}_x$ be their corresponding imaginary parts.
These variables are given by
\be
\begin{split}
&X^{\text{effective}}_{i=1,2}=-\tilde{X}^{\text{effective}}_{i=1,2}=-\f{h(k)\left[F(u_{i=1,2},k^2)+F(\overline{u}_{i=1,2},k^2)\right]}{4 K(k^2)},\\
&X^{\text{effective}}_{i=3,4}=-X^{\text{effective}}_{i=3,4}=\f{h(k)\left[F(u_{i=3,4},k^2)+F(\overline{u}_{i=3,4},k^2)\right]}{4 K(k^2)},\\
&\tau^{\text{Real}}_{i=1,2,3,4}=\f{h(k)}{2},\tilde{\tau}^{\text{Real}}_{i=1,2,3,4}=\f{(4n-1)h(k)}{2}, \nonumber \\
&\tau^{\text{Imaginary}}_{i=1,2}=-\tilde{\tau}^{\text{Imaginary}}_{i=1,2}=-\f{h(k)\left[F(u_{i=1,2},k^2)-F(\overline{u}_{i=1,2},k^2)\right]}{4 K(k^2)},\\
&\tau^{\text{Imaginary}}_{i=3,4}=-\tilde{\tau}^{\text{Imaginary}}_{i=3,4}=\f{h(k)\left[F(u_{i=3,4},k^2)-F(\overline{u}_{i=3,4},k^2)\right]}{4 K(k^2)},\nonumber \\ 
&F(u,k^2)=\int^{u}_0\f{dx}{\sqrt{(1+x^2)(1+k^2x^2)}}
\end{split}
\ee
where $\tilde{z}_{i=1,2}=-i u_{i=1,2}$, $\overline{\tilde{z}}_{i=1,2}=i \overline{u}_{i=1,2}$, $\tilde{z}_{i=3,4}=i u_{i=3,4}$, and $\overline{\tilde{z}}_{i=3,4}=-i \overline{u}_{i=3,4}$.
In the following, let us take the range of the real part of $\mathcal{W}$ to be $0 \le \Re[\mathcal{W}] < 2h(k)$.
Then, in terms of $\mathcal{W}$ and $\overline{\mathcal{W}}$, the locations of the original operators and their images are given by
\be
\begin{split}
   & \mathcal{W}_x=\f{h(x)}{2}+i \left(\tau^{\text{Imaginary}}_{x}+X^{\text{effective}}_x\right),~\overline{\mathcal{W}}_x=\f{h(x)}{2}+i \left(\tau^{\text{Imaginary}}_{x}-X^{\text{effective}}_x\right), \\
   &\tilde{\mathcal{W}}_x=\f{3h(x)}{2}-i \left(\tau^{\text{Imaginary}}_{x}+X^{\text{effective}}_x\right),~\overline{\tilde{\mathcal{W}}}_x=\f{3h(x)}{2}-i \left(\tau^{\text{Imaginary}}_{x}-X^{\text{effective}}_x\right). \\
\end{split}
\ee
Thus, the original operators are placed at $\Re[\mathcal{W}]=\Re[\overline{\mathcal{W}}]=\f{h(x)}{2}$, while their images are positioned on $\Re[\mathcal{W}]=\Re[\overline{\mathcal{W}}]=\f{3h(x)}{2}$.
Only their locations along $\Im[\mathcal{W}]$ and $\Im[\overline{\mathcal{W}}]$ depend on the times $t_1$ and $t_2$, and their trajectories are determined by
\be
\begin{split}
   &\Im[\mathcal{W}_{i=1,2}]=-\Im[\overline{\tilde{\mathcal{W}}}_{i=1,2}]=-\f{h(k)F\left(u_{i=1,2},k^2\right)}{2K(k^2)}, \Im[\overline{\mathcal{W}}_{i=1,2}]=-\Im[\tilde{\mathcal{W}}_{i=1,2}]=\f{h(k)F\left(\overline{u}_{i=1,2},k^2\right)}{2K(k^2)},\\
    &\Im[\mathcal{W}_{i=3,4}]=-\Im[\overline{\tilde{\mathcal{W}}}_{i=3,4}]=\f{h(k)F\left(u_{i=3,4},k^2\right)}{2K(k^2)}, \Im[\overline{\mathcal{W}}_{i=3,4}]=-\Im[\tilde{\mathcal{W}}_{i=3,4}]=-\f{h(k)F\left(\overline{u}_{i=3,4},k^2\right)}{2K(k^2)}.\\
\end{split}
\ee
The asymptotic behaviors of $\mathcal{W}$ and $\overline{\mathcal{W}}$ are reported in Appendix \ref{sec:asym-beh-large-limit}. 
As in Fig.\ \ref{trajectory_of_operators}, the imaginary parts of $\mathcal{W}$ are monotonically decreasing functions of $t_1$ or $t_2$, while those of are the monotonically increasing functions.
\if[0]
$\Im[\overline{\mathcal{W}}_{i=1,2}]=-\Im[\tilde{\mathcal{W}}_{i=1,2}]$ and $\Im[\mathcal{W}_{i=3,4}]=-\Im[\overline{\tilde{\mathcal{W}}}_{i=3,4}]$ are monotonically increasing functions of $t_1$ while the imaginary parts $\Im[\mathcal{W}_{i=1,2}]=-\Im[\tilde{\mathcal{W}}_{i=1,2}]$ and $\Im[\overline{\mathcal{W}}_{i=1,2}]=-\Im[\overline{\tilde{\mathcal{W}}}_{i=1,2}]$ are monotonically decreasing functions of $t_2$.
\fi
In the large $t_1$-limit, $t_1 \rightarrow \infty$, $\Im[\mathcal{W}_{i=3,4}]=-\Im[\overline{\tilde{\mathcal{W}}}_{i=3,4}]$ is approximately given by $0$, while $ \Im[\overline{\mathcal{W}}_{i=3,4}]=-\Im[\tilde{\mathcal{W}}_{i=3,4}]$ is approximately given by $-\pi$.
In the large $t_2$-limit where $t_2 \rightarrow \infty$, $\Im[\mathcal{W}_{i=1,2}]=-\Im[\overline{\tilde{\mathcal{W}}}_{i=1,2}]$ is approximately given by $-\pi$, while $\Im[\overline{\mathcal{W}}_{i=1,2}]=-\Im[\tilde{\mathcal{W}}_{i=1,2}]$ is approximately given by $0$.

\begin{figure}[htbp]
  \begin{tabular}{cc}
 \begin{minipage}{0.50\hsize}
  \begin{center}
   \includegraphics[width=75mm]{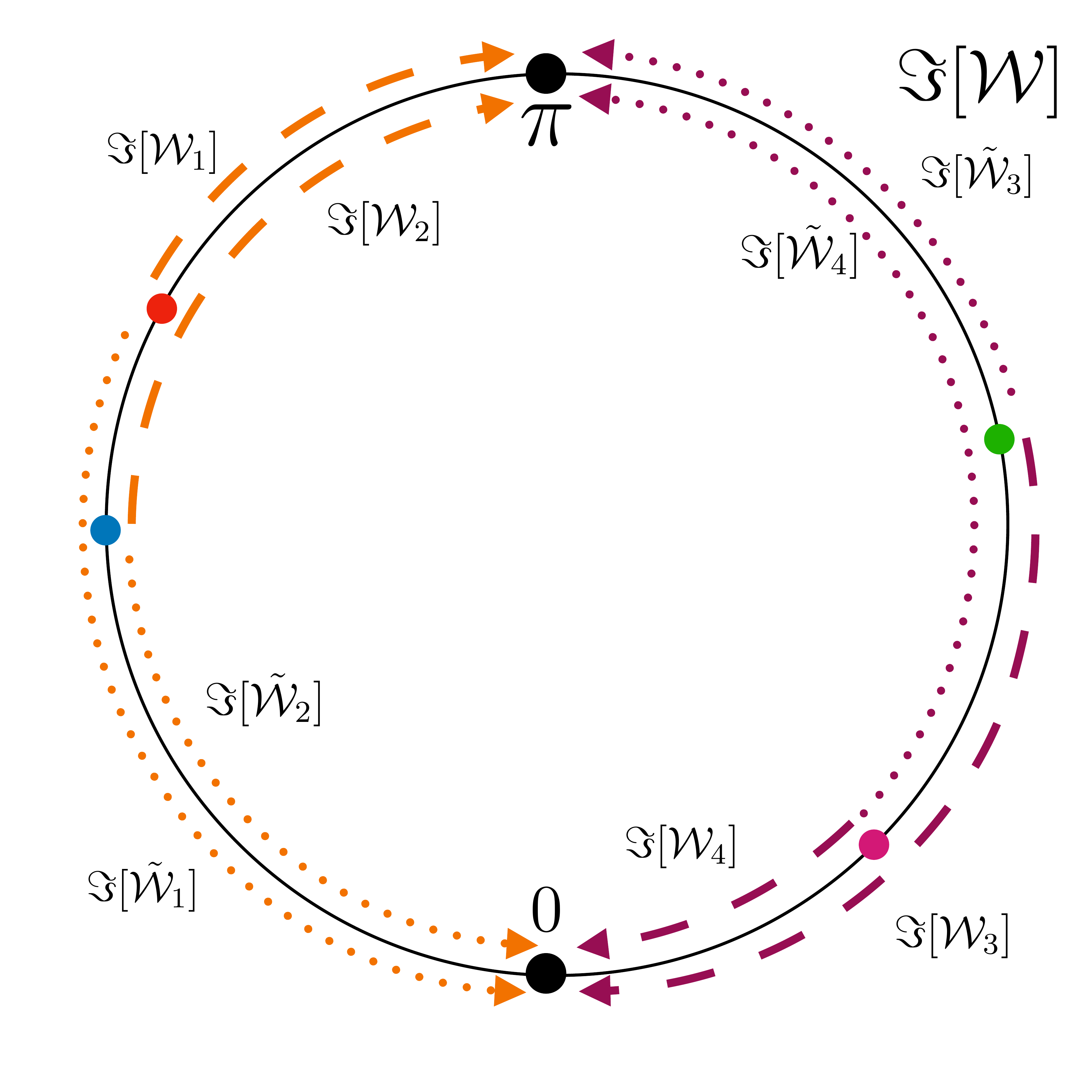}
   
   [a] $\Im[\mathcal{W}]$
   
  \end{center}
 \end{minipage}

 \begin{minipage}{0.50\hsize}
  \begin{center}
   \includegraphics[width=75mm]{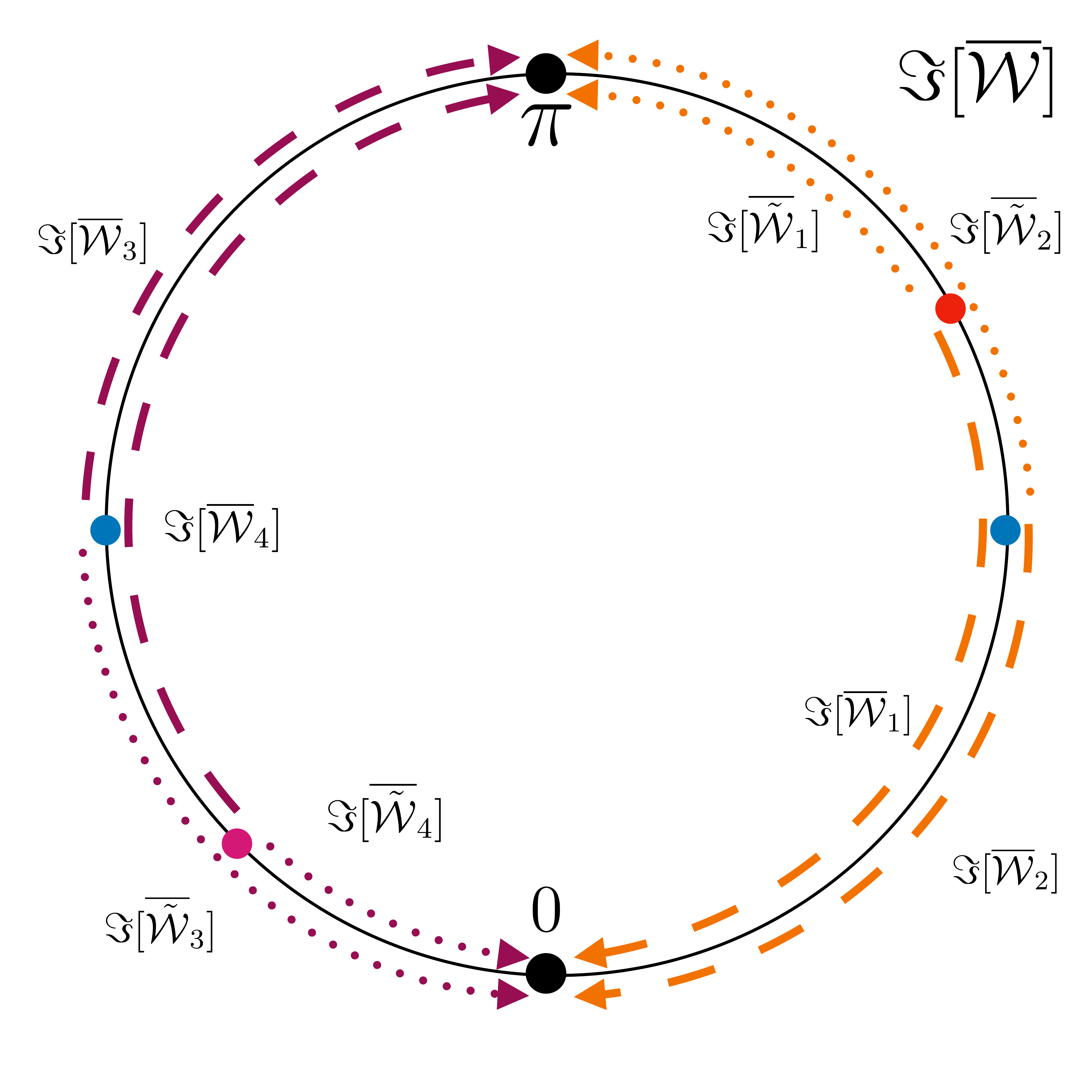}

  [b] $\Im[\overline{\mathcal{W}}]$

  \end{center}
 \end{minipage}
\end{tabular}
 \caption{A cartoon of the trajectory of the operator. Orange dashed and dotted arrows illustrate the trajectory in $t_2$ of the operators, while the purple dashed and dotted arrows illustrate that the trajectory in $t_1$. In the limits where $t_1 \rightarrow \infty$ or $t_2 \rightarrow \infty$, operators approach $0$ or $\pi$. }
 \label{trajectory_of_operators}
\end{figure}

\subsection{Bipartite operator mutual information (BOMI)}
With the formulas for the OEE at hand, we can proceed to compute the bipartite operator mutual information (BOMI) which is defined as
\be
I_{A,B}=S_A+S_B-S_{A\cup B}.
\ee
The BOMI, $I_{A,B}$, measures the non-local correlation between $A$ of $\mathcal{H}_1$ and $B$ of $\mathcal{H}_2$.
Put differently, $I_{A,B}$ measures how much an observer in $B$ is able to learn about $A$. By the subadditivity of entropy, the value of $I_{A,B}$ has to be greater than or equal to $0$. For the subsystems considered in this paper, $I_{A,B}$ is independent of the UV cutoff. The BOMI can be written in terms of twist and anti-twist operators as
\be
I_{A,B}= \lim_{n\rightarrow 1}\f{1}{1-n}\log{\left[\f{\left\langle \sigma_n (\mathcal{W}_1, \overline{\mathcal{W}}_1)\overline{\sigma}_n (\mathcal{W}_2, \overline{\mathcal{W}}_2)  \right \rangle_{\text{Cylinder}}\left \langle \overline{\sigma}_n (\mathcal{W}_3, \overline{\mathcal{W}}_3)\sigma_n (\mathcal{W}_4, \overline{\mathcal{W}}_4)  \right \rangle_{\text{Cylinder}}}{\left \langle \sigma_n (\mathcal{W}_1, \overline{\mathcal{W}}_1)\overline{\sigma}_n (\mathcal{W}_2, \overline{\mathcal{W}}_2) \overline{\sigma}_n (\mathcal{W}_3, \overline{\mathcal{W}}_3)\sigma_n (\mathcal{W}_4, \overline{\mathcal{W}}_4) \right \rangle_{\text{Cylinder}}}\right]},
\ee
where the contributions from the conformal factor cancel out each other so that $I_{A,B}$ depends only on the non-universal pieces of the OEEs.

\section{Dynamics of OEE and BOMI \label{Section:OEE_and_BOMI}}
In this section, we present the time-dependence of the OEEs and the BOMI. 

\subsection{Time-dependence of OEE \label{Section:OEE}}



Let us look closer at the time-dependence of $S^{(n)}_{\mathcal{V}=A,B}$ for $A$, $B$, and $A\cup B$ in $2$d holographic CFTs. As explained above, $S_{\mathcal{V}=A,B,A\cup B}$ is given by a sum of the universal piece, the contribution that is theory-independent, and the non-universal piece, which does depend on the particular CFT considered:
\be
S_{\mathcal{V}=A,B,A\cup B} = \f{h_{n}}{1-n}\log{\left[\Pi_i \f{d \mathcal{W}(w_i)}{dw_i}\f{d \overline{\mathcal{W}}(w_i)}{d\overline{w}_i}\right]} +\text{non-universal piece}.
\ee

The contribution from the conformal factor of a single twist or anti-twist operator is given by
\be
\begin{split}
&\f{h_{n}}{1-n}\log{\left[ \f{d \mathcal{W}(w_i)}{dw_i}\f{d \overline{\mathcal{W}}(w_i)}{d\overline{w}_i}\right]}=-\f{c(n+1)}{12n} \log{\left[\f{\pi}{K(1-k^2)}\right]}\\
&~~~~~~~~~~~-\f{c(n+1)}{48n}\log{\left[\f{1}{\left(1-\tilde{z}^2_i\right)\left(1-\overline{\tilde{z}}^2_i\right)\left(1-k^2\tilde{z}^2_i\right)\left(1-k^2\overline{\tilde{z}}^2_i\right)}\right]}\\
&~~~~~~~~~~~~=-\f{c(n+1)}{12n} \log{\left[\f{\pi}{K(1-k^2)}\right]}-\f{c(n+1)}{48n}\log{\left[\f{\left(\f{\pi}{4\epsilon}\right)^4 e^{\f{\pi(P_1-P_2)}{\epsilon}}}{2^4\Pi_{j=1,2}\left|e^{\f{\pi(w_i-P_j)}{2\epsilon}}\sinh{\left[\f{\pi(w_i-P_j)}{2\epsilon}\right]}\right|}\right]}
\end{split}
\ee
where the first term of the last line depends on only global structure of torus, while the second term depends on the distance between the locations of the operators and the edges of projective measurement.

After performing the analytic continuation as in (\ref{ac}), the conformal factors from each operator is given by
\be
\begin{split}
    &\lim_{n\rightarrow 1}\f{h_{n}}{1-n}\log{\left[ \f{d \mathcal{W}(w_i)}{dw_i}\f{d \overline{\mathcal{W}}(w_i)}{d\overline{w}_i}\right]}=-\f{c(n+1)}{12n} \log{\left[\f{\pi}{K(1-k^2)}\right]}\\
    &-\f{c(n+1)}{48n}
    \times\begin{cases}
    \log{\left[\f{\left(\f{\pi}{4\epsilon}\right)^4 e^{\f{\pi(P_1-P_2)}{\epsilon}}}{ \Pi^{2}_{j=1}\cosh{\left[\f{\pi(X_{i=1,2}+t_2-P_j)}{2\epsilon}\right]}\cosh{\left[\f{\pi(X_{i=1,2}-t_2-P_j)}{2\epsilon}\right]}}\right]} & ~\text{for}~ i=1,2\\
    \log{\left[\f{\left(\f{\pi}{4\epsilon}\right)^4 e^{\f{\pi(P_1-P_2)}{\epsilon}}}{2^4 \Pi^{2}_{j=1}e^{\f{\pi(Y_{i=1,2}-P_j)}{\epsilon}}\cosh{\left[\f{\pi(Y_{i=1,2}+t_1-P_j)}{2\epsilon}\right]}\cosh{\left[\f{\pi(Y_{i=1,2}-t_1-P_j)}{2\epsilon}\right]}}\right]} & ~\text{for}~ i=3,4\\
    \end{cases}.
\end{split}
\ee
Taking the replica limit $n\rightarrow 1$, the asymptotic forms in the small and large measurement region limits are respectively given by
\be
\begin{split}
    &\f{h_{n}}{1-n}\log{\left[ \f{d \mathcal{W}(w_i)}{dw_i}\f{d \overline{\mathcal{W}}(w_i)}{d\overline{w}_i}\right]}\approx\f{c}{6}\times\begin{cases}
\log{\left[\f{P_1-P_2}{2\epsilon}\right]} & ~\text{for}~ P_1-P_2 \gg \epsilon\\
\log{\left[\f{1}{2}\right]} & ~\text{for}~  \epsilon \gg P_1-P_2\\
\end{cases}\\
&+\f{c}{24}
    \times\begin{cases}
    \log{\left[\left(\f{\pi}{4\epsilon}\right)^{-4} e^{\f{-\pi(P_1-P_2)}{\epsilon}} \Pi^{2}_{j=1}\cosh{\left[\f{\pi(X_{i=1,2}+t_2-P_j)}{2\epsilon}\right]}\cosh{\left[\f{\pi(X_{i=1,2}-t_2-P_j)}{2\epsilon}\right]}\right]} & ~\text{for}~ i=1,2\\
    \log{\left[\left(\f{\pi}{4\epsilon}\right)^{-4} e^{\f{-\pi(P_1-P_2)}{\epsilon}} \Pi^{2}_{j=1}\cosh{\left[\f{\pi(Y_{i=1,2}+t_1-P_j)}{2\epsilon}\right]}\cosh{\left[\f{\pi(Y_{i=1,2}-t_1-P_j)}{2\epsilon}\right]}\right]} & ~\text{for}~ i=3,4\\
    \end{cases}.
\end{split}
\ee

\subsection{OEEs in $2$d holographic CFTs}
Let us take a closer look at the non-universal pieces of OEE for the single intervals in $2$d holographic CFTs.
In the von Neumann limit, the non-universal pieces are given by geodesic lengths.
By employing the method of images and then performing the analytic continuation in (\ref{ac}), the original twist and anti-twist operators sit on the earlier Euclidean time slice, $\tau=\f{h(k)}{2}$, while their images are located on the later Euclidean time slice, $\tau=\f{3h(k)}{2}$. 
There are two types of geodesics: one kind of geodesic connects points on the same Euclidean time slice, which we denote by $\mathcal{L}_\text{con}$, while the other kind of geodesic connects points located on different Euclidean time slices, which we denote by $\mathcal{L}_\text{dis}$. Let $S_{\alpha,\text{con}}$ and $S_{\alpha,\text{dis}}$ denote the contributions coming solely from $\mathcal{L}_\text{con}$ and $\mathcal{L}_\text{dis}$ respectively.

\subsection{Case-study on the time evolution of OEE}
From now, we take $X_i$ to be the same as $Y_i$, for simplicity.
To elucidate the properties of strong scrambling dynamics with projective measurements of finite size, we study the time-dependence of $S_{\alpha=A,B}$ in five cases:
\be
\begin{split}
\begin{cases}
P_1=X_1=Y_1, P_2=X_2=Y_2 & ~\text{for~Case 1}\\
P_1>P_2>X_1=Y_1>X_2=Y_2>0 &   ~\text{for~Case 2}\\
X_1=Y_1>P_1>P_2>X_2=Y_2>0, ~X_1-P_2>P_1-X_2>X_1-P_1>P_2-X_2>0 &  ~\text{for~Case 3}\\
P_1>Y_1=X_1>P_2>Y_2=X_2>0, P_1-X_2>P_1-X_1>P_2-X_2>X_1-P_2>0  &  ~\text{for~Case 4}\\
P_1>X_1>X_2>P_2>0, P_1-X_2>X_1-P_2>P_1-X_1>X_2-P_2>0 &  ~\text{for~Case 5}\\
\end{cases}
\end{split}
\ee
where we take the size of the region the projective measurement is done to be  $P_1-P_2 \gg \epsilon$.
\if[0]
\textcolor{red}{In Case 6, the locations and sizes of the subsystems and projective measurement are taken to be
\be
\begin{split}
    &X_1=Y_1>P_1>P_2>X_2=Y_2>0,~ X_1-P_1> P_2-X_2, \\
    &X_1-P_1 \approx X_1-P_2, ~X_2-P_1\approx X_2-P_2,\epsilon \gg P_1-P_2 ~\text{for~Case 6}\\
\end{split}
\ee $\rightarrow$ {\bf MN: Probably, we should remove.}}
\fi
The time-dependence OEEs in the fourth case sufficiently exhibits the characteristic properties of quantum chaotic dynamics that includes a finite-sized projective measurement. Therefore, we present the behavior in time of OEEs in only Case 4 here, and the results for the other cases are reported in Appendix. \ref{app:oee}.


\subsubsection{Case 4}
Here, we present the time-dependence of OEE in Case 4.
Replacing $t_2$ of $S_{B}$ with $t_1$ in Case 4 gives the time-dependence of $S_A$. Therefore, in the following, we shall only present the results on $S_B$. The time-dependence of the universal piece is given by
\be
\begin{split}
     & \f{c}{24}\log{\left[\Pi_{i=1,2} \f{d \mathcal{W}(w_i)}{dw_i}\f{d \overline{\mathcal{W}}(w_i)}{d\overline{w}_i}\right]}\approx-\f{c}{3}\log{\left(\f{\pi}{2\epsilon}\right)}+\f{c}{6}\log{\left(\f{P_1-P_2}{\epsilon}\right)}\\
    &+\f{c}{24}\times\begin{cases}
    \f{2\pi\left(P_2-X_2\right)}{\epsilon} & X_1-P_2>t_2>0\\
    \f{\pi (2(P_2-X_2)-(X_1-P_2)+t_2)}{\epsilon} & P_2-X_2>t_2>X_1-P_2 \\
    \f{\pi (-(X_1-P_2)+P_2-X_2+2t_2)}{\epsilon} & P_1-X_1>t_2>P_2-X_2 \\
    \f{\pi (-(P_1-P_2)+P_2-X_2+3t_2)}{\epsilon} & P_1-X_2 >t_2>P_1-X_1 \\
    \f{2\pi(-(P_1-P_2)+2t_2)}{\epsilon} & t_2>P_1-X_2
    \end{cases}.\\
\end{split}
\ee
The time-dependence of the non-universal piece is obtained by minimizing the geodesic length so 
\be
S_{B,\text{Non-uni.}} =\text{Min}\left[S_{B,\text{con}}, S_{B,\text{dis}}\right].
\ee
where  $S_{B,\text{con}}$ is independent of $t_2$ and is given by
\be \label{S_Bcon}
S_{B,\text{con}} =-\f{c}{6}\log{\left(\f{P_1-P_2}{\epsilon}\right)}.
\ee
On the other hand, the other geodesic length gives rise to a time-dependent non-universal piece:
\be
\begin{split}
    &S_{\text{B,dis}} \approx -\f{c}{6}\log{\left(\f{P_1-P_2}{\epsilon}\right)}+\f{c \pi}{24\epsilon}\times\begin{cases}
   2(X_1-P_2) & X_1-P_2>t_2>0\\
   3(X_1-P_2)-t_2 & P_2-X_2>t_2>X_1-P_2\\
   X_1-X_2+2(X_1-P_2)-2t_2 & P_1-X_1>t_2>P_2-X_2\\
   P_1-X_2+2(X_1-P_2)-3t_2 & P_1-X_2>t_2>P_1-X_1\\
   2(X_1-X_2+P_1-P_2-2t_2) & t_2>P_1-X_2\\
    \end{cases}.\\
\end{split}
\ee
The early time evolution of $S_{B,\text{Non-uni}}$ is governed by the time-dependence of $S_{B,\text{con}}$ which is a constant. After $t_2=t_{P}$, $S_{B,\text{dis}}$ is smaller than $S_{B,\text{con}}$ so that the time evolution of $S_{B,\text{Non-uni}}$ beyond this time is determined by  $S_{B,\text{dis}}$. The value of $t_{PE}$ is determined by $X_1-P_2$:
\be
\begin{split}
    t_{PE}=\begin{cases}
     3(X_1-P_2)&~\text{for}~P_2-X_2>3(X_1-P_2)\\
   \f{3(X_1-P_2)+P_2-X_2}{2}&~\text{for}~ 2(P_1-X_1)-(P_2-X_2)>3(X_1-P_2)>P_2-X_2\\
   P_1-X_1 &~\text{for}~ 3(X_1-P_2)>2(P_1-X_1)-(P_2-X_2)
    \end{cases}.
\end{split}
\ee
The dependence of the transition time $t_{PE}$ on $X_1-P_2$ can be explained by the line tension picture which will be discussed later in Section \ref{Section:line_tension_picture}.

As a consequence, the time-dependence of $S_B$ for the three subcases of Case 4 are given by 
\be
\begin{split}
    &\underline{P_2-X_2>3(X_1-P_2)>0:}\\
    &S_B=\f{c}{3}\log{\left(\f{2\epsilon}{\pi}\right)}+\f{c\pi}{24\epsilon} \times \begin{cases}
    2(P_2-X_2) & X_1-P_2>t_2>0 \\
    \left[2(P_2-X_2)-(X_1-P_2)+t_2\right] & 3(X_1-P_2)>t_2>X_1-P_2\\
    2(X_1-X_2) & t_2>3(X_1-P_2)\\
    \end{cases},\\
     &\underline{2(P_1-X_1)-(P_2-X_2)>3(X_1-P_2)>P_2-X_2:}\\
    &S_B=\f{c}{3}\log{\left(\f{2\epsilon}{\pi}\right)}+\f{c\pi}{24\epsilon} \times \begin{cases}
    2(P_2-X_2) & X_1-P_2>t_2>0 \\
    \left[2(P_2-X_2)-(X_1-P_2)+t_2\right] & P_2-X_2>t_2>X_1-P_2\\
     \left[-(X_1-P_2)+(P_2-X_2)+2t_2 \right] & \f{3(X_1-P_2)+P_2-X_2}{2}>t_2>P_2-X_2\\
      2(X_1-X_2) & t_2>\f{3(X_1-P_2)+P_2-X_2}{2}
    \end{cases},\\
     &\underline{ 3(X_1-P_2)>2(P_1-X_1)-(P_2-X_2):}\\
    &S_B=\f{c}{3}\log{\left(\f{2\epsilon}{\pi}\right)}+\f{c\pi}{24\epsilon} \times \begin{cases}
    2(P_2-X_2) & X_1-P_2>t_2>0 \\
    \left[2(P_2-X_2)-(X_1-P_2)+t_2\right] & P_2-X_2>t_2>X_1-P_2\\
     \left[-(X_1-P_2)+(P_2-X_2)+2t_2 \right] & P_1-X_1>t_2>P_2-X_2\\
      \left[-(P_1-P_2)+P_2-X_2+3t_2\right] & \f{P_1-X_2+2(X_1-P_2)}{3}>t_2>P_1-X_1\\
      2(X_1-X_2) & t_2 >\f{P_1-X_2+2(X_1-P_2)}{3}
    \end{cases},\\
\end{split}
\ee

The projective measurement in the subregion of $B$ defined by $X_1>x>P_2$ reduces the amount of the entanglement entropy for $B$ because such measurements replace the entanglement structure for the measurement region $V$ with that of a boundary state which is known to be approximately unentangled \cite{Miyaji:2014mca}. 
Therefore, the initial value of $S_B$ is given by the thermal entropy for the subregion of $B$ that is not included in the region where the projective measurement is implemented, namely $[P_2,X_2]$. After $t_1=X_1-P_2$, the scrambling effect of the time evolution operator recovers the entanglement that is removed by the finite-sized projective measurement, so that $S_B$ increases with time. Within certain time intervals,  $S_B$ growslinearly with $t_2$: 
\be \label{fractional_grwoth}
S_B \sim \f{1}{2}\times\f{c\pi t_2}{12 \epsilon},~ 1\times \f{c\pi t_2}{12 \epsilon},~\text{or}~ \f{3}{2}\times\f{c\pi t_2}{12 \epsilon}.
\ee

The fractional coefficient of the linear time growth of OEE as in $ \f{1}{2}\times\f{c\pi t_2}{12 \epsilon}$ and $\f{3}{2}\times\f{c\pi t_2}{12 \epsilon}$ does not occur when the quantum circuit does not involve the finite-sized projective measurement. By contrast, when a projective measurement is introduced, these fractional coefficients can occur. For example, in Case 4, if $3(X_1-P_2)>2(P_1-X_1)-(P_2-X_2)$, the linear time growth of $S_B$ is accompanied by fractional coefficients in the time intervals $P_2-X_2>t_2>X_1-P_2$ and $\f{P_1-X_2+2(X_1-P_2)}{3}>t_2>P_1-X_1$. We will describe these linear time growths with fractional coefficients in terms of a line tension model (see Section \ref{Section:line_tension_picture}) and a dual geometry involving an EOW brane that corresponds to the projective measurement (see Section \ref{Section:Gravitational_description}).  
At late times, when $t_2>3(X_1-P_2)$ for $P_2-X_2>3(X_1-P_2)>0$, $t_2>\f{3(X_1-P_2)+P_2-X_2}{2}$ for $2(P_1-X_1)-(P_2-X_2)>3(X_1-P_2)>P_2-X_2$, and $t_2 >\f{P_1-X_2+2(X_1-P_2)}{3}$ for $ 3(X_1-P_2)>2(P_1-X_1)-(P_2-X_2)$, the scrambling effect completely recovers the entanglement that is removed by the projective measurement so that $S_B$ becomes the OEE for the thermal state with inverse temperature $4\epsilon$.

\subsection{Time dependence of BOMI \label{Section:BOMI}}
Let us study how the finite-sized projective measurement affects the non-local correlation. We will study the time-dependence of $I_{A,B}$ in Case 4. For simplicity, we take $t_1=t_2=t$. 
Since the conformal factors cancel out, only the non-universal pieces of $S_A$, $S_B$ and $S_{A\cup B}$ contribute to the time-dependence of $I_{A,B}$. 
The time-dependence of $S_{A,\text{Non-uni.}}$ and $S_{B,\text{Non-uni.}}$ is the same as that in Section \ref{Section:OEE}. 
Therefore, let us focus on the time-dependence of $S_{A\cup B, \text{Non-uni.}}$ in Case 4.
%
\subsubsection{$S_{A\cup B, \text{Non-uni.}}$ in Case 4}
In the method of images, one of the candidates for the minimal geodesic length is given by
\be \label{Scand1}
S^{\text{cand.1}}_{A\cup B, \text{Non-uni.}}=S_{A,\text{Non-uni}}+S_{B,\text{Non-uni}}.
\ee
When $S_{A\cup B, \text{Non-uni}}$ is given by $S^{\text{cand.}1}_{A\cup B, \text{Non-uni.}}$, the value of $I_{A,B}$ is zero.
The other candidate which gives $S^{\text{cand.2}}_{A\cup B}$ is determined by the summation of the lengths of only the geodesics that correspond to one- or two-point functions in the $2$d boundary CFT. More explicitly, the time-dependence of $S^{\text{cand.2}}_{A\cup B}$ is given by
\be
\begin{split}
    &S^{\text{cand.2}}_{A\cup B} \approx -\f{c}{3}\log{\left(\f{P_1-P_2}{\epsilon}\right)}+ f_1(t)+f_2(t),\\
    &f_1(t) = \text{Min}\left[0,L_1(t)\right], ~f_2(t) = \text{Min}\left[0,L_2(t)\right],\\
    &L_1(t)=\f{c}{12} \log{\left[-\sinh{\left[\f{sn^{-1}(\tilde{z}_1,k^2)-sn^{-1}(\tilde{z}_3,k^2)}{2}\right]}\sinh{\left[\f{sn^{-1}(\overline{\tilde{z}}_1,k^2)-sn^{-1}(\overline{\tilde{z}}_3,k^2)}{2}\right]}\right]}\\
    &\approx \f{c \pi}{24 \epsilon} \times \begin{cases}
    \left[2(X_1-P_2)\right] & X_1-P_2>t>0\\
    \left[X_1-P_2+t\right] & P_1-X_1>t>X_1-P_2\\
    \left[P_1-P_2\right] & P_1-X_1>t>X_1-P_2\\
    \end{cases},\\
    &L_2(t)=\f{c}{12} \log{\left[-\sinh{\left[\f{sn^{-1}(\tilde{z}_2,k^2)-sn^{-1}(\tilde{z}_4,k^2)}{2}\right]}\sinh{\left[\f{sn^{-1}(\overline{\tilde{z}}_2,k^2)-sn^{-1}(\overline{\tilde{z}}_4,k^2)}{2}\right]}\right]}\\
    &\approx \f{c \pi}{24 \epsilon} \begin{cases}
    2\left[t-(P_2-X_2)\right] & P_2-X_2>t>0\\
    \left[t-(P_2-X_2)\right] & P_1-X_2>t>P_2-X_2\\
    [P_1-P_2] & t>P_1-X_2
    \end{cases},\\
\end{split}
\ee
where since $L_1(t)$ is positive, $f_1(t)$ is given by $0$.
When $f_1(t)=f_2(t)=0$, $S^{\text{cand.2}}_{A\cup B}$ reduces to  $S^{\text{cand.1}}_{A\cup B}$.
In the early-time region, $P_2-X_2>t>0$, the value of $L_2(t)$ is negative, while in the time region $t>P_2-X_2$, it is positive. 
Therefore, the time evolution of $S^{\text{cand.2}}_{A\cup B}$ is given by 
\be
\begin{split}
    S^{\text{cand.2}}_{A\cup B} \approx \f{c\pi }{12\epsilon}\begin{cases}
    2[t-(P_2-X_2)] & P_2-X_2>t>0\\
    0 & t> P_2-X_2
    \end{cases}.
\end{split}
\ee

\subsubsection{$I_{A,B}$ in Case 4}
In Case 4, after $t=T_{PE}$, the value of $S^{\text{cand.2}}_{A\cup B}$ is larger than that of $S^{\text{cand.1}}_{A\cup B}$.
The value of $T_{PE}$ depends on $X_1-P_2$, and its dependence on $X_1-P_2$ is given by
\be
\begin{split}
T_{PE}= \begin{cases}
\f{3(X_1-P_2)+P_2-X_2}{2} & P_2-X_2>3(X_1-P_2)\\
P_2-X_2 & 3(X_1-P_2) >P_2-X_2
\end{cases}.
\end{split}
\ee
Thus, the time evolution of $I_{A,B}$ for two subcases of Case 4 is given by
\be
\begin{split}
   &\underline{P_2-X_2>3(X_1-P_2):}
   ~~ I_{A,B} \approx \f{c\pi}{12 \epsilon}\times\begin{cases}
    \left[3(X_1-P_2)+(P_2-X_2)-2t\right] &T_{PE}>t>0 \\
    0 & t>T_{PE}
    \end{cases},\\
     &\underline{3(X_1-P_2)>P_2-X_2:}
     ~~ I_{A,B} \approx \f{c\pi}{12 \epsilon}\times\begin{cases}
    \left[(P_2-X_2)-t\right] &T_{PE}>t>0 \\
    0 & t>T_{PE}
    \end{cases}.\\
\end{split}
\ee

\subsubsection{Effect of projective measurement}

Finally, let us compare the BOMI of the time evolution operator with and without a single projective measurement of finite size.
The time-dependence of the BOMI without a single projective measurement may be approximated by the BOMI in Case 2. We now assume that $P_2-X_2$ is much larger than $X_1-X_2>0$. Under this assumption, the time-dependence of BOMI is given by
\be
\begin{split}
    I_{A,B} \approx\f{c \pi}{6\epsilon}\times\begin{cases}
   \left[ \left(X_1-X_2\right)-t\right] &X_1-X_2>t>0 \\
    0 & t>X_1-X_2
    \end{cases}.
\end{split}
\ee
One possible interpretation for this behavior of $I_{A,B}$ is that after $t=X_1-X_2$, $\rho_{A\cup B}$ factorizes into $\rho_A$ and $\rho_B$.  
The time for $I_{A,B}$ without the projective measurement to vanish is larger than in Case 4.
We see here that the projective measurement in a quantum circuit may promote the factorization of a density matrix $\rho_{A\cup B}$.
Note that after $\rho_{A\cup B}$ factorizes into $\rho_A$ and $\rho_B$, the OEE does not immediately follow the Page curve \cite{Page_1993,1996PhRvL..77....1S}.
The time for $\rho_{\alpha=A,B,A\cup B}$ to be approximately given by a typical one is larger than the time for $\rho_{A\cup B}$ to factorize into $\rho_A$ and $\rho_B$. 

\section{Line tension picture}\label{Section:line_tension_picture}
Now, let us introduce an effective description of the entanglement dynamics of the non-unitary holographic channel considered in this paper.
\subsection{Line-tension picture in a unitary circuit}\label{LineTensionPictureSection}
\begin{figure}[h!!]
\begin{center}
\includegraphics[scale=0.3]{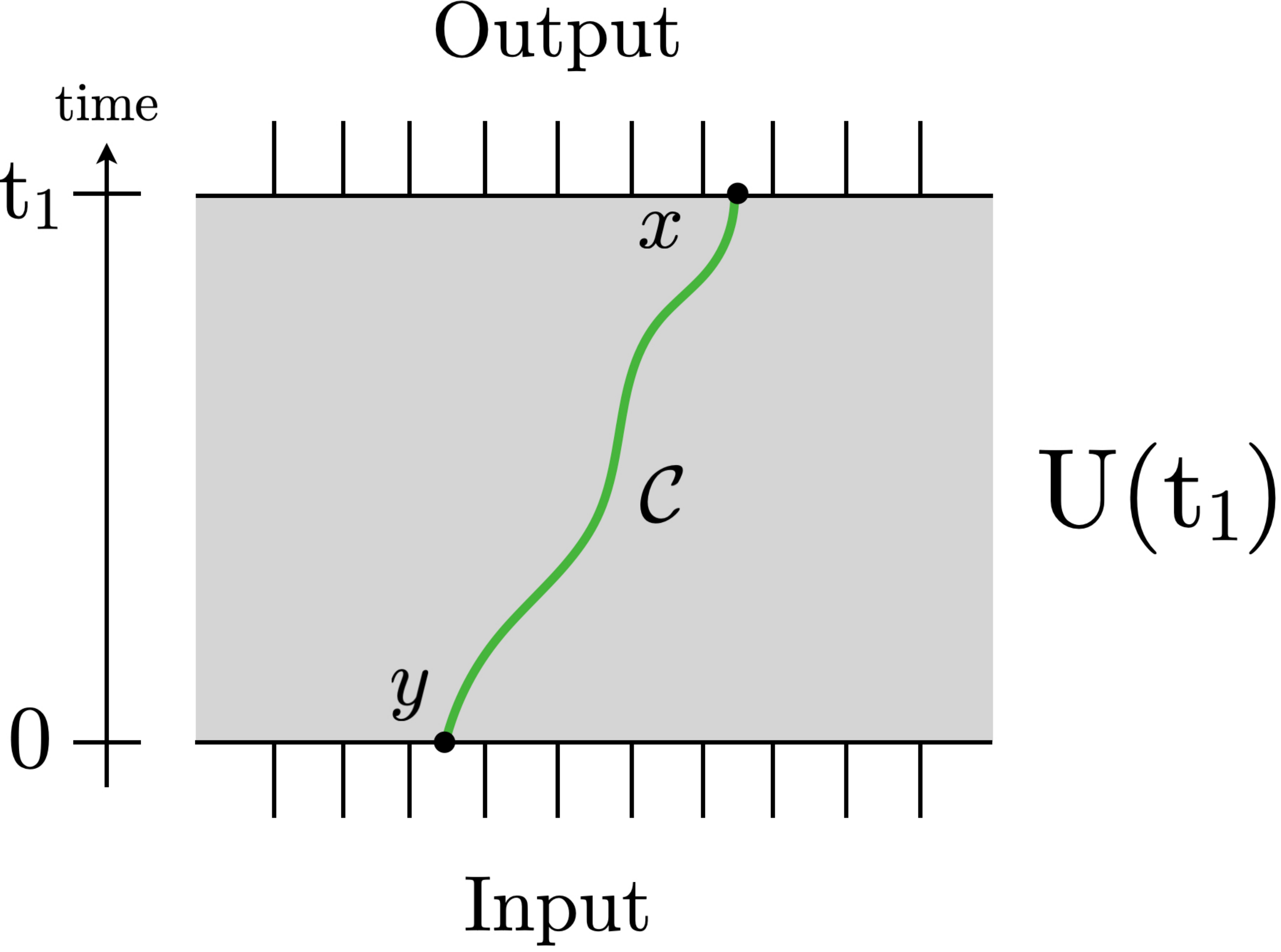}
\caption{A curve ${\cal C}$ that divides the unitary circuit into two parts. In the line-tension picture, the entanglement entropy $S_U(x,y,t_1)$ is given by the integral of the line tension ${\cal T}(v)$ along the curve ${\cal C}$. }
\label{LT1}
\end{center}
\end{figure}
We propose a generalization of the so-called line-tension picture to a random unitary circuit with a projective measurement.
It is known that in a chaotic system, the hydrodynamics of entanglement production is well described by the line-tension picture introduced in \cite{
2018arXiv180300089J,2017PhRvX...7c1016N,PhysRevD.98.106025,PhysRevX.8.031058,PhysRevX.8.021013}.  
As the simplest example, we assume that the spatial direction is homogeneous and infinite, and no projective measurement is performed. These assumptions can be relaxed to allow for a compact space as well as the presence of spatial inhomogeneity \cite{OMISSD}. For simplicity, we assume that the system is time-evolved by the unitary operator $U(t_1)$ from $t=0$ to $t=t_1$. We divide the infinite line where the system lives into two pieces at position $x$ at $t=t_1$. We also divide the line at position $y$ at $t=0$.  The entanglement entropy $S_U(x,y,t_1)$ of the unitary operator is computed in  the line-tension picture as
\ba
S_U(x,y,t_1)={\rm min}_{\cal C}\int_{\cal C}dt\,  {\cal T}(v)\, ,
\ea
 where the minimization is taken over all the possible curves ${\cal C}$ that connects the point $(x,t_1)$ and $(y,0)$. 
The symbol ${\cal T}(v)$ is the line-tension associated to a curve ${\cal C}$ that connects the point $(x,t_1)$ and $(y,0)$ as in Fig. \ref{LT1}. 
The curve ${\cal C}$ in spacetime has a velocity $v=dx/dt$ and a line-tension ${\cal T}(v)$ that depends on $v$. In our case, when the spacetime is uniform, the minimal curve is given by a straight line with a constant velocity $v=(x-y)/t_1$.

The details of the function ${\cal T}(v)$ depend on the system and are estimated in a chaotic system using random unitary circuits which illustrate the phenomenon of quantum information scrambling. In the scaling limit and in the limit of large bond dimension $q$, the line-tension is simply given by counting the number of bonds cut which is
\begin{align}
\mathcal{T}(v)= \begin{cases}\log q & v<1 \\ v \log q & v>1\, .\end{cases}
\end{align}
To compute the entanglement of the unitary operator in a holographic CFT using the line-tension picture, we need to identify the bond dimension (the local Hilbert space dimension) $q$ in the random unitary circuit. This can be accomplished by comparing the rates at which the information gets scrambled.
While the entanglement entropy grows at a rate of $\log q$ in random unitary circuits, it is known that in holographic CFTs the entanglement of the unitary operator (computed as the entanglement between two CFTs in the time-evolved thermofield double state) grows at a rate of $\f{c\pi}{3\beta}$. Here, $\beta$ is dimensionless as it has been written in units of the lattice spacing.
Therefore, we make the identification
\ba
q\sim e^{\f{c\pi}{3\beta}}\, .
\ea
Notice that $\log q$ simply equal to the entropy density given by the Cardy formula $S_{\rm Cardy}/(2\pi R)=\f{c\pi}{3\beta}$. Using this, one can correctly reproduce the growth of the entanglement in holographic CFTs which is
\ba
S_U(x,y,t_1)\sim \f{c\pi}{3\beta}t_1\, .
\ea
\begin{figure}[h]
\begin{center}
\includegraphics[scale=0.4]{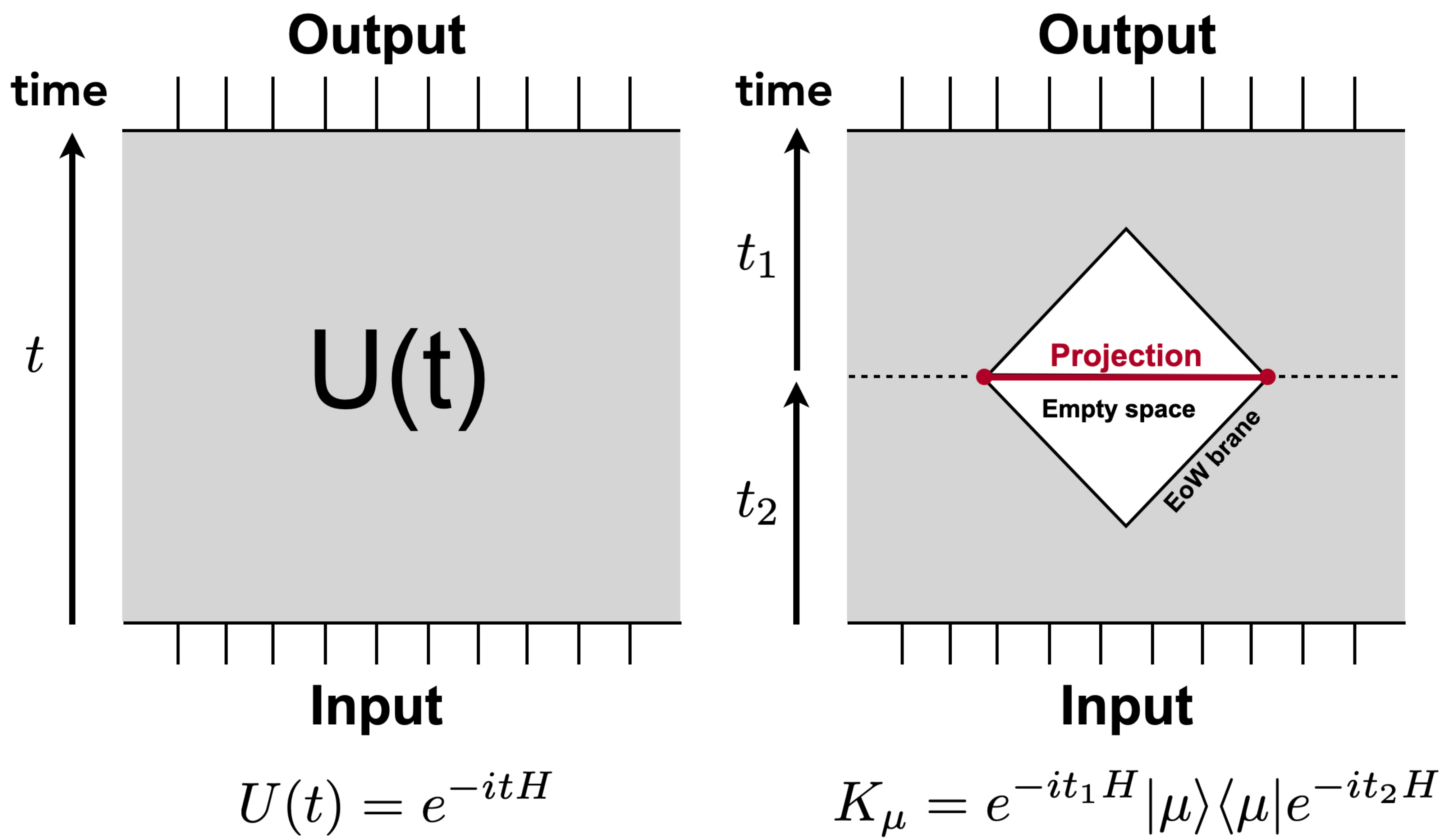}
\caption{Quantum circuits with (right)/without (left) a projective measurement.  }
\label{LTp}
\end{center}
\end{figure}
{\bf Line-Tension Picture with a Projective Measurement}\\
Here we will explain a generalization of the simplest example of the line-tension picture described above to the system with a projective measurement. In this paper, the state is projected to a boundary state with no spatial entanglement in the coarse-grained limit $\epsilon \rightarrow 0$. In a unitary circuit, the state with no entanglement can be represented as an empty space. Since we expect that the unitary time-evolution makes the entanglement spread causally, the empty space with no entanglement shrinks according to the light-cone associated to the projected interval, thus the non-unitary quantum circuit consisting of two unitary time evolution operators and a projective measurement
\begin{align}
    K_\mu=e^{-it_1H}|\mu\rangle\langle \mu|e^{-it_2H}
\end{align}
can be represented as Fig. \ref{LTp}. 

\subsection{Time dependence of OEE}
\begin{figure}[h]
\begin{center}
\includegraphics[scale=0.2]{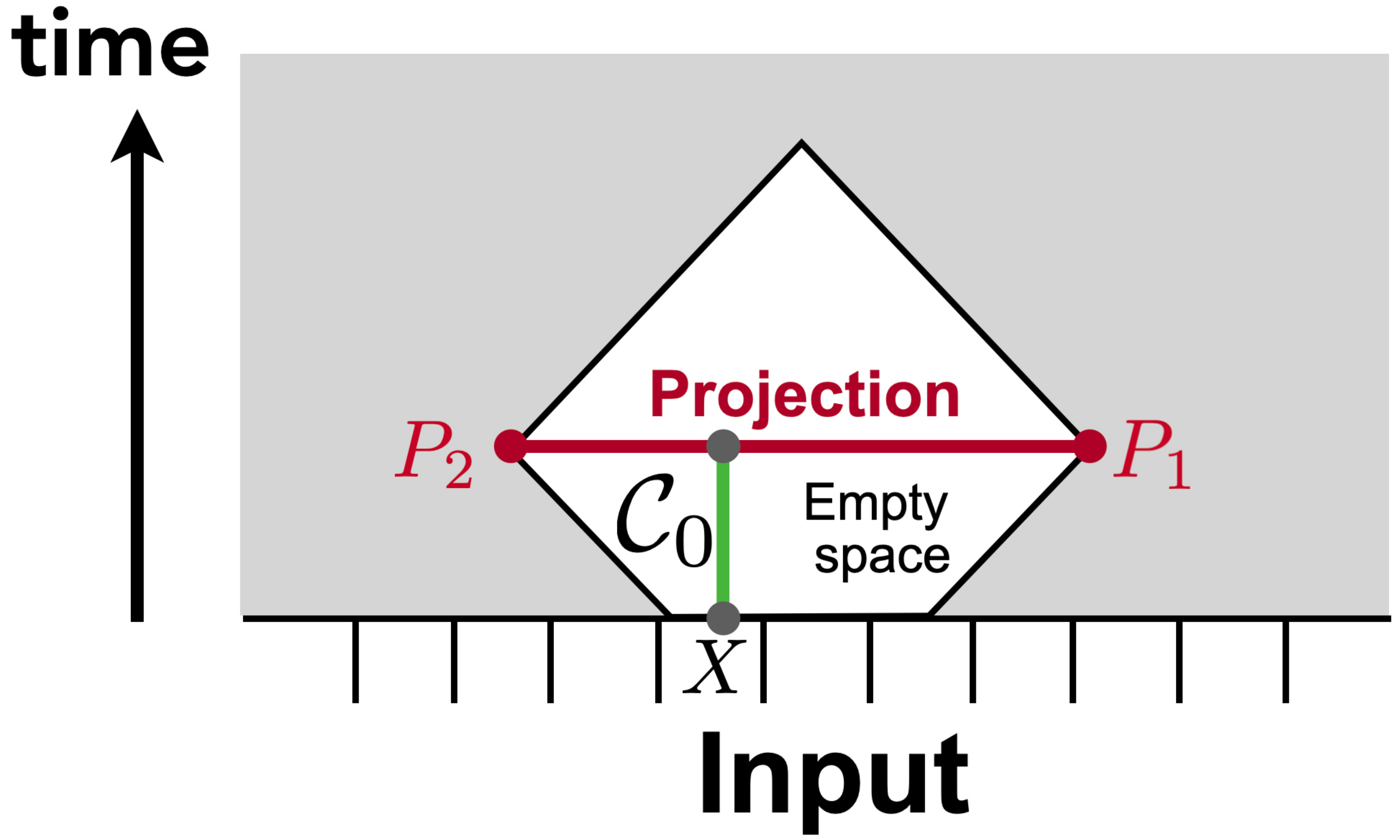}
\caption{Configuration with trivial entanglement in line-tension picture.}
\label{LEoW2}
\end{center}
\end{figure}
Here we will explain how the operator entanglement computed using the line-tension picture that we proposed in this paper can reproduce the holographic results.
Employing the line-tension picture in the unitary circuit proposed above (Fig. \ref{LTp}), we can reproduce the behavior of the entanglement computed in the previous section in the coarse-grained limit. There are non-trivial configurations of the minimal curves due to the existence of the projected region in the  unitary circuit. In the case where the endpoint $X$ of the subregion at $t=0$ is inside the projected region with no entanglement, the entanglement associated to a curve ${\cal C}_0$ anchored at $X$ as in Fig. \ref{LEoW2} becomes zero 
\begin{align}\label{Sconst1}
    S_{{\cal C}_0}=0\, .
\end{align}
\begin{figure}[h]
\begin{center}
\includegraphics[scale=0.4]{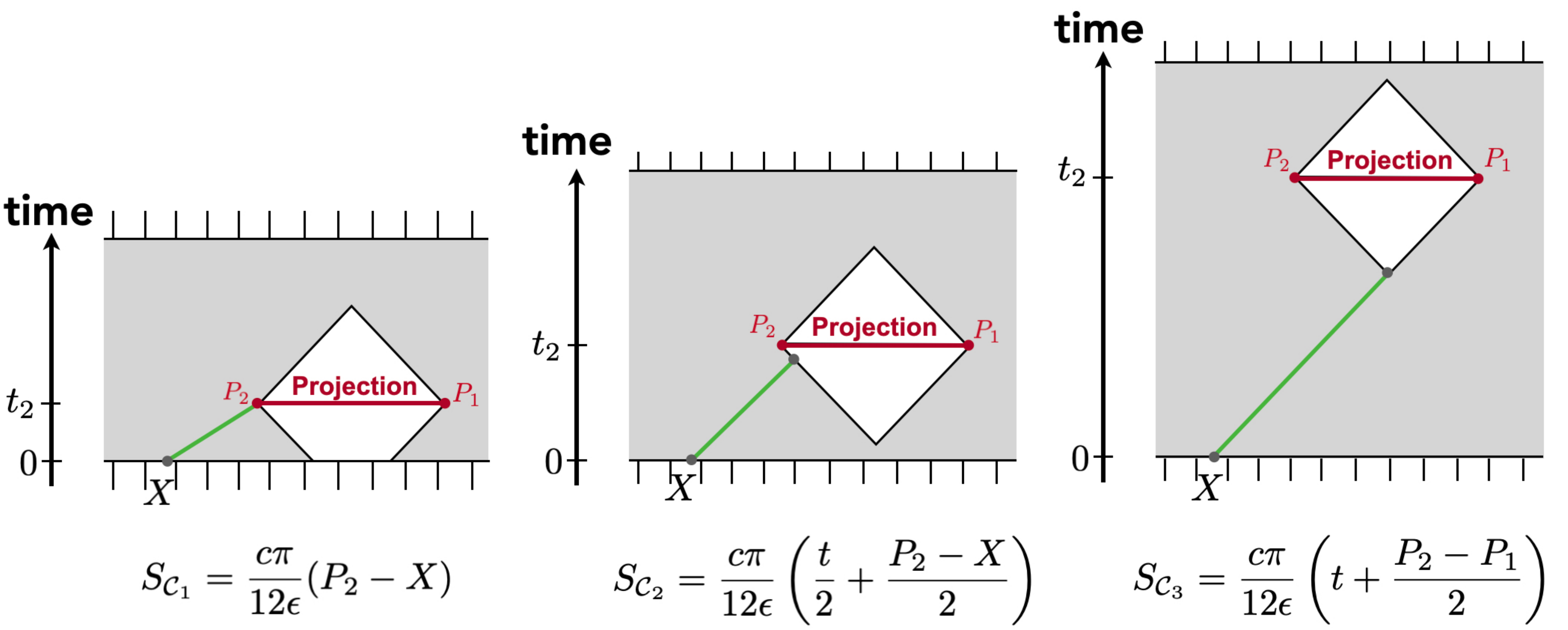}
\caption{Three configurations in line-tension picture.}
\label{LEoW}
\end{center}
\end{figure}
There are also several characteristic configurations of the minimal curves in the case where the endpoint $X$ of the subregion at $t=0$ is outside the projected region. In Fig. \ref{LEoW}, we drew such configurations ${\cal C}_1,{\cal C}_2$,and ${\cal C}_3$ anchored at $X$, which lies outside the projected region at $t=0$, and the``end-of-the-world brane", the boundary of the projected region. The entanglement $S_{C_i}$ associated to a curve ${\cal C}_i$ can be computed as 
\begin{align}
    S_{{\cal C}_1}&=\frac{c\pi}{12\epsilon}(P_2-X)\,, \label{Sconst2}\\
    S_{{\cal C}_2}&=\frac{c\pi}{12\epsilon}\left(\frac{t}{2}+\frac{P_2-X}{2}\right)\,, \\
    S_{{\cal C}_3}&=\frac{c\pi}{12\epsilon}\left(t+\frac{P_2-P_1}{2}\right)\,. 
\end{align}
In particular, the time-growth $ S_{{\cal C}_2}=\frac{c\pi}{6\beta}t$ is a characteristic behavior not seen in the absence of the projection.

\begin{figure}[h]
\begin{center}
\includegraphics[scale=0.35]{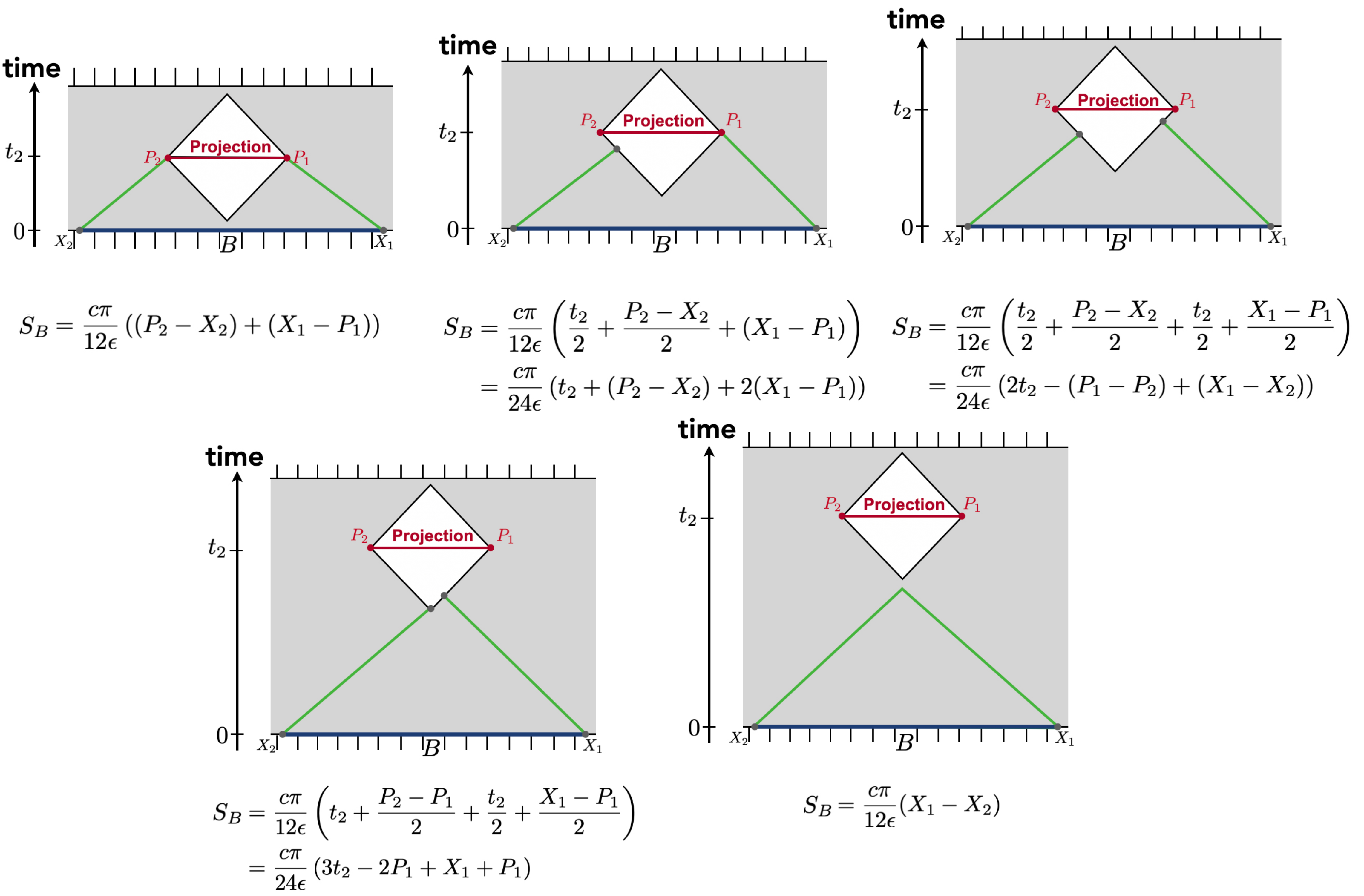}
\caption{Diagrams in the line-tension picture in Case 3.}
\label{OEE}
\end{center}
\end{figure}

\begin{figure}[h]
\begin{center}
\includegraphics[scale=0.33]{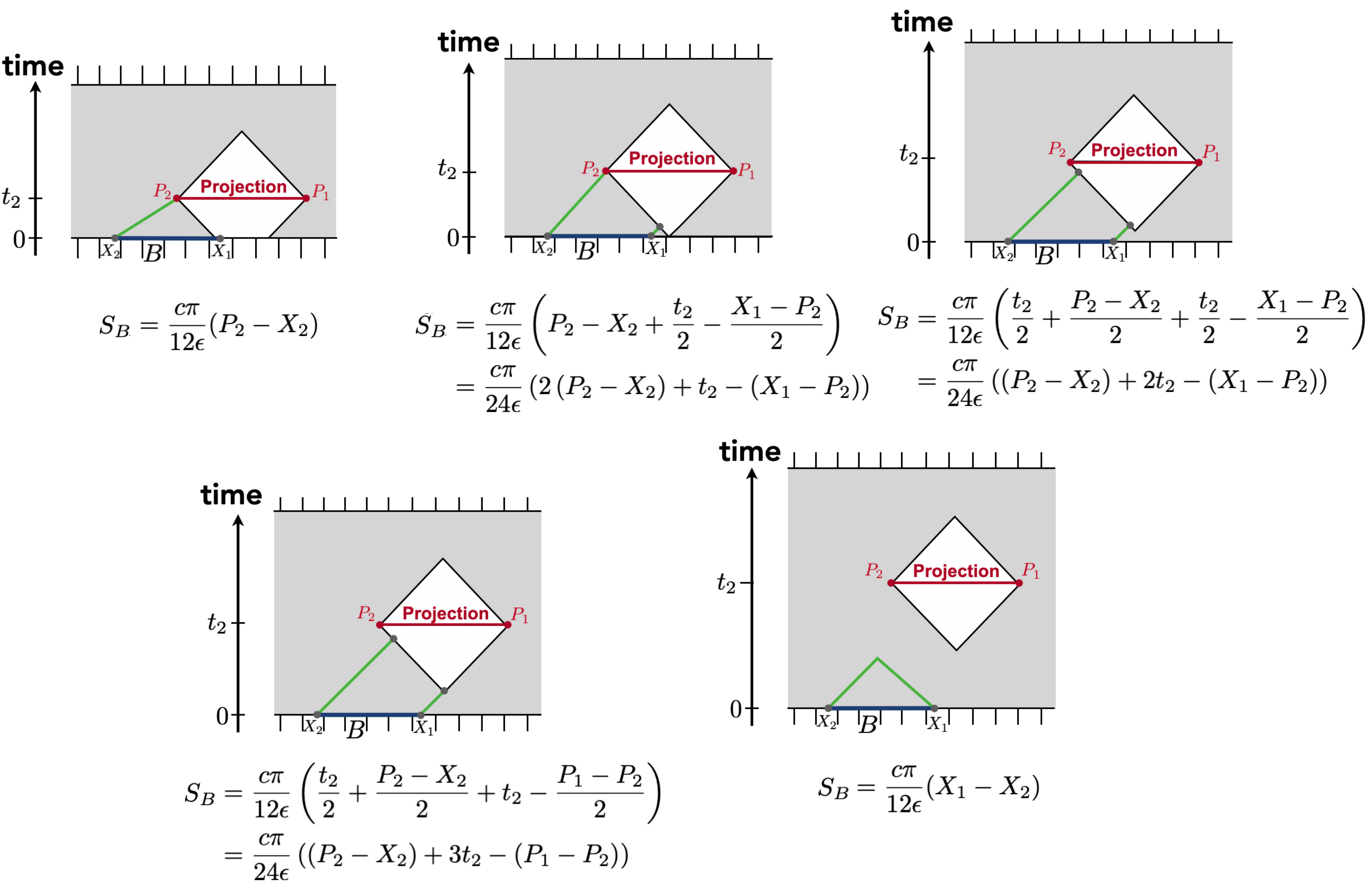}
\caption{Diagrams for the line-tension picture in Case 4.}
\label{OEE2}
\end{center}
\end{figure}

Here we will demonstrate how the holographic calculations of the operator entanglement entropy are correctly reproduced for Cases 3 and 4.\\
{\bf Case 3}\\
Here, the subsystem $B$ and the projective measurement are located such that
\begin{align}
0<X_{2}<P_{2}<P_{1}<X_{1}, X_{1}-P_{2}>P_{1}-X_{2}>X_{1}-P_{1}>P_{2}-X_{2}>0
\end{align}
In this configuration, the time evolution of $S_B$ computed in the line-tension picture (Fig. \ref{OEE}) is
\begin{align}
S_{B} \approx \begin{cases}\frac{c \pi\left[\left(X_{1}-X_{2}\right)-\left(P_{1}-P_{2}\right)\right]}{12 \epsilon}& P_{2}-X_{2}>t_{2}>0 \\ \frac{c \pi\left[t_{2}-2\left(P_{1}-X_{1}\right)-\left(X_{2}-P_{2}\right)\right]}{24 \epsilon}& X_{1}-P_{1}>t_{2}>P_{2}-X_{2} \\ \frac{c \pi\left[2 t_{2}-\left(P_{1}-P_{2}\right)+\left(X_{1}-X_{2}\right)\right]}{24 \epsilon}& P_{1}-X_{2}>t_{2}>X_{1}-P_{1} \\ \frac{c \pi\left[3 t_{2}-2 P_{1}+X_{1}+P_{2}\right]}{24 \epsilon}& \frac{X_{1}+2 P_{1}-2 X_{2}-P_{2}}{3}>t_{2}>P_{1}-X_{2} \\ \frac{c \pi\left[X_{1}-X_{2}\right]}{12 \epsilon} & t_{2}>\frac{X_{1}+2 P_{1}-2 X_{2}-P_{2}}{3}\end{cases}
\end{align}
This correctly reproduces the leading order behavior of the entanglement entropy in the coarse-grained limit $\ep\rightarrow 0$. 
\\
{\bf Case 4}\\
 We take the locations of $B$  and the projective measurement to be
\begin{align}
P_{1}>X_{1}>P_{2}>X_{2}>0, P_{1}-X_{1}>0, X_{1}-P_{2}>0,\\ \nonumber P_{1}-X_{2}>P_{1}-X_{1}>P_{2}-X_{2}>X_{1}-P_{2}>0
\end{align}
In this configuration, the time evolution of $S_B$ computed in the line-tension picture (Fig. \ref{OEE2}) is
\begin{align}
S_{B} \approx  \begin{cases}\ \frac{\pi\left(P_{2}-X_{2}\right)}{12 \epsilon} & X_{1}-P_{2}>t_{2}>0 \\ \frac{c \pi\left[2\left(P_{2}-X_{2}\right)+t_{2}-\left(X_{1}-P_{2}\right)\right]}{24 \epsilon} & P_{2}-X_{2}>t_{2}>X_{1}-P_{2} \\ \frac{c \pi\left[\left(P_{2}-X_{2}\right)+2 t_{2}-\left(X_{1}-P_{2}\right)\right]}{24 \epsilon} & P_{1}-X_{1}>t_{2}>P_{2}-X_{2} \\ \frac{c \pi\left[\left(P_{2}-X_{2}\right)+3 t_{2}-\left(P_{1}-P_{2}\right)\right]}{24 \epsilon} & \frac{2\left(X_{1}-P_{2}\right)+P_{1}-X_{2}}{3}>t_{2}>P_{1}-X_{1} \\ \frac{c \pi\left(X_{1}-X_{2}\right)}{12 \epsilon} & t_{2}>\frac{2\left(X_{1}-P_{2}\right)+P_{1}-X_{2}}{3}\end{cases}
\end{align}
for the case $2\left(P_{1}-X_{1}\right)-\left(P_{2}-X_{2}\right)<3\left(X_{1}-P_{2}\right)$. The line-tension picture once again correctly reproduces the leading order behavior of the entanglement entropy in the coarse-grained limit $\epsilon \rightarrow 0$.

\section{Gravitational description \label{Section:Gravitational_description}}
We have discussed operator entanglement for projection operators in holographic CFTs. In this section, we investigate the bulk spacetime which is relevant to our projection operator state. The gravity picture is also consistent with the light-cone structure of EOW brane discussed in Section \ref{Section:line_tension_picture}. 
\subsection{Projection and the AdS/BCFT correspondence}

Recall that operator states are defined as maximally entangled states acted upon by the operator of interest,
\begin{align}
    |{\mathcal{O}_A}\rangle\equiv\mathcal{O}_A\otimes 1_B\sum_{n}|{n_A n^\ast_B}\rangle,
\end{align}
where we omit the normalization constant.
Note that the maximally entangled state is equivalent to the (infinite temperature) TFD state,
\begin{align}
   |{\text{TFD}_\beta}\rangle=\sum_{n} \mathrm{e}^{-\frac{\beta}{2}E_n}|n_A n^\ast_B\rangle,
\end{align}
which is holographically dual to an eternal black hole in AdS. Therefore, our projection operator state (with UV regulator $\epsilon$) can be written as
\begin{align}
    |{K_{\mu}}\rangle=\mathrm{e}^{-\epsilon H}\mathcal{P}_{\mu}|{\text{TFD}_{2\epsilon}}\rangle,
\end{align}
where $\mathcal{P}_{\mu}$ is the projection operator acting on a certain spatial subsystem $V$ and $H$ is the Hamiltonian of the system. We need to take into account the effect of the spatial projection $\mathcal{P}_{\mu}$ in the eternal black hole geometry. 
Since the Euclidean path integral treats the spatial projections as slits, we can discuss the dual gravity picture based on the AdS/BCFT correspondence \cite{Takayanagi:2011zk,Fujita:2011fp}. 
In light of AdS/BCFT, we expect that in the gravity picture, such spatial projections introduce an end-of-the-world (EOW) brane in the bulk. 

We are interested in Einstein gravity on a bulk spacetime $M$ coupled to matter fields on a dynamical brane $Q$,
\begin{align}
S=\dfrac{1}{16\pi G_N}\int_M \sqrt{-g}(R-2\Lambda)+\dfrac{1}{8\pi G_N}\int_{\partial M\cup Q} \sqrt{-h}K+S_Q,
\end{align}
where $S_Q$ denotes matter contributions localized on the dynamical brane $Q$. The second term is a Gibbons-Hawking boundary term for $\partial M$ and $Q$. In contrast with the standard Dirichlet boundary condition on the asymptotic boundary $\delta g_{\mu\nu}=0$, we will impose the Neumann-type boundary condition on $Q$ so that the metric on $Q$ remains dynamical,
\begin{align}
    K_{ij}-h_{ij}K= -\dfrac{16\pi G_N}{\sqrt{-h}}\dfrac{\delta S_Q}{\delta h^{ij}}\equiv T^Q_{ij}. \label{eq:bdryeq}
\end{align}
In principle, we would solve \eqref{eq:bdryeq} for a given matter action $S_Q$ and determine the properties of the brane $Q$. In bottom up models of AdS/BCFT, we assume that
\begin{align}
T^Q_{ij}=-T h_{ij},
\end{align}
where $T$ is a constant tension for the brane. 
In this way, we can preserve the boundary conformal symmetry. The brane $Q$ is often called the end-of-the-world (EOW) brane. 

As in the previous section, we will be particularly interested in the situation where the EOW brane is tensionless, i.e. $T=0$. Since the extrinsic curvature becomes trivial in this case, $K_{ij}=0$, the EOW brane becomes a totally geodesic surface. After the conformal map discussed in Section \ref{Section:Preliminary}, our computation reduces to the holographic entanglement entropy for the standard Hartman-Maldacena geometry. One can check that this reproduces the CFT results in Section \ref{Section:OEE_and_BOMI}.

\subsection{Holographic entanglement entropy in a simpler setup}\label{subsec:hee}
We may view our setup as a generalization of the Hartman-Maldacena geometry. From this point of view, the new ingredient is a finite size EOW brane inside the horizon. This quantitatively changes the time evolution of the entanglement entropy. As we have already seen, the linear growth in time can become milder than in the stardard Hartman-Maldacena geometry in the middle time regime. It is given by
\begin{align}
S_{\text{dis.}}=\dfrac{\alpha_0}{2}t+\cdots,
\end{align}
where
\begin{align}
\alpha_0=\dfrac{c\pi}{3\beta},
\end{align}
is the coefficient in the ordinary Hartman-Maldacena geometry. Here, we took the large black hole limit, i.e. the high temperature limit, so that the linear growth is manifest. The remaining terms denoted by dots are the sub-leading terms in this limit. 

\begin{figure}
\centering
\resizebox{100mm}{!}{\includegraphics{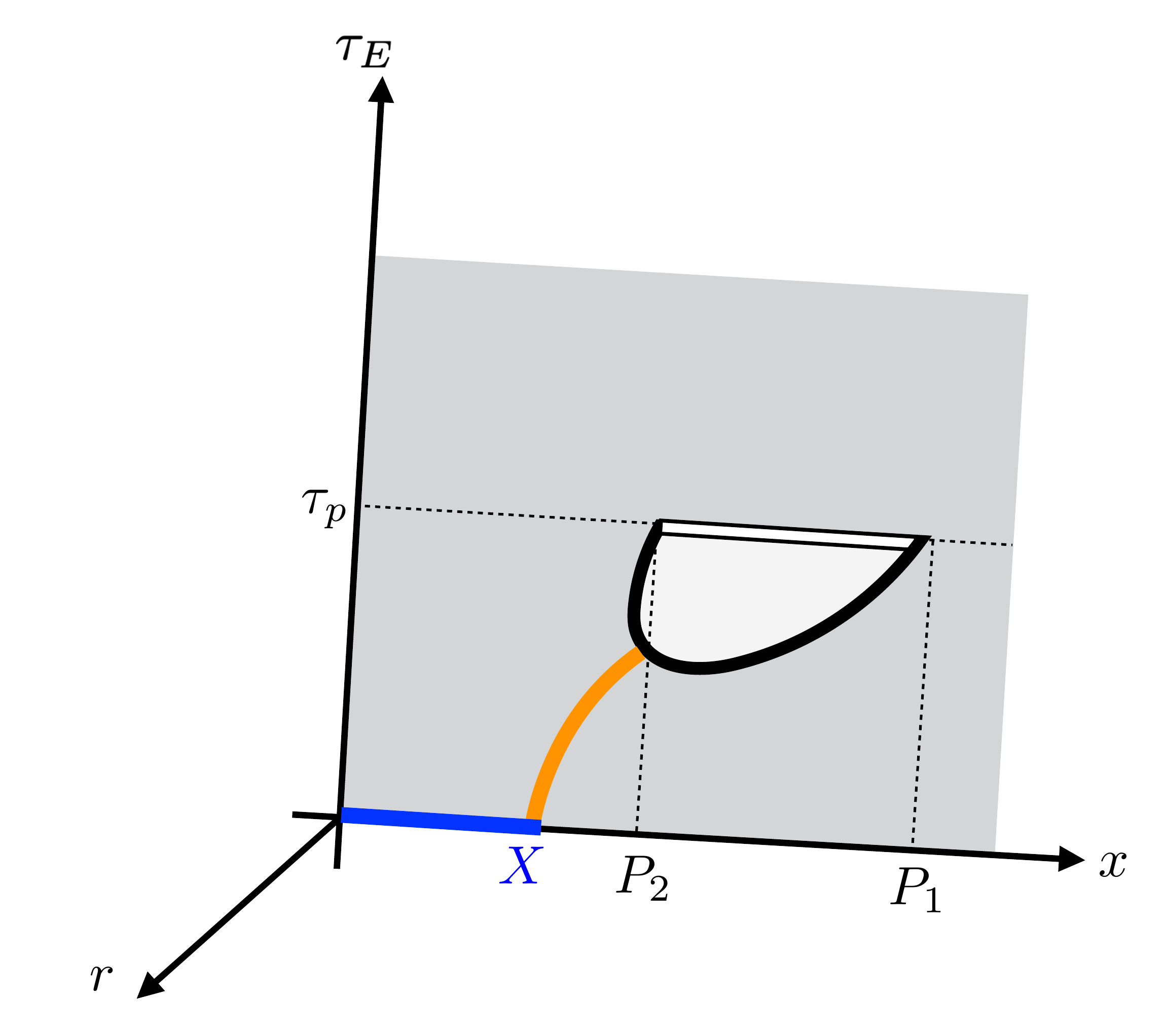}}
    \caption{Euclidean geometry for computing the minimal area surface in Section \ref{subsec:hee}. We omit the lower half of the Euclidean geometry ($\tau_E<0$ part) as the geometry is time-reflection symmetric. }
    \label{fig:asetup}
\end{figure}

In light of the line-tension picture, we understand that this difference comes from the light-cone structure of the EOW brane.  
Here, we attempt to understand the origin of the additional factor of $1/2$ in a simpler setup in the bulk. To accomplish this, let us consider the single-sided setup (see Figure \ref{fig:asetup}). Note that if we take the large black hole limit, the difference between the previous setup and the current setup is sub-leading. 
\subsubsection*{Setup and computation of the minimal area surface}
For simplicity, we focus on the entanglement entropy of a semi-infinite line $A=[-\infty,X]$. Based on the prescription of holographic entanglement entropy, we would like to compute the extremal surface connecting the EOW brane and a boundary point $X$. 
Since the tensionless EOW brane is a totally geodesic submanifold, we can view the domain of the EOW brane as geodesics anchored on the boundary of the projection. Using this fact, one can find the extremal surface as the minimal geodesics between EOW ({\it i.e.}, geodesics anchored on the boundary of the projection) and the boundary point. After the computation, we will analytically continue to the Lorenzian time. 

For this purpose, it is useful to utilize the embedding formalism. Namely, we view vacuum solutions in three dimensional Einstein gravity with the negative cosmological constant as submanifolds in $\mathbb{R}^{1,3}$,
\begin{align}
    X^2&\equiv X\cdot X=-1,\\
    \eta_{AB}&=\mathrm{diag}(-1,1,1,1),
\end{align}
and its quotient. Here, we introduced the inner product in the embedding space as
\begin{align}
X\cdot Y\equiv \eta_{AB}X^AY^B.
\end{align}
The spacelike geodesics $\gamma_{ij}$ in the embedding space $\mathbb{R}^{1,3}$ are given by
\begin{align}
    X_{ij}^A(\lambda)=\dfrac{P_i^A\mathrm{e}^{-\lambda}+P_j^A\mathrm{e}^{\lambda}}{\sqrt{-2P_i\cdot P_j}},
\end{align}
where $\lambda$ is ``proper time'' which runs from $-\infty$ to $\infty$ and $P_{i,j}^A$ are the geodesic boundary coordinates in the embedding space,
\begin{align}
    P_{i,j}^A&=(\cosh r_+P_{i,j},\sin r_+\tau_{i,j},\sinh r_+P_{i,j},\cos r_+\tau_{i,j}).
\end{align}
Now we can use this spacelike geodesics as the coordinates of the EOW brane. We then extremize the geodesic length between the EOW brane and $X$ with respect to $\lambda$. 
Assuming $X$ is on the cutoff surface of the asymptotic AdS, we obtain the geodesic length as
\begin{align}
    L(\lambda_{\text{ext}})\simeq\log(-2X_{ij}(\lambda_{\text{ext}})\cdot X),
\end{align}
where
\begin{align}
    -2X_{ij}(\lambda_{\text{ext.}})\cdot X=2\sqrt{\dfrac{(-2P_1\cdot X)(-2P_2\cdot X)}{(-2P_1\cdot P_2)(-2P_1\cdot P_2)}}.
\end{align}
Here, $P_1$ and $P_2$ are the embedding coordinates of the boundary of the projection, and $X$ is the boundary of the subsystem $A$ (on the bulk cutoff surface),
\begin{align}
    X^A&\simeq\dfrac{r_{\infty}}{r_+}(\cosh r_+X,\sin r_+\tau_{X},\sinh r_+X,\cos r_+\tau_{X}).
\end{align}
Note that we are using the bulk counterpart of the boundary coordinates $X^A$ (which is on the cutoff surface $r=r_\infty$). 

\begin{figure}
\centering
    \subfigure[]{
    \resizebox{70mm}{!}{\includegraphics{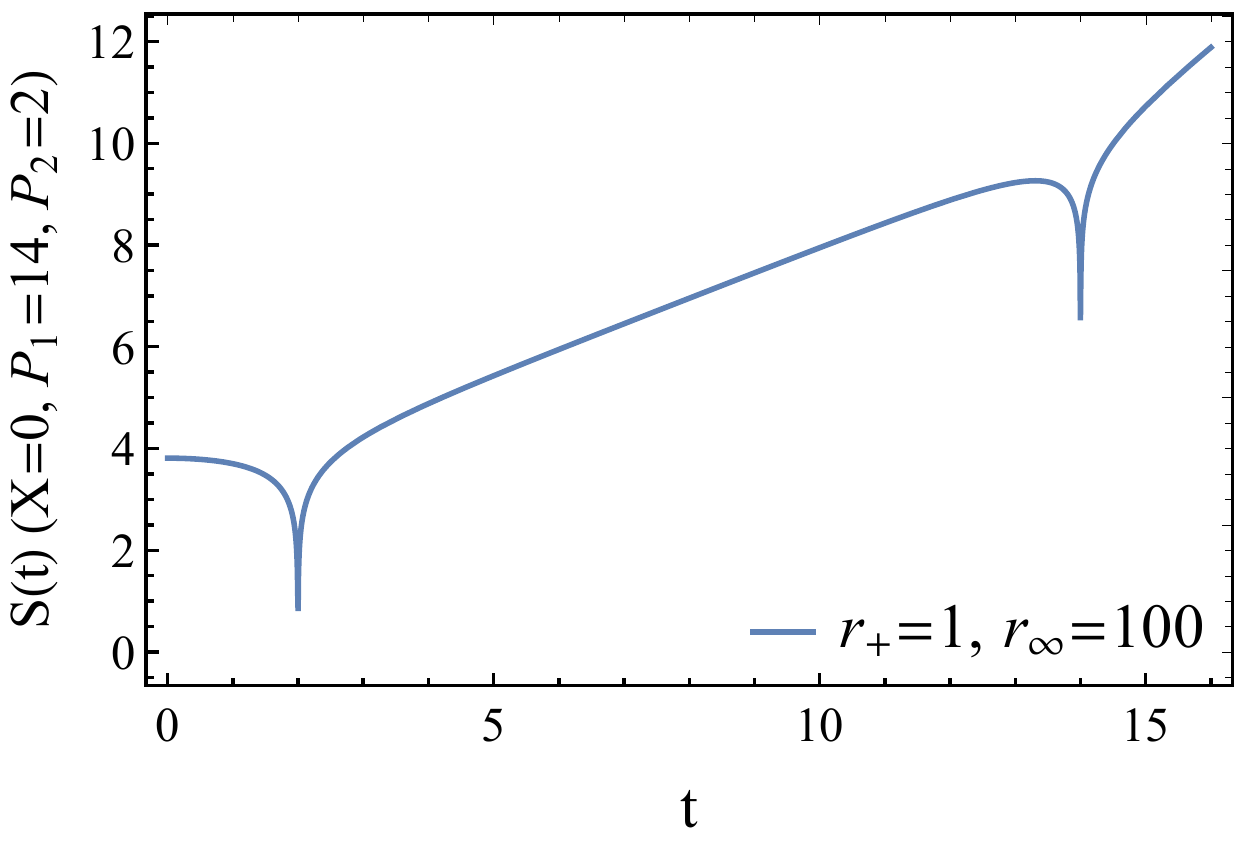}}
    }
    \subfigure[]{
    \resizebox{70mm}{!}{\includegraphics{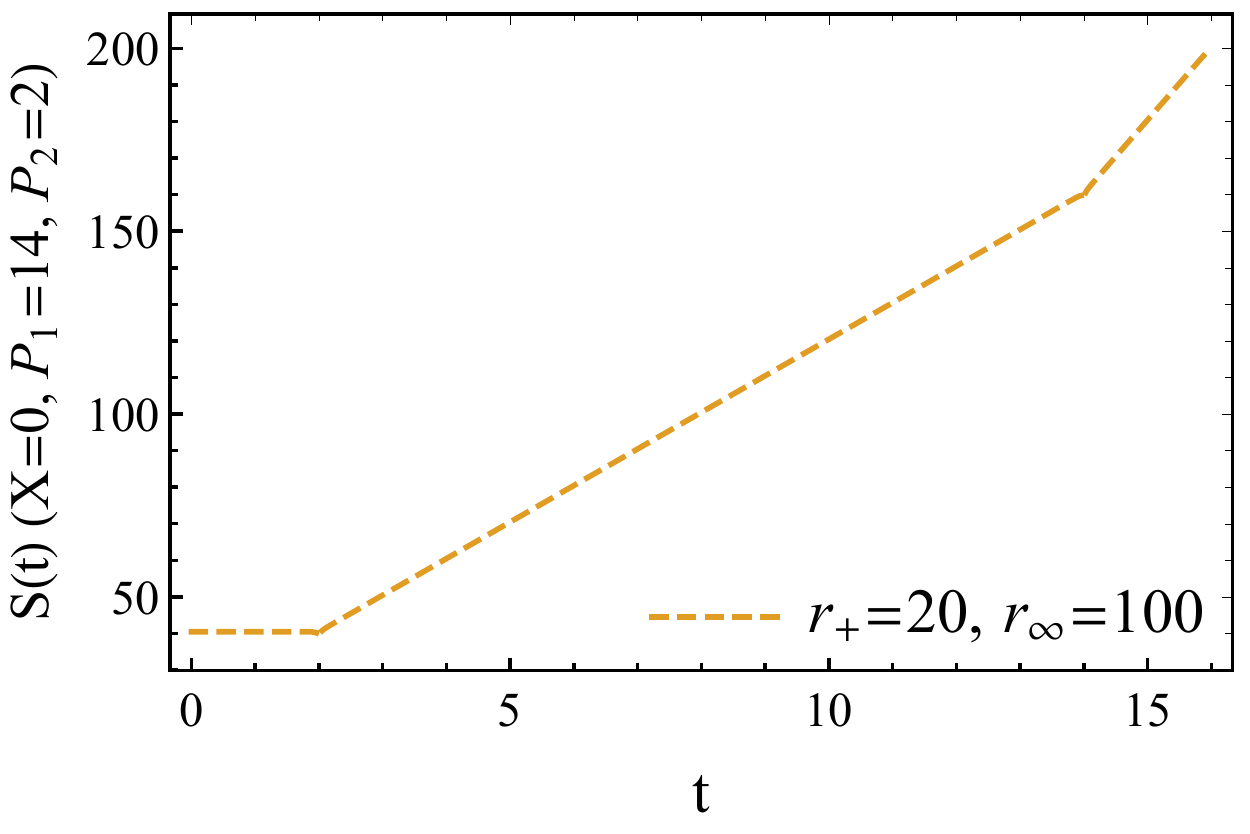}}
    }
    \caption{Time dependence of the holographic entanglement entropy given by \eqref{eq:length} (with $4G_N=1$) for small and large $r_+$, as a function of $t$. Here we took $X=0, P_1=14, P_2=2$ and $r_\infty=100$. For the plot, we recover the constant term $\frac{1}{2}\log(2r_\infty/r_+)$.}
    \label{fig:hee}
\end{figure}
After the analytic continuation to the Lorenzian signature $\tau\rightarrow it$, we obtain
\begin{align}
    L&\simeq\frac{1}{2}\log \left|\frac{\sinh\left(\frac{1}{2} r_+ (-P_1+t+X)\right)\sinh \left(\frac{1}{2}r_+ (-P_1-t+X)\right)}{\sinh \left(\frac{1}{2} r_+ (P_1-P_2)\right)}\right|\nonumber\\
    &+\,\frac{1}{2} \log \left|\frac{\sinh \left(\frac{1}{2} r_+ (-P_2-t+X)\right) \sinh \left(\frac{1}{2} r_+ (-P_2+t+X)\right)}{\sinh \left(\frac{1}{2} r_+ (P_1-P_2)\right)}\right| \label{eq:length}
\end{align}
where we omit a constant term $\frac{1}{2}\log(2r_\infty/r_+)$ as we are interested in the large $r_+$ limit. This expression gives the time-evolution of the holographic entanglement entropy (see Figure \ref{fig:hee}).
\subsubsection*{Time dependence of the minimal surface area}
\begin{figure}
\centering
\resizebox{100mm}{!}{\includegraphics{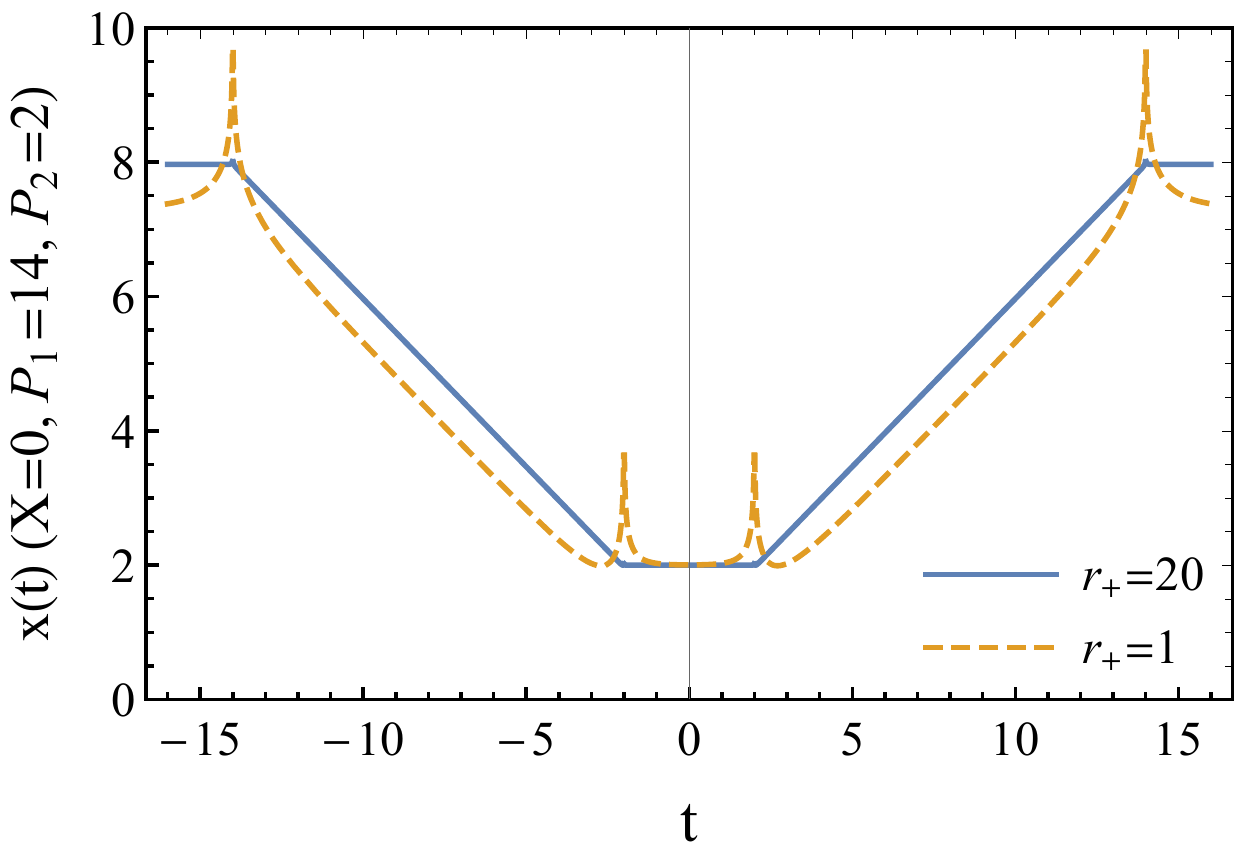}}
    \caption{Time dependence of trajectory of the minimal surface with $X=0, P_1=14, P_2=2$ as a function of $t$. We plot how the $x$-coordinates of endpoint of the minimal area surface move under the time-evolution. The large $r_+$ regime (solid blue curve) is consistent with the line-tension picture. In this setup, it starts at $X=P_2$ (edge) and finally stops at $X=(P_1+P_2)/2$ (center).}
    \label{fig:proj_traj}
\end{figure}

The leading piece in the large $r_+$ limit of the area of the minimal surface \eqref{eq:length} is given by
\begin{align}
    L&\simeq\dfrac{r_+}{4}(|X-t-P_1|+|X+t-P_1|+|X-t-P_2|+|X+t-P_2|-2|P_1-P_2|).
\end{align}
As a concrete example, let us assume that $X<P_2<P_1$. In this case, we obtain
\begin{align}
    S_A=\frac{L}{4G_N}\simeq \begin{cases}
    \frac{c\pi}{3\beta}(P_2-X) & (0<t<P_2-X), \\
    \frac{c\pi}{6\beta}(t+P_2-X) & (P_2-X<t<P_1-X), \\
    \frac{c\pi}{6\beta}\left(2t+P_2-P_1\right) & (P_1-X<t).
    \end{cases}
\end{align}
Here, we used the standard AdS/CFT dictionary,
\begin{align}
    \frac{r_+}{4G_N}=\dfrac{c\pi}{3\beta}. 
\end{align}
By identifying $\beta=4\epsilon$, we reproduce the results of the holographic CFT and the line-tension picture calculations. 
Furthermore, one can confirm that the motion of the minimal surface is also consistent with the line-tension picture (see Figure \ref{fig:proj_traj}).

Note that this simple setup should be universal in any $2$d CFT. In particular, the increase of entanglement can be understood as the propagation of EPR pairs from the each point of the projection. 

\section{Discussions and Future directions}
In this paper, we discussed the effect of a single projective measurement on the dynamics of $2$d holographic CFTs. It is shown that when the projective measurement is performed at a certain time in the holographic CFTs, the dual gravitational bulk is given by an AdS spacetime with an end-of-the-world brane at the corresponding time slice. We also proposed a line tension model that effectively describes the effect of projective measurement on entanglement in a  chaotic system. It is an interesting future direction to extend our analysis to the case of multiple projective measurements. It can be seen from (\ref{Sconst1}) and (\ref{Sconst2}) that a single projective measurement stops the non-trivial time evolution of the subsystem for a certain finite time, depending on the size of the subsystem and its location relative to the measurement. If the line tension picture for a single projective measurement we proposed in this paper is straightforwardly applicable to the case of multiple projective measurements, it is expected that by keeping the time interval between each projective measurement sufficiently short, the subsystem will not evolve in time at all.  This can be thought of as a realization of the quantum Zeno effect in quantum mechanics, in which repeated observations cause the system to freeze.  If this line tension picture is correct and reflects the geometrical structure in the gravity dual, the corresponding spacetime is expected to be the AdS spacetime with multiple branes inserted. However, the backreaction due to the insertion of multiple measurements cannot be ignored in general, and the AdS spacetime itself may be deformed significantly\footnote{We thank Tadashi Takayanagi for pointing out this possibility.}. We will leave it to future studies to elucidate what spacetime is realized and what line tension picture appropriately reflects the spacetime in the case of multiple measurements.

\subsection*{Future directions}
Here are a few possible future directions:
\begin{itemize}
    \item {\bf Single projective measurement in free CFTs}: In this paper, we studied how a single projective measurement of finite size affects quantum dynamics with maximal information scrambling. We expect that the effect of the projective measurement on the unitary time evolution will depend on the degree of information scrambling generated by the dynamics.
    The dynamics of the unitary time evolution in the $2$d free fermion can be described by relativistic propagation of quasi-particles \cite{2018arXiv181200013N,Goto:2021gve}. 
    It would be interesting to study how this effective picture is modified when a finite-size projective measurement is included in the quantum circuit.
    \item {\bf Relation to the random projective measurement}: In a quantum circuit with a random arrangement of projective measurements acting on a single site, it was discovered that an area law-volume law phase transition of the OEE occurs, depending on the frequency of these projections.
    However, such a phase transition did not occur in a $2$d holographic system containing only one projective measurement of finite size. It would be interesting to study whether such a phase transition occurs when a $2$d holographic system contains randomly placed projective measurements.
\end{itemize}

\section*{Acknowledgements}

We thank useful discussions with Jonah Kudler-Flam, Shinsei Ryu, and Tadashi Takayanagi.
K.G.~is supported by JSPS KAKENHI Grant-in-Aid for Early-Career Scientists (21K13930) and Research Fellowships of Japan Society for the Promotion of Science for Young Scientists (22J00663).
M.N.~is supported by funds from the University of Chinese Academy of Sciences (UCAS), funds from the Kavli Institute for Theoretical Sciences (KITS).
K.T.~is supported by JSPS KAKENHI Grant No.~21K13920 and MEXT KAKENHI Grant No.~22H05265. This material is based upon work supported by the National Science Foundation under Grant No. NSF-DMR 2018358 and by an
appointment to the YST Program at the APCTP through the Science and Technology Promotion Fund and Lottery Fund of the Korean Government, as well as the Korean Local Governments -
Gyeongsangbuk-do Province and Pohang City (MT).


\appendix

\section{The time-dependence of OEE for various subsystems \label{app:oee}}
In Appendix \ref{app:oee}, we will explain the time evolution of $S_B$ for cases excluding Case 4.
By replacing $t_2$ of $S_B$ with $t_1$, the time evolution of $S_A$ is given by that of $S_B$.  

\subsection{Case 1}
Let us first consider the simplest case where the location of $A,B$ and the projective measurement at $t_1=t_2=0$ are on the top of each other:
\be
P_1=X_1=Y_1, P_2=X_2=Y_2
\ee
\if[0]
\subsubsection*{The spatial and temporal locations of operators in $P_1-P_2 \gg \epsilon$}
In the large projection limit, $P_1-P_2 \gg \epsilon$, the time evolution of the spatial and temporal locations of operators is given by
\be
\begin{split}
   & X^{\text{effective}}_{X_1}\approx\begin{cases}
    -\f{\pi(2X_1-2X_2-t_{i=1,2})}{2(X_1-X_2)} & X_1-X_2>t_{i=1,2}>0 \\
    -\f{\pi}{2} & t_{i=1,2}>X_1-X_2\\
    \end{cases},
     X^{\text{effective}}_{X_2}\approx\begin{cases}
    -\f{\pi t_{i=1,2}}{2} & X_1-X_2>t_{i=1,2}>0 \\
    -\f{\pi}{2} & t_{i=1,2}>X_1-X_2\\
    \end{cases},\\
     & \tau^{\text{effective}}_{X_1}\approx \f{2\pi \epsilon}{P_1-P_2}+\begin{cases}
    -i\f{\pi t_{i=1,2}}{2(X_1-X_2)} & X_1-X_2>t_{i=1,2}>0 \\
    -i\f{\pi}{2} & t_{i=1,2}>X_1-X_2\\
    \end{cases},\\
    & \tau^{\text{effective}}_{X_2}\approx \f{2\pi \epsilon}{P_1-P_2}+\begin{cases}
    -i\f{\pi t_{i=1,2}}{2(X_1-X_2)} & X_1-X_2>t_{i=1,2}>0 \\
    -i\f{\pi}{2} & t_{i=1,2}>X_1-X_2\\
    \end{cases},\\
\end{split}
\ee

With the condition in (\ref{bc1}), the locations of images are given by
\be
\begin{split}
    &\tilde{X}^{\text{effective}}_{x}= X^{\text{effective}}_{x},~\tilde{\tau}^{\text{effective}}_{x}=2h-\tau^{\text{effective}}_{x}.\\
\end{split}
\ee
With the condition in (\ref{bc2}), the locations of images are given by
\be
\begin{split}
    &\tilde{X}^{\text{effective}}_{x}= X^{\text{effective}}_{x},~\tilde{\tau}^{\text{effective}}_{x}=2h+\tau^{\text{effective}}_{x}.\\
\end{split}
\ee
\fi
\subsubsection{The time-dependence of OEE}

In this case, let us look closely at the time-dependence of OEE in the large measurement limit, $P_1-P_2\gg \epsilon$.
In this limit, the contribution from the conformal factor is approximated by 
\be
\begin{split}
    &\lim_{n\rightarrow 1}\f{h_{n}}{1-n}\log{\left[\prod_{i=1,2} \f{d \mathcal{W}(w_i)}{dw_i}\f{d \overline{\mathcal{W}}(w_i)}{d\overline{w}_i}\right]}\\
   &\approx\f{c}{6}\log{\left[\f{X_1-X_2}{2\epsilon}\right]}-\f{c}{3}\log{\left(\f{\pi}{2\epsilon}\right)}+\f{c}{12}\times\begin{cases}
    \f{\pi t_{2}}{\epsilon} ~&~\text{for} ~X_1-X_2>t_{2} \ge 0\\
     \f{\pi (2t_{2}-(X_1-X_2)}{\epsilon} ~&~\text{for} ~t_{2} >X_1-X_2
    \end{cases}
\end{split}
\ee

Then, consider the time-dependence of the non-universal pieces of the OEE for the single interval.
As in Case 4, $S_{B,\text{con}}$ is independent of $t_2$, while the time-dependence of $S_{B,\text{dis}}$ is given by
\be
\begin{split}
&S_{\text{B,dis}} \approx -\f{c}{6}\log{\left(\f{X_1-X_2}{\epsilon}\right)} +\f{c}{12}\times\begin{cases}
\f{\pi(X_1-X_2- t_2)}{\epsilon} & ~\text{for}~ X_1-X_2>t_2>0\\
\f{-2\pi (t_2-(X_1-X_2))}{\epsilon} & ~\text{for}~ t_2>X_1-X_2\\
\end{cases}.\\
\end{split}
\ee

The time-dependence of non-universal piece of $S_B$ is given by
\be
\begin{split}
     &S_{\text{B,Non-uni.}}=\text{Min}\left[S_{\text{B,con}},S_{\text{B,dis}}\right]\approx -\f{c}{6}\log{\left(\f{X_1-X_2}{\epsilon}\right)}+\begin{cases}
    0 & X_1-X_2>t_2>0 \\
    \f{-2\pi (t_2-(X_1-X_2))}{\epsilon} & t_2>X_1-X_2
    \end{cases},\\
\end{split}
\ee
where in the early-time region, $X_1-X_2>t_{2}>0$, the non-universal piece of $S_{B}$ is determined by $S_{B,\text{con}}$ while in the late time region, $t_{2}>X_1-X_2$, that of $S_{B}$ is determined by $S_{B,\text{dis}}$.

Consequently, the time evolution of $S_B$ is given by 
\be
\begin{split}
     &S_B \approx \f{c}{3}\log{\left(\f{2\epsilon}{\pi}\right)}+\f{c\pi}{12 \epsilon} \times\begin{cases}
    t_2 & X_1-X_2>t_2>0 \\
    X_1-X_2 & t_2>X_1-X_2\\
    \end{cases}.\\
\end{split}
\ee
In this case, the projective measurement acts on the whole region of $B$. Then, the initial value of $S_B$ is approximately zero, except for $\f{c}{3}\log{\left(\f{2\epsilon}{\pi}\right)}$. 
Since the size of the projective measurement is the same as that of $B$, the scrambling effect of the time evolution operator increases the amount of $S_B$ from the beginning, $t_2=0$. 
In the late-time region, $t_2>X_1-X_2$,   $S_B$ becomes the entanglement entropy of the thermal state where the inverse temperature is $4\epsilon$.
\subsection{Case 2}
Second, let us study how the projective measurement far from the subsystems contributes to the time evolution of OEE for those subsystems.
To do so, consider the case where the spatial location of the projective measurement is far from that of the subsystems, $A$ and $B$, at $t_1=t_2=0$:
\be
P_1>P_2>X_1=Y_1>X_2=Y_2>0.
\ee

\subsubsection*{The time-dependence of OEE}
Here, we turn to the analysis on the time-dependence of the universal pieces in two subcases: $P_1-P_2>X_1-X_2>0$ and $X_1-X_2>P_1-P_2$. 
\subsubsection{$P_1-P_2>X_1-X_2>0$}
In the case of $P_1-P_2>X_1-X_2>0$, the contribution from the global structure of the torus should be in the large measurement limit since we are taking the coarse grained limit.
Thus, the time evolution of the universal piece is given by
\be
\begin{split}
    &\lim_{n\rightarrow 1}\f{h_{n}}{1-n}\log{\left[\Pi_{i=1,2} \f{d \mathcal{W}(w_i)}{dw_i}\f{d \overline{\mathcal{W}}(w_i)}{d\overline{w}_i}\right]}=\f{c}{6}\log{\left(\f{P_1-P_2}{\epsilon}\right)}
\end{split}
\ee
In the limit, $P_1-P_2\gg \epsilon$, the contribution from the conformal factor is approximated by 
\be
\begin{split}
&\lim_{n\rightarrow 1}\f{h_{n}}{1-n}\log{\left[\Pi_{i=1,2} \f{d \mathcal{W}(w_i)}{dw_i}\f{d \overline{\mathcal{W}}(w_i)}{d\overline{w}_i}\right]}\approx \f{c}{6}\log{\left(\f{P_1-P_2}{\epsilon}\right)}-\f{c}{3}\log{\left(\f{\pi}{2\epsilon}\right)}\\
&+\f{c}{24}\times\begin{cases}
\left(\f{2\pi(2P_2-X_1-X_2)}{\epsilon}\right) & P_2-X_1>t_2>0 \\
\f{\pi (3P_2-X_1-2X_2+t_2)}{\epsilon} & P_2-X_2>t_2>P_2-X_1 \\
\f{\pi (2P_2-X_1-X_2+2t_2)}{\epsilon} & P_1-X_1>t_2>P_2-X_2 \\
\f{\pi (-P_1+2P_2-X_2+3t_2)}{\epsilon} & P_1-X_2>t_2>P_1-X_1 \\
\f{2\pi (-P_1+P_2+2t_2)}{\epsilon} & t_2>P_1-X_2 \\
\end{cases}.\\
\end{split}
\ee

The value of $S_{B,\text{con}}$ is independent of $t_2$ and is given by (\ref{S_Bcon}), while
the time-dependence of  $S_{\text{B,dis}}$ is given by
\be
\begin{split}
    &S_{\text{B,dis}} \approx -\f{c}{6}\log{\left(\f{P_1-P_2}{\epsilon}\right)}+\f{c}{12}\times\begin{cases}
   \left( \f{-2\pi(P_2-X_1)}{\epsilon}\right) & P_2-X_1>t_2>0\\
   \left(\f{-\pi (t_2+3(P_2-X_1))}{2\epsilon}\right) & P_2-X_2>t_2>P_2-X_1\\
   \left(\f{\pi\left[X_1-X_2-2(P_2-X_1)-2t_2\right]}{2\epsilon}\right) & P_1-X_1>t_2>P_2-X_2 \\
   \left(\f{\pi\left[P_1-X_2-2(P_2-X_1)-3t_2\right]}{2\epsilon}\right) & P_1-X_2>t_2>P_1-X_1 \\
   \left(\f{\pi \left[P_1-X_2-(P_2-X_1)-2t_2\right]}{\epsilon}\right) & t_2>P_1-X_2
    \end{cases},\\
\end{split}
\ee
where the value of $S_{B,\text{con}}$ is always larger than that of $S_{B,\text{dis}}$.
Therefore, $S_{B,\text{Non-uni.}}$ is determined by $S_{B,\text{dis}}$.
Thus, $S_B$ is independent of $t_1$ and $t_2$, and its value is given by
\be \label{EE_thermal}
\begin{split}
    S_B\approx\f{c}{3}\log{\left(\f{2\epsilon}{\pi}\right)}+\f{c\pi (X_1-X_2)}{12\epsilon}.
\end{split}
\ee

\subsubsection*{$X_1-X_2>P_1-P_2>0$}
In the case of $X_1-X_2>P_1-P_2>0$, the universal piece in the large projection limit, $P_1-P_2 \gg \epsilon$, is given by
\be
\begin{split}
&\lim_{n\rightarrow 1}\f{h_{n}}{1-n}\log{\left[\Pi_{i=1,2} \f{d \mathcal{W}(w_i)}{dw_i}\f{d \overline{\mathcal{W}}(w_i)}{d\overline{w}_i}\right]}\approx -\f{c}{3}\log{\left(\f{\pi}{2\epsilon}\right)}+\f{c}{6}\times\begin{cases}
    \log{\left(\f{P_1-P_2}{\epsilon}\right)} & P_1-P_2 \gg \epsilon \\
    \log{\left(\f{1}{2}\right)} & \epsilon \gg P_1-P_2 >0
    \end{cases}\\
&+\f{c}{24}\times\begin{cases}
\left(\f{2\pi(2P_2-X_1-X_2)}{\epsilon}\right) & P_2-X_1>t_2>0 \\
\f{\pi (3P_2-X_1-2X_2+t_2)}{\epsilon} & P_1-X_1>t_2>P_2-X_1 \\
\f{\pi (-P_1+3P_2-2X_2+2t_2)}{\epsilon} & P_2-X_2>t_2>P_1-X_1 \\
\f{\pi (-P_1+2P_2-X_2+3t_2)}{\epsilon} & P_1-X_2>t_2>P_2-X_2 \\
\f{2\pi (-P_1+P_2+2t_2)}{\epsilon} & t_2>P_1-X_2 \\
\end{cases}.\\
\end{split}
\ee

In this case, the value of $S_{B,\text{con}}$ is equal to (\ref{S_Bcon}) so the late time evolution of $S_{\text{B,dis}}$ depends on $t_2$ and is given by 
\be
\begin{split}
    & S_{\text{B,dis}} \approx -\f{c}{6}\log{\left(\f{P_1-P_2}{\epsilon}\right)}+\f{c}{12}\times\begin{cases}
   \left( \f{-2\pi(P_2-X_1)}{\epsilon}\right) & P_2-X_1>t_1>0\\
   \left(\f{-\pi (t_2+3(P_2-X_1))}{2\epsilon}\right) & P_1-X_1>t_1>P_2-X_1\\
   \left(\f{\pi\left[P_1-P_2-2(P_2-X_1)-2t_2\right]}{2\epsilon}\right) & P_2-X_2>t_1>P_1-X_1 \\
   \left(\f{\pi\left[P_1-X_2-2(P_2-X_1)-3t_2\right]}{2\epsilon}\right) & P_1-X_2>t_1>P_2-X_2 \\
   \left(\f{\pi \left[P_1-X_2-(P_2-X_1)-2t_2\right]}{\epsilon}\right) & t_1>P_1-X_2 \\
    \end{cases},\\
    \end{split}
\ee
where the value of  $S_{B,\text{con}}$ is always larger than that of $S_{B,\text{dis}}$, so that the time evolution of non-universal piece is determined by $S_{B,\text{dis}}$.
As a consequence, the values of OEE are independent of $t_1$ and $t_2$, and their values are given by (\ref{EE_thermal}).

\subsection{Case 3}
Next, we study how the projective measurement changes the time-dependence of OEE when the projective measurement is a subregion of the subsystems at $t_1=t_2=0$.
More precisely, we will study the time-dependence of OEE in the following configuration:
\be
X_1=Y_1>P_1>P_2>X_2=Y_2>0, ~X_1-P_2>P_1-X_2>X_1-P_1>P_2-X_2>0.
\ee

\subsubsection{Time-dependence of OEE}
The time-dependence of the universal piece in Case 3 is given by
\be
\begin{split}
   & \f{c}{24}\log{\left[\Pi_{i=1,2} \f{d \mathcal{W}(w_i)}{dw_i}\f{d \overline{\mathcal{W}}(w_i)}{d\overline{w}_i}\right]}\approx-\f{c}{3}\log{\left(\f{\pi}{2\epsilon}\right)}+\f{c}{6}\times\begin{cases}
    \log{\left(\f{P_1-P_2}{\epsilon}\right)} & P_1-P_2 \gg \epsilon \\
    \log{\left(\f{1}{2}\right)} & \epsilon \gg P_1-P_2 >0
    \end{cases}\\
    &+\f{c}{24}\times\begin{cases}
\left(\f{2(X_1-X_2)-2(P_1-P_2))}{\epsilon}\right) & P_2-X_2>t_2>0 \\
\f{\pi (P_2-X_2+2(X_1-P_1)+t_2)}{\epsilon} & X_1-P_1>t_2>P_2-X_2 \\
\f{\pi (X_1-X_2-(P_1-P_2)+2t_2)}{\epsilon} & P_1-X_2>t_2>X_1-P_1 \\
\f{\pi (X_1-2P_1+P_2+3t_2)}{\epsilon} & X_1-P_2>t_2>P_1-X_2 \\
\f{2\pi (-P_1+P_2+2t_2)}{\epsilon} & t_2>X_1-P_2 \\
\end{cases}.\\
\end{split}
\ee

In the large projection limit, the value of $S_{B,\text{con}}$ is equal to (\ref{S_Bcon}), while the time-dependence of $S_{B,\text{dis}}$ is given by
\be
\begin{split}
    & S_{\text{B,dis}} \approx -\f{c}{6}\log{\left(\f{P_1-P_2}{\epsilon}\right)}+\f{c}{24}\times\begin{cases}
   \left( \f{2\pi(P_1-P_2)}{\epsilon}\right) & P_2-X_2>t_2>0\\
   \left(\f{\pi ((P_1-X_2)+(P_1-P_2)-t_2)}{\epsilon}\right) & X_1-P_1>t_1>P_2-X_2\\
   \left(\f{\pi\left((X_1-X_2)+(P_1-P_2)-2t_2\right)}{\epsilon}\right) & P_1-X_2>t_2>X_1-P_1 \\
   \left(\f{\pi\left(X_1-P_2+2(P_1-X_2)-3t_2\right)}{\epsilon}\right) & X_1-P_2>t_2>P_1-X_2 \\
   \left(\f{2\pi \left(P_1-X_2+(X_1-P_2)-2t_2\right)}{\epsilon}\right) & t_1>X_1-P_2 \\
    \end{cases},\\
    \end{split}
\ee
where $S_{B,\text{dis}}$ decreases with $t_{2}$. 
In this case, in the early time region, $\f{X_1-P_2+2(P_1-X_2)}{3}>t_2>0$, the non-universal piece of OEE is determined by $S_{B,\text{con}}$.
In the late time region, $t_2 >\f{X_1-P_2+2(P_1-X_2)}{3}$, the non-universal piece is determined by $S_{B,\text{dis}}$.
Thus, the time evolution of $S_{B}$ is given by 
\be
\begin{split}
    & S_{B} \approx \f{c}{3}\log{\left(\f{2\epsilon}{\pi}\right)}+\f{c\pi}{24 \epsilon} \times \begin{cases}
   2\left[(X_1-X_2)-(P_1-P_2)\right] & P_2-X_2>t_2>0\\
   \left[P_2-X_2+2(X_1-P_1)+t_2\right] & X_1-P_1>t_2>P_2-X_2\\
   \left[ X_1-X_2-(P_1-P_2)+2t_2 \right]& P_1-X_2>t_2>X_1-P_1\\
    \left[X_1-P_1+P_2-P_1+3t_2\right]& \f{X_1-P_2+2(P_1-X_2)}{3}>t_2>P_1-X_2 \\
    2(X_1-X_2) & t_2>\f{X_1-P_2+2(P_1-X_2)}{3}.
    \end{cases}
\end{split}
\ee

In the early time region, $P_2-X_2>t_2>0$, the projective measurement reduces the amount of OEE for $B$, so that $S_B$ at $\mathcal{O}\left(\f{1}{\epsilon}\right)$ is proportional to $X_1-X_2-(P_1-P_2)$, the size of the subregion excluding the region where the projective measurement acts.
For $t_2>P_2-X_2$, the scrambling effect of the dynamics increases the amount of $S_B$ so that at $\mathcal{O}\left(\f{1}{\epsilon}\right)$, $S_B$ is proportional to the size of $B$.
In the time interval, $\f{X_1-P_2+2(P_1-X_2)}{3}>t_1>P_2-X_2$, the time-dependence of $S_B$ exhibits the linear growth with time with a fractional coefficient as in (\ref{fractional_grwoth}).
\subsection{Case 5}
Finally, consider the case where the subsystems $A$ and $B$ are in the region where the projective measurement is performed.
The parameter region which we consider is 
\be
P_1>X_1>X_2>P_2>0, P_1-X_2>X_1-P_2>P_1-X_1>X_2-P_2>0.
\ee

\subsubsection{Time evolution of OEE}

In the large projection limit, the time evolution of the universal piece is given by
\be
\begin{split}
      & \f{c}{24}\log{\left[\Pi_{i=1,2} \f{d \mathcal{W}(w_i)}{dw_i}\f{d \overline{\mathcal{W}}(w_i)}{d\overline{w}_i}\right]}\approx-\f{c}{3}\log{\left(\f{\pi}{2\epsilon}\right)}+\f{c}{6}\times\begin{cases}
    \log{\left(\f{P_1-P_2}{\epsilon}\right)} & P_1-P_2 \gg \epsilon \\
    \log{\left(\f{1}{2}\right)} & \epsilon \gg P_1-P_2 >0
    \end{cases}\\
    &+\f{c}{24}\times\begin{cases}
 0 & X_2-P_2>t_2>0\\
    \f{\pi (P_2-X_2+t_2)}{\epsilon} & P_1-X_1>t_2>X_2-P_2 \\
    \f{\pi (X_1-X_2-(P_1-P_2)+2t_2)}{\epsilon} & X_1-P_2>t_2>P_1-X_1 \\
    \f{\pi (-P_1+2P_2-X_2+3t_2)}{\epsilon} & P_1-X_2 >t_2>X_1-P_2 \\
    \f{2\pi(2t_2-(P_1-P_2))}{\epsilon} & t_2>P_1-X_2
    \end{cases}.\\
\end{split}
\ee
In this large projection limit, the value of $S_{B,\text{con}}$ is given by (\ref{S_Bcon}), and $S_{B,\text{dis}}$ is given by 
\be
\begin{split}
    & S_{B,\text{dis}} \approx -\f{c}{6}\log{\left(\f{P_1-P_2}{\epsilon}\right)}+\f{c\pi}{24\epsilon}\times\begin{cases}
    2(X_1-X_2) & X_2-P_2>t_2>0\\
   X_1-X_2+X_1-P_2-t_2 & P_1-X_1>t_2>X_2-P_2\\
   X_1-X_2+(P_1-P_2)-2t_2 & X_1-P_2>t_2>P_1-X_1\\
   P_1-X_2+2(X_1-P_2)-3t_2 & P_1-X_2>t_2>X_1-P_2\\
   2(X_1-X_2+P_1-P_2-2t_2) & t_2>P_1-X_2\\
    \end{cases}.\\
    \end{split}
\ee

In this case, $S_{B,\text{Non-uni}}$ in the early time region, $\f{(P_1-X_2)+2(X_1-P_2)}{3}>t_{2}>0$, is determined by $S_{B,\text{con}}$. 
In the late time region, $t_{2}>\f{(P_1-X_2)+2(X_1-P_2)}{3}$, $S_{B,\text{Non-uni}}$ is determined by $S_{B,\text{dis}}$.
Consequently, the time evolution of $S_{B}$ is given by
\be
\begin{split}
    S_{B}\approx \f{c}{3}\log{\left(\f{2\epsilon}{\pi}\right)}+\f{c\pi}{24\epsilon}\times \begin{cases}
    0 & X_2-P_2>t_2>0\\
    \left[P_2-X_2+t_2\right] & P_1-X_1>t_2>X_2-P_2\\
     \left[X_1+X_2-(P_1-P_2)+2t_2\right] &X_1-P_2>t_2>P_1-X_1\\
     \left[-(X_2-P_2)-(P_1-P_2)+3t_2\right] & \f{(P_1-X_2)+2(X_1-P_2)}{3}>t_2>X_1-P_2 \\
     2(X_1-X_2) & t_2 >\f{(P_1-X_2)+2(X_1-P_2)}{3}
    \end{cases}.
\end{split}
\ee

\if[0]
\subsection{Case.6 }
\textcolor{red}{\bf We may remove this subsection}
Here, we consider the time evolution of OEE in the small projection limit, $\epsilon \gg P_1-P_2>0$. 
The spatial location of subsystems and the projection are given by
\be
X_1=Y_1>P_1>P_2>X_2=Y_2>0,~ X_1-P_1> P_2-X_2, ~X_1-P_1 \approx X_1-P_2, ~X_2-P_1\approx X_2-P_2.
\ee
In this case, the time evolution of the contribution from the conformal factor is approximated by
\be
\begin{split}
    \lim_{n\rightarrow 1}\f{h_n}{1-n}\log{\left[\prod_{i=1,2}\f{d\mathcal{W}(w_i)}{dw_i}\f{d\overline{\mathcal{W}}(\overline{w}_i)}{d\overline{w}_i}\right]} &\approx -\f{c}{3}\log{\left(\f{\pi}{2\epsilon}\right)}\\
   &+\f{c\pi}{12 \epsilon} \times \begin{cases}
    (X_1-X_2) & P_2-X_2>t_2>0\\
    (t_2+X_1-P_2) & X_1-P_2>t_2>P_2-X_2\\
    2t_2 & t_2>X_1-P_2 \\
    \end{cases}.\\
\end{split}
\ee
Next, consider the time-evolution of the non-universal piece of $S_{B}$.
Since $S_{B,\text{dis}}$ is smaller than $S_{B,\text{con}}$ in the small projection limit, the non-universal piece of $S_{B}$ is determined by $S_{B,\text{dis}}$.
The value of $S_{B}$ is a stationary constant and it is approximated by
\be
S_{\alpha=A,B} \approx -\f{c}{3}\log{\left(\f{\pi}{4}\right)}+\f{c\pi (X_1-X_2)}{12\epsilon}.
\ee
\fi
\if[0]
\section{The $I_{A,B}$ in Case.3}

\textcolor{red}{\bf MN: We may remove this section.}

In the method of image, one of the candidates for the minimal geodesic length is given by
\be \label{Scand1}
S^{\text{cand.1}}_{A\cup B, \text{Non-uni.}}=S_{A,\text{Non-uni}}+S_{B,\text{Non-uni}}.
\ee
When $S_{A\cup B, \text{Non-uni}}$ is given by $S^{\text{cand.}}_{A\cup B, \text{Non-uni.}}$, the value of $I_{A,B}$ is zero.
The other candidate is $S^{\text{cand.2}}_{A\cup B}$.
This candidate is given by the summation of length of only geodesics corresponding to one- or two-point function in $$2$d$ boundary CFT.
The time evolution of $S^{\text{cand.2}}_{A\cup B}$ is given by 
\be
\begin{split}
    &S^{\text{cand.2}}_{A\cup B} \approx -\f{c}{3}\log{\left(\f{P_1-P_2}{\epsilon}\right)}+ f_1(t)+f_2(t),\\
    &f_1(t) = \text{Min}\left[0,L_1(t)\right], ~f_2(t) = \text{Min}\left[0,L_2(t)\right],\\
    &L_1(t)=\f{c}{12} \log{\left[-\sinh{\left[\f{sn^{-1}(\tilde{z}_1,k^2)-sn^{-1}(\tilde{z}_3,k^2)}{2}\right]}\sinh{\left[\f{sn^{-1}(\overline{\tilde{z}}_1,k^2)-sn^{-1}(\overline{\tilde{z}}_3,k^2)}{2}\right]}\right]}\\
    &\approx \f{c \pi}{24 \epsilon} \times \begin{cases}
    2(P_1-P_2) & X_1-P_1>t>0\\
    X_1-P_2+(P_1-P_2)-t & X_1-P_2>t>X_1-P_1\\
    P_1-P_2 & t>X_1-P_2
    \end{cases}\\
    &L_2(t)=\f{c}{12} \log{\left[-\sinh{\left[\f{sn^{-1}(\tilde{z}_2,k^2)-sn^{-1}(\tilde{z}_4,k^2)}{2}\right]}\sinh{\left[\f{sn^{-1}(\overline{\tilde{z}}_2,k^2)-sn^{-1}(\overline{\tilde{z}}_4,k^2)}{2}\right]}\right]}\\
    &\approx \f{c \pi}{24 \epsilon} \times \begin{cases}
    2\left[t-(P_2-X_2)\right] & P_2-X_2>t>0\\
    \left[t-(P_2-X_2)\right] & P_1-X_2>t>P_2-X_2\\
    P_1-P_2 & t>P_1-X_2
    \end{cases},\\
\end{split}
\ee
where when $f_1(t)=f_2(t)=0$, $S^{\text{cand.2}}_{A\cup B}$ reduces to  $S^{\text{cand.1}}_{A\cup B}$. \footnote{\textcolor{red}{\bf MN: Probably, we can removes this equation.}}
The value of $L_1(t)$ is always positive, so that $f_1(t)$ should be zero.
The value of $L_2(t)$ is negative in the early-time region, $P_2-Y_2>t>0$, while  $L_2(t)$ is positive in the late-time region, $P_2-Y_2>t>0$.
In summary, the time evolution of $S^{\text{cand.2}}_{A\cup B}$ is given by
\be
\begin{split}
   &S^{\text{cand.2}}_{A\cup B} \approx -\f{c}{3}\log{\left(\f{P_1-P_2}{\epsilon}\right)}+  \f{c\pi}{12 \epsilon} \begin{cases}
   \left[t-(P_2-X_2)\right] & P_2-X_2>t>0 \\
   0 & t>P_2-X_2
   \end{cases}.
\end{split}
\ee
\text{red}{There are other candidates for minimum geodesics length, but the contributions from them will be sub-dominant}

\subsubsection{$I_{A,B}$ in Case.3}
In Case.3, the time evolution of $I_{A,B}$ is given by
\be
\begin{split}
    I_{A,B} \approx \f{c\pi}{12 \epsilon}\times\begin{cases}
    \left[(P_2-X_2)-t\right] &P_2-X_2>t>0 \\
    0 & t>P_2-X_2
    \end{cases}.
\end{split}
\ee
\fi

\section{Early and late time-expansions of $\mathcal{W}$ and $\overline{\mathcal{W}}$ in the large projection limit\label{sec:asym-beh-large-limit}}
Now, we describe the early and late time-expansion of the variables analytically-continued to real times in the large projection limit where $P_1-P_2 \gg \epsilon$.
After the analytic continuation in (\ref{ac}), the variables $\tilde{z}$ and $\overline{\tilde{z}}$ are purely imaginary, so the imaginary part of $\mathcal{W}$ and $\overline{\mathcal{W}}$  are given by the following integrals:
\be
sn^{-1}(\pm i Z ,k^2)= \int^{Z}_0 \f{\pm i dq}{\sqrt{(1+q^2)(1+k^2q^2)}},
\ee
where $\pm i Z $ denote the analytically-continued variables.
\subsection*{$sn^{-1}(\tilde{z}_{1}, k^2)$}

The early time-expansion used in this paper of $sn^{-1}(\tilde{z}_{1}, k^2)$ is 
\be
\begin{split}
sn^{-1}(\tilde{z}_1,k^2)=-i\int ^{e^{\f{\pi}{2\epsilon}(X_1-P_2+t_2)}}_0 \f{dq}{\sqrt{(1+q^2)(1+k^2 q^2)}} 
= -i \int ^{e^{\f{\pi}{2\epsilon}(X_1-P_1+t_2)}}_0 \f{d\tilde{q}}{\sqrt{(k^2+\tilde{q}^2)(1+\tilde{q}^2)}},
\end{split}
\ee
where the new variable $\tilde{q}$ is defined as $k q = \tilde{q}$. In the early time-region, $X_1-P_1+t_2<0$, the function $e^{\f{\pi}{2\epsilon}(X_1-P_1+t_2)}$ may be very small, $e^{\f{\pi}{2\epsilon}(X_1-P_1+t_2)} \ll 1$. In this time-regime, $sn^{-1}( i Z ,k^2)$ may be approximated by
\be
\begin{split}
sn^{-1}(\tilde{z}_1,k^2)&\approx -i \int ^{e^{\f{\pi}{2\epsilon}(X_1-P_1+t_2)}}_0 \f{d\tilde{q}}{\sqrt{(k^2+\tilde{q}^2)}}\\
&=-i \left[\log{\left(e^{\f{\pi}{2\epsilon}(X_1-P_1+t_2)}+\sqrt{e^{\f{-\pi}{\epsilon}(P_1-P_2)}+e^{\f{\pi}{\epsilon}(X_1-P_1+t_2)}}\right)}-\log{e^{\f{-\pi}{2\epsilon}(P_1-P_2)}}\right] \\
&\approx \begin{cases}
-i e^{\f{\pi}{2\epsilon}(X_1-P_2+t_2)} & X_1-P_2+t_2<0\\
-i \f{\pi}{2\epsilon}(X_1-P_2+t_2) & X_1-P_2+t_2>0 \\
\end{cases}.
\end{split}
\ee

For the late time-regime, $X_1-P_1+t_2>0$, the inverse of Jacobian elliptic sn function may be approximated by 
\be
\begin{split}
sn^{-1}(\tilde{z}_1,k^2)&= -i \left[ \int ^{\infty}_{0}\f{d \bar{V}}{\sqrt{(k^2+\bar{V}^2)(1+\bar{V}^2)}}-\int ^{\infty}_{e^{\f{\pi}{2\epsilon}(X_1-P_1+t_2)}}\f{d \bar{V}^2}{\sqrt{(k^2+\bar{V})(1+\bar{V}^2)}}\right]\\
&\approx -i \left[ \int ^{\infty}_{0}\f{d \bar{V}}{\sqrt{(k^2+\bar{V}^2)(1+\bar{V}^2)}}-\int ^{\infty}_{e^{\f{\pi}{2\epsilon}(X_1-P_1+t_2)}}\f{d \bar{V}}{\bar{V}^2}\right]\\
&= -i\left[K(1-k^2)-e^{\f{-\pi}{2\epsilon}(X_1-P_1+t_2)}\right] \approx -i\f{\pi}{2\epsilon}(P_1-P_2).
\end{split}
\ee
This is the late time-expansion used in this paper.
\subsection*{$sn^{-1}(\tilde{z}_{2}, k^2)$}
In this paper, the early time-expansion of $sn^{-1}(\tilde{z}_{2}, k^2)$ that is used is 
\be
\begin{split}
sn^{-1}(\tilde{z}_2,k^2)=-i\int ^{e^{\f{\pi}{2\epsilon}(X_2-P_2+t_2)}}_0 \f{dq}{\sqrt{(1+q^2)(1+k^2 q^2)}} 
= -i \int ^{e^{\f{\pi}{2\epsilon}(X_2-P_1+t_2)}}_0 \f{d\tilde{q}}{\sqrt{(k^2+\tilde{q}^2)(1+\tilde{q}^2)}},
\end{split}
\ee
where the new variable $\tilde{q}$ is defined as $k q = \tilde{q}$. In the early time-regime, $X_2-P_1+t_2<0$, the function $e^{\f{\pi}{2\epsilon}(X_2-P_1+t_2)}$ may be very small, $e^{\f{\pi}{2\epsilon}(X_2-P_1+t_2)} \ll 1$. In this time-regime, $sn^{-1}( i Z ,k^2)$ may be approximated by
\be
\begin{split}
sn^{-1}(\tilde{z}_2,k^2)&\approx -i \int ^{e^{\f{\pi}{2\epsilon}(X_2-P_1+t_2)}}_0 \f{d\tilde{q}}{\sqrt{(k^2+\tilde{q}^2)}}\\
&=-i \left[\log{\left(e^{\f{\pi}{2\epsilon}(X_2-P_1+t_2)}+\sqrt{e^{\f{-\pi}{\epsilon}(P_1-P_2)}+e^{\f{\pi}{\epsilon}(X_2-P_1+t_2)}}\right)}-\log{e^{\f{-\pi}{2\epsilon}(P_1-P_2)}}\right] \\
&\approx \begin{cases}
-i e^{\f{\pi}{2\epsilon}(X_2-P_2+t_2)} & X_2-P_2+t_2<0\\
-i \f{\pi}{2\epsilon}(X_2-P_2+t_2) & X_2-P_2+t_2>0 \\
\end{cases}
\end{split}
\ee

For the late time regime, $X_2-P_1+t_2>0$, the inverse of the Jacobian elliptic sn function may be approximated by 
\be
\begin{split}
sn^{-1}(\tilde{z}_2,k^2)&= -i \left[ \int ^{\infty}_{0}\f{d \bar{V}}{\sqrt{(k^2+\bar{V}^2)(1+\bar{V}^2)}}-\int ^{\infty}_{e^{\f{\pi}{2\epsilon}(X_2-P_1+t_2)}}\f{d \bar{V}^2}{\sqrt{(k^2+\bar{V})(1+\bar{V}^2)}}\right]\\
&\approx -i \left[ \int ^{\infty}_{0}\f{d \bar{V}}{\sqrt{(k^2+\bar{V}^2)(1+\bar{V}^2)}}-\int ^{\infty}_{e^{\f{\pi}{2\epsilon}(X_2-P_1+t_2)}}\f{d \bar{V}}{\bar{V}^2}\right]\\
&= -i\left[K(1-k^2)-e^{\f{-\pi}{2\epsilon}(X_2-P_1+t_2)}\right] \approx -i\f{\pi}{2\epsilon}(P_1-P_2).
\end{split}
\ee
This is the late time-expansion used in this paper.

\subsection*{$sn^{-1}(\tilde{z}_{3}, k^2)$}
The early time-expansion used in this paper of $sn^{-1}(\tilde{z}_{3}, k^2)$ is 
\be
\begin{split}
sn^{-1}(\tilde{z}_{3},k^2)=i\int ^{e^{\f{\pi}{2\epsilon}(Y_1-P_2-t_1)}}_0 \f{dq}{\sqrt{(1+q^2)(1+k^2 q^2)}} 
= i \int ^{e^{\f{\pi}{2\epsilon}(Y_1-P_1-t_1)}}_0 \f{d\tilde{q}}{\sqrt{(k^2+\tilde{q}^2)(1+\tilde{q}^2)}},
\end{split}
\ee
where the new variable $\tilde{q}$ is defined as $k q = \tilde{q}$. In the early time-regime, $Y_1-P_1-t_1<0$, the function $e^{\f{\pi}{2\epsilon}(Y_1-P_1-t_1)}$ may be very small, $e^{\f{\pi}{2\epsilon}(Y_1-P_1-t_1)} \ll 1$. In this time-regime, $sn^{-1}( i Z ,k^2)$ may be approximated by
\be
\begin{split}
sn^{-1}(\tilde{z}_3,k^2)&\approx i \int ^{e^{\f{\pi}{2\epsilon}(Y_1-P_1-t_1)}}_0 \f{d\tilde{q}}{\sqrt{(k^2+\tilde{q}^2)}}\\
&=i \left[\log{\left(e^{\f{\pi}{2\epsilon}(Y_1-P_1-t_1)}+\sqrt{e^{\f{-\pi}{\epsilon}(P_1-P_2)}+e^{\f{\pi}{\epsilon}(Y_1-P_1-t_1)}}\right)}-\log{e^{\f{-\pi}{2\epsilon}(P_1-P_2)}}\right] \\
&\approx \begin{cases}
i e^{\f{\pi}{2\epsilon}(Y_1-P_2-t_1)} & Y_1-P_2-t_1<0\\
i \f{\pi}{2\epsilon}(Y_1-P_2-t_1) & Y_1-P_2-t_1>0 \\
\end{cases}
\end{split}
\ee

For the late time-regime, $Y_1-P_1-t_1>0$, the inverse of Jacobian elliptic sn function may be approximated by 
\be
\begin{split}
sn^{-1}(\tilde{z}_3,k^2)&= i \left[ \int ^{\infty}_{0}\f{d \bar{V}}{\sqrt{(k^2+\bar{V}^2)(1+\bar{V}^2)}}-\int ^{\infty}_{e^{\f{\pi}{2\epsilon}(Y_1-P_1-t_1)}}\f{d \bar{V}^2}{\sqrt{(k^2+\bar{V})(1+\bar{V}^2)}}\right]\\
&\approx i \left[ \int ^{\infty}_{0}\f{d \bar{V}}{\sqrt{(k^2+\bar{V}^2)(1+\bar{V}^2)}}-\int ^{\infty}_{e^{\f{\pi}{2\epsilon}(Y_1-P_1-t_1)}}\f{d \bar{V}}{\bar{V}^2}\right]\\
&= \left[K(1-k^2)-e^{\f{-\pi}{2\epsilon}(Y_1-P_1-t_1)}\right] \approx i\f{\pi}{2\epsilon}(P_1-P_2).
\end{split}
\ee
This is the late time-expansion used in this paper.
\subsection*{$sn^{-1}(\tilde{z}_{4}, k^2)$}
The early time-expansion used in this paper of $sn^{-1}(\tilde{z}_{4}, k^2)$ is 
\be
\begin{split}
sn^{-1}(\tilde{z}_{4},k^2)=i\int ^{e^{\f{\pi}{2\epsilon}(Y_2-P_2-t_1)}}_0 \f{dq}{\sqrt{(1+q^2)(1+k^2 q^2)}} 
= i \int ^{e^{\f{\pi}{2\epsilon}(Y_2-P_1-t_1)}}_0 \f{d\tilde{q}}{\sqrt{(k^2+\tilde{q}^2)(1+\tilde{q}^2)}},
\end{split}
\ee
where the new variable $\tilde{q}$ is defined as $k q = \tilde{q}$. In the early time-regime, $Y_2-P_1-t_1<0$, the function $e^{\f{\pi}{2\epsilon}(Y_2-P_1-t_1)}$ may be very small, $e^{\f{\pi}{2\epsilon}(Y_2-P_1-t_1)} \ll 1$. In this time region, $sn^{-1}( i Z ,k^2)$ may be approximated by
\be
\begin{split}
sn^{-1}(\tilde{z}_4,k^2)&\approx i \int ^{e^{\f{\pi}{2\epsilon}(Y_2-P_1-t_1)}}_0 \f{d\tilde{q}}{\sqrt{(k^2+\tilde{q}^2)}}\\
&=i \left[\log{\left(e^{\f{\pi}{2\epsilon}(Y_2-P_1-t_1)}+\sqrt{e^{\f{-\pi}{\epsilon}(P_1-P_2)}+e^{\f{\pi}{\epsilon}(Y_2-P_1-t_1)}}\right)}-\log{e^{\f{-\pi}{2\epsilon}(P_1-P_2)}}\right] \\
&\approx \begin{cases}
i e^{\f{\pi}{2\epsilon}(Y_2-P_2-t_1)} & Y_2-P_2-t_1<0\\
i \f{\pi}{2\epsilon}(Y_2-P_2-t_1) & Y_2-P_2-t_1>0 \\
\end{cases}
\end{split}
\ee

For the late time regime, $Y_2-P_1-t_1>0$, the inverse of Jacobian elliptic sn function may be approximated by 
\be
\begin{split}
sn^{-1}(\overline{\tilde{z}}_4,k^2)&= i \left[ \int ^{\infty}_{0}\f{d \bar{V}}{\sqrt{(k^2+\bar{V}^2)(1+\bar{V}^2)}}-\int ^{\infty}_{e^{\f{\pi}{2\epsilon}(Y_1-P_1-t_1)}}\f{d \bar{V}^2}{\sqrt{(k^2+\bar{V})(1+\bar{V}^2)}}\right]\\
&\approx i \left[ \int ^{\infty}_{0}\f{d \bar{V}}{\sqrt{(k^2+\bar{V}^2)(1+\bar{V}^2)}}-\int ^{\infty}_{e^{\f{\pi}{2\epsilon}(Y_1-P_1-t_1)}}\f{d \bar{V}}{\bar{V}^2}\right]\\
&= i\left[K(1-k^2)-e^{\f{-\pi}{2\epsilon}(Y_1-P_1-t_1)}\right] \approx i\f{\pi}{2\epsilon}(P_1-P_2).
\end{split}
\ee
This is the late time-expansion used in this paper.


\subsection*{$sn^{-1}(\overline{\tilde{z}}_{1}, k^2)$}
The early time-expansion used in this paper of $sn^{-1}(\overline{\tilde{z}}_{1}, k^2)$ is 
\be
\begin{split}
sn^{-1}(\overline{\tilde{z}}_1,k^2)=i\int ^{e^{\f{\pi}{2\epsilon}(X_1-P_2-t_2)}}_0 \f{dq}{\sqrt{(1+q^2)(1+k^2 q^2)}} 
= i \int ^{e^{\f{\pi}{2\epsilon}(X_1-P_1-t_2)}}_0 \f{d\tilde{q}}{\sqrt{(k^2+\tilde{q}^2)(1+\tilde{q}^2)}},
\end{split}
\ee
where the new variable $\tilde{q}$ is defined as $k q = \tilde{q}$. In the early time region, $X_1-P_1-t_2<0$, the function $e^{\f{\pi}{2\epsilon}(X_1-P_1-t_2)}$ may be very small, $e^{\f{\pi}{2\epsilon}(X_1-P_1-t_2)} \ll 1$. In this time regime, $sn^{-1}( i Z ,k^2)$ may be approximated by
\be
\begin{split}
sn^{-1}(\overline{\tilde{z}}_1,k^2)&\approx i \int ^{e^{\f{\pi}{2\epsilon}(X_1-P_1-t_2)}}_0 \f{d\tilde{q}}{\sqrt{(k^2+\tilde{q}^2)}}\\
&=i \left[\log{\left(e^{\f{\pi}{2\epsilon}(X_1-P_1-t_2)}+\sqrt{e^{\f{-\pi}{\epsilon}(P_1-P_2)}+e^{\f{\pi}{\epsilon}(X_1-P_1-t_2)}}\right)}-\log{e^{\f{-\pi}{2\epsilon}(P_1-P_2)}}\right] \\
&\approx \begin{cases}
i e^{\f{\pi}{2\epsilon}(X_1-P_2-t_2)} & X_1-P_2-t_2<0\\
i \f{\pi}{2\epsilon}(X_1-P_2-t_2) & X_1-P_2-t_2>0 \\
\end{cases}
\end{split}
\ee

For the late time regime, $X_1-P_1-t_2>0$, the inverse of Jacobian elliptic sn function may be approximated by 
\be
\begin{split}
sn^{-1}(\overline{\tilde{z}}_1,k^2)&= i \left[ \int ^{\infty}_{0}\f{d \bar{V}}{\sqrt{(k^2+\bar{V}^2)(1+\bar{V}^2)}}-\int ^{\infty}_{e^{\f{\pi}{2\epsilon}(X_1-P_1+t_2)}}\f{d \bar{V}^2}{\sqrt{(k^2+\bar{V})(1+\bar{V}^2)}}\right]\\
&\approx i \left[ \int ^{\infty}_{0}\f{d \bar{V}}{\sqrt{(k^2+\bar{V}^2)(1+\bar{V}^2)}}-\int ^{\infty}_{e^{\f{\pi}{2\epsilon}(X_1-P_1-t_2)}}\f{d \bar{V}}{\bar{V}^2}\right]\\
&= i\left[K(1-k^2)-e^{\f{-\pi}{2\epsilon}(X_1-P_1-t_2)}\right] \approx i\f{\pi}{2\epsilon}(P_1-P_2).
\end{split}
\ee
This is the late time expansion used in this paper.
\subsection*{$sn^{-1}(\overline{\tilde{z}}_{2}, k^2)$}
The early time-expansion used in this paper of $sn^{-1}(\overline{\tilde{z}}_{2}, k^2)$ is 
\be
\begin{split}
sn^{-1}(\overline{\tilde{z}}_2,k^2)=i\int ^{e^{\f{\pi}{2\epsilon}(X_2-P_2-t_2)}}_0 \f{dq}{\sqrt{(1+q^2)(1+k^2 q^2)}} 
= i \int ^{e^{\f{\pi}{2\epsilon}(X_2-P_1-t_2)}}_0 \f{d\tilde{q}}{\sqrt{(k^2+\tilde{q}^2)(1+\tilde{q}^2)}},
\end{split}
\ee
where the new variable $\tilde{q}$ is defined as $k q = \tilde{q}$. In the early time region, $X_2-P_1-t_2<0$, the function $e^{\f{\pi}{2\epsilon}(X_2-P_1-t_2)}$ may be very small, $e^{\f{\pi}{2\epsilon}(X_2-P_1-t_2)} \ll 1$. In this time regime, $sn^{-1}( i Z ,k^2)$ may be approximated by
\be
\begin{split}
sn^{-1}(\overline{\tilde{z}}_2,k^2)&\approx i \int ^{e^{\f{\pi}{2\epsilon}(X_2-P_1-t_2)}}_0 \f{d\tilde{q}}{\sqrt{(k^2+\tilde{q}^2)}}\\
&=i \left[\log{\left(e^{\f{\pi}{2\epsilon}(X_2-P_1-t_2)}+\sqrt{e^{\f{-\pi}{\epsilon}(P_1-P_2)}+e^{\f{\pi}{\epsilon}(X_2-P_1-t_2)}}\right)}-\log{e^{\f{-\pi}{2\epsilon}(P_1-P_2)}}\right] \\
&\approx \begin{cases}
i e^{\f{\pi}{2\epsilon}(X_2-P_2-t_2)} & X_2-P_2-t_2<0\\
i \f{\pi}{2\epsilon}(X_2-P_2-t_2) & X_2-P_2-t_2>0 \\
\end{cases}
\end{split}
\ee

For the late time region, $X_2-P_1-t_2>0$, the inverse of Jacobian elliptic sn function may be approximated by 
\be
\begin{split}
sn^{-1}(\overline{\tilde{z}}_2,k^2)&= i \left[ \int ^{\infty}_{0}\f{d \bar{V}}{\sqrt{(k^2+\bar{V}^2)(1+\bar{V}^2)}}-\int ^{\infty}_{e^{\f{\pi}{2\epsilon}(X_2-P_1+t_2)}}\f{d \bar{V}^2}{\sqrt{(k^2+\bar{V})(1+\bar{V}^2)}}\right]\\
&\approx i \left[ \int ^{\infty}_{0}\f{d \bar{V}}{\sqrt{(k^2+\bar{V}^2)(1+\bar{V}^2)}}-\int ^{\infty}_{e^{\f{\pi}{2\epsilon}(X_2-P_1-t_2)}}\f{d \bar{V}}{\bar{V}^2}\right]\\
&= i\left[K(1-k^2)-e^{\f{-\pi}{2\epsilon}(X_2-P_1-t_2)}\right] \approx i\f{\pi}{2\epsilon}(P_1-P_2).
\end{split}
\ee

This is the late time expansion used in this paper.

\subsection*{$sn^{-1}(\overline{\tilde{z}}_{3}, k^2)$}
The early time-expansion used in this paper of $sn^{-1}(\tilde{z}_{3}, k^2)$ is 
\be
\begin{split}
sn^{-1}(\overline{\tilde{z}}_3,k^2)=-i\int ^{e^{\f{\pi}{2\epsilon}(Y_1-P_2+t_1)}}_0 \f{dq}{\sqrt{(1+q^2)(1+k^2 q^2)}} 
= -i \int ^{e^{\f{\pi}{2\epsilon}(Y_1-P_1+t_1)}}_0 \f{d\tilde{q}}{\sqrt{(k^2+\tilde{q}^2)(1+\tilde{q}^2)}},
\end{split}
\ee
where the new variable $\tilde{q}$ is defined as $k q = \tilde{q}$. In the early time-regime, $Y_1-P_1+t_1<0$, the function $e^{\f{\pi}{2\epsilon}(Y_1-P_1+t_1)}$ may be very small, $e^{\f{\pi}{2\epsilon}(Y_1-P_1+t_1)} \ll 1$. In this time-regime, $sn^{-1}( i Z ,k^2)$ may be well-approximated by
\be
\begin{split}
sn^{-1}(\overline{\tilde{z}}_3,k^2)&\approx -i \int ^{e^{\f{\pi}{2\epsilon}(Y_1-P_1+t_1)}}_0 \f{d\tilde{q}}{\sqrt{(k^2+\tilde{q}^2)}}\\
&=-i \left[\log{\left(e^{\f{\pi}{2\epsilon}(Y_1-P_1+t_1)}+\sqrt{e^{\f{-\pi}{\epsilon}(P_1-P_2)}+e^{\f{\pi}{\epsilon}(Y_1-P_1+t_1)}}\right)}-\log{e^{\f{-\pi}{2\epsilon}(P_1-P_2)}}\right] \\
&\approx \begin{cases}
-i e^{\f{\pi}{2\epsilon}(Y_1-P_2+t_1)} & Y_1-P_2+t_1<0\\
-i \f{\pi}{2\epsilon}(Y_1-P_2+t_1) & Y_1-P_2+t_1>0 \\
\end{cases}
\end{split}
\ee

For the late time-region, $Y_1-P_1+t_1>0$, the inverse of Jacobian elliptic sn function may be approximated by 
\be
\begin{split}
sn^{-1}(\overline{\tilde{z}}_3,k^2)&= -i \left[ \int ^{\infty}_{0}\f{d \bar{V}}{\sqrt{(k^2+\bar{V}^2)(1+\bar{V}^2)}}-\int ^{\infty}_{e^{\f{\pi}{2\epsilon}(Y_1-P_1+t_1)}}\f{d \bar{V}^2}{\sqrt{(k^2+\bar{V})(1+\bar{V}^2)}}\right]\\
&\approx -i \left[ \int ^{\infty}_{0}\f{d \bar{V}}{\sqrt{(k^2+\bar{V}^2)(1+\bar{V}^2)}}-\int ^{\infty}_{e^{\f{\pi}{2\epsilon}(Y_1-P_1+t_1)}}\f{d \bar{V}}{\bar{V}^2}\right]\\
&= -i\left[K(1-k^2)-e^{\f{-\pi}{2\epsilon}(Y_1-P_1+t_1)}\right] \approx -i\f{\pi}{2\epsilon}(P_1-P_2).
\end{split}
\ee
This is the late-time expansion used in this paper. 
\subsection*{$sn^{-1}(\overline{\tilde{z}}_{4}, k^2)$}
The early time-expansion of $sn^{-1}(\tilde{z}_{4}, k^2)$ is 
\be
\begin{split}
sn^{-1}(\overline{\tilde{z}}_4,k^2)=-i\int ^{e^{\f{\pi}{2\epsilon}(Y_2-P_2+t_1)}}_0 \f{dq}{\sqrt{(1+q^2)(1+k^2 q^2)}} 
= -i \int ^{e^{\f{\pi}{2\epsilon}(Y_2-P_1+t_1)}}_0 \f{d\tilde{q}}{\sqrt{(k^2+\tilde{q}^2)(1+\tilde{q}^2)}},
\end{split}
\ee
where the new variable $\tilde{q}$ is defined as $k q = \tilde{q}$. In the early time region, $Y_2-P_1+t_1<0$, the function $e^{\f{\pi}{2\epsilon}(Y_2-P_1+t_1)}$ may be very small, $e^{\f{\pi}{2\epsilon}(Y_2-P_1+t_1)} \ll 1$. In this time regime, $sn^{-1}( i Z ,k^2)$ may be approximated by
\be
\begin{split}
sn^{-1}(\overline{\tilde{z}}_4,k^2)&\approx -i \int ^{e^{\f{\pi}{2\epsilon}(Y_2-P_1+t_1)}}_0 \f{d\tilde{q}}{\sqrt{(k^2+\tilde{q}^2)}}\\
&=-i \left[\log{\left(e^{\f{\pi}{2\epsilon}(Y_2-P_1+t_1)}+\sqrt{e^{\f{-\pi}{\epsilon}(P_1-P_2)}+e^{\f{\pi}{\epsilon}(Y_2-P_1+t_1)}}\right)}-\log{e^{\f{-\pi}{2\epsilon}(P_1-P_2)}}\right] \\
&\approx \begin{cases}
-i e^{\f{\pi}{2\epsilon}(Y_2-P_2+t_1)} & Y_2-P_2+t_1<0\\
-i \f{\pi}{2\epsilon}(Y_2-P_2+t_1) & Y_2-P_2+t_1>0 \\
\end{cases}
\end{split}
\ee

For the late time regime, $Y_2-P_1+t_1>0$, the inverse of Jacobian elliptic sn function may be approximated by 
\be
\begin{split}
sn^{-1}(\overline{\tilde{z}}_4,k^2)&= -i \left[ \int ^{\infty}_{0}\f{d \bar{V}}{\sqrt{(k^2+\bar{V}^2)(1+\bar{V}^2)}}-\int ^{\infty}_{e^{\f{\pi}{2\epsilon}(Y_2-P_1+t_1)}}\f{d \bar{V}^2}{\sqrt{(k^2+\bar{V})(1+\bar{V}^2)}}\right]\\
&\approx -i \left[ \int ^{\infty}_{0}\f{d \bar{V}}{\sqrt{(k^2+\bar{V}^2)(1+\bar{V}^2)}}-\int ^{\infty}_{e^{\f{\pi}{2\epsilon}(Y_2-P_1+t_1)}}\f{d \bar{V}}{\bar{V}^2}\right]\\
&= -i\left[K(1-k^2)-e^{\f{-\pi}{2\epsilon}(Y_2-P_1+t_1)}\right] \approx -i\f{\pi}{2\epsilon}(P_1-P_2).
\end{split}
\ee

\if(
\section{Early and late time expansions of $\mathcal{W}$ and $\overline{\mathcal{W}}$ in small projection limit $P_1-P_2 \ll \epsilon$ \label{sec:asym-beh-small-limit}}
Let us now turn to the analysis on the early and late time expansions of $\mathcal{W}$ and $\overline{\mathcal{W}}$, in the in small projection limit $P_1-P_2 \ll \epsilon$.
\subsection*{$sn^{-1}(\tilde{z}_1,k^2)$}
After the analytic continuation to real time, $sn^{-1}(\tilde{z}_1,k^2)$ is given by
\be
sn^{-1}(\tilde{z}_1,k^2) = -i \int ^{e^{\f{\pi}{2\epsilon}(X_1-P_2+t_2)}}_{0}\f{dq}{\sqrt{(1+q^2)(1+k^2 q^2)}}\stackrel{\tilde{q}=kq}{=}-i\int^{e^{\f{\pi}{2\epsilon}(X_1-P_1+t_2)}}_0 \f{d\tilde{q}}{\sqrt{(k^2+\tilde{q}^2)(1+\tilde{q}^2)}}.
\ee

\subsubsection*{In the early time regime, $X_1-P_1+t_2<0$}
Here, we may approximate $sn^{-1}(\tilde{z}_1,k^2)$ in the early time region, $X_1-P_1+t_2<0$, as 
\be
\begin{split}
    sn^{-1}(\tilde{z}_1,k^2) \approx \f{-i}{k} \int^{e^{\f{\pi}{2\epsilon}(X_1-P_1+t_2)}}_{0}d\tilde{y}=-i e^{\f{\pi}{2\epsilon}(X_1-P_2+t_2)},
\end{split}
\ee
where we assume that $e^{\f{\pi}{2\epsilon}(X_1-P_2+t_2)} \ll 1$.
\subsubsection*{In the late time regime, $X_1-P_1+t_2>0$}
Let us approximate $sn^{-1}(\tilde{z}_1,k^2)$ in the late time regime, $X_1-P_1+t_2>0$, as
\be
\begin{split}
    sn^{-1}(\tilde{z}_1,k^2) &=-i\left[\int^{\infty}_{0}\f{d\tilde{q}}{\sqrt{(k^2+\tilde{q}^2)(1+\tilde{q}^2)}}-\int^{\infty}_{e^{\f{\pi(X_1-P_1+t_2)}{2\epsilon}}} \f{d\tilde{q}}{\sqrt{(k^2+\tilde{q}^2)(1+\tilde{q}^2)}}\right]\\
  &  \approx -i\left[\int^{\infty}_{0}\f{d\tilde{q}}{\sqrt{(k^2+\tilde{q}^2)(1+\tilde{q}^2)}}-\int^{\infty}_{e^{\f{\pi(X_1-P_1+t_2)}{2\epsilon}}} \f{d\tilde{q}}{\tilde{q}^2}\right]\\
    &=-i \left[K(1-k^2)-e^{\f{-\pi}{2\epsilon}(X_1-P_1+t_2)}\right],
\end{split}
\ee
where we assumed that $e^{\f{-\pi}{2\epsilon}(X_1-P_1+t_2)} \ll 1$.
\subsection*{$sn^{-1}(\tilde{z}_2,k^2)$}
After the analytic continuation to real time, $sn^{-1}(\tilde{z}_2,k^2)$ is given by 
\be
sn^{-1}(\tilde{z}_2,k^2) = -i \int ^{e^{\f{\pi}{2\epsilon}(X_2-P_2+t_2)}}_{0}\f{dq}{\sqrt{(1+q^2)(1+k^2 q^2)}}\stackrel{\tilde{q}=kq}{=}-i\int^{e^{\f{\pi}{2\epsilon}(X_2-P_2+t_2)}}_0 \f{d\tilde{q}}{\sqrt{(k^2+\tilde{q}^2)(1+\tilde{q}^2)}}.
\ee
\subsubsection*{In the early time regime, $X_2-P_1+t_2<0$}
Here, we may approximate $sn^{-1}(\tilde{z}_2,k^2)$ in the early time regime, $X_2-P_1+t_2<0$, as 
\be
\begin{split}
    sn^{-1}(\tilde{z}_2,k^2) \approx \f{-i}{k} \int^{e^{\f{\pi}{2\epsilon}(X_2-P_1+t_2)}}_{0}d\tilde{y}=-i e^{\f{\pi}{2\epsilon}(X_2-P_2+t_2)},
\end{split}
\ee
where we assume that $e^{\f{\pi}{2\epsilon}(X_2-P_2+t_2)} \ll 1$. 
\subsubsection*{In the late time regime, $X_2-P_1+t_2>0$}
Here, we may approximate $sn^{-1}(\tilde{z}_2,k^2)$ in the late time-regime, $X_2-P_1+t_2>0$, as 
\be
\begin{split}
    sn^{-1}(\tilde{z}_2,k^2) &=-i\left[\int^{\infty}_{0}\f{d\tilde{q}}{\sqrt{(k^2+\tilde{q}^2)(1+\tilde{q}^2)}}-\int^{\infty}_{e^{\f{\pi(X_2-P_1+t_2)}{2\epsilon}}} \f{d\tilde{q}}{\sqrt{(k^2+\tilde{q}^2)(1+\tilde{q}^2)}}\right]\\
  &  \approx -i\left[\int^{\infty}_{0}\f{d\tilde{q}}{\sqrt{(k^2+\tilde{q}^2)(1+\tilde{q}^2)}}-\int^{\infty}_{e^{\f{\pi(X_2-P_1+t_2)}{2\epsilon}}} \f{d\tilde{q}}{\tilde{q}^2}\right]\\
    &=-i \left[K(1-k^2)-e^{\f{-\pi}{2\epsilon}(X_2-P_1+t_2)}\right],
\end{split}
\ee
where we assumed that $e^{\f{-\pi}{2\epsilon}(X_2-P_1+t_2)}\ll 1$.

\subsection*{$sn^{-1}(\tilde{z}_3,k^2)$}
After the analytic continuation to real time, the function $sn^{-1}(\tilde{z}_3,k^2)$ is given by 
\be
sn^{-1}(\tilde{z}_3,k^2) = i \int ^{e^{\f{\pi}{2\epsilon}(X_1-P_2-t_1)}}_{0}\f{dq}{\sqrt{(1+q^2)(1+k^2 q^2)}}\stackrel{\tilde{q}=kq}{=}i\int^{e^{\f{\pi}{2\epsilon}(X_1-P_1-t_1)}}_0 \f{d\tilde{q}}{\sqrt{(k^2+\tilde{q}^2)(1+\tilde{q}^2)}}.
\ee
\subsubsection*{In the early time regime, $X_1-P_1>t_1>0$}
Here, we may approximate $sn^{-1}(\tilde{z}_2,k^2)$ in the early time region, $X_1-P_1-t_1>0$, as 
\be
\begin{split}
    sn^{-1}(\tilde{z}_3,k^2) &=i\left[\int^{\infty}_{0}\f{d\tilde{q}}{\sqrt{(k^2+\tilde{q}^2)(1+\tilde{q}^2)}}-\int^{\infty}_{e^{\f{\pi(X_1-P_1-t_1)}{2\epsilon}}} \f{d\tilde{q}}{\sqrt{(k^2+\tilde{q}^2)(1+\tilde{q}^2)}}\right]\\
  &  \approx i\left[\int^{\infty}_{0}\f{d\tilde{q}}{\sqrt{(k^2+\tilde{q}^2)(1+\tilde{q}^2)}}-\int^{\infty}_{e^{\f{\pi(X_1-P_1-t_1)}{2\epsilon}}} \f{d\tilde{q}}{\tilde{q}^2}\right]\\
    &=i \left[K(1-k^2)-e^{-\f{\pi}{2\epsilon}(X_1-P_1-t_1)}\right],
\end{split}
\ee
where we assumed that  $e^{\f{-\pi}{2\epsilon}(X_1-P_1-t_1)}\ll 1$.
\subsubsection*{In the late time regime, $t_1>X_1-P_1$}
Here, we may approximate $sn^{-1}(\tilde{z}_3,k^2)$ in the early time region, $t_1>X_1-P_1$, as 
\be
\begin{split}
    sn^{-1}(\tilde{z}_3,k^2) \approx \f{i}{k} \int^{e^{\f{\pi}{2\epsilon}(X_1-P_1+t_1)}}_{0}d\tilde{y}=i e^{\f{\pi}{2\epsilon}(X_1-P_2-t_1)},
\end{split}
\ee
where we assumed that $e^{\f{\pi}{2\epsilon}(X_1-P_2-t_1)} \ll 1$. 
\subsection*{$sn^{-1}(\tilde{z}_4,k^2)$}
After the analytic continuation to real time, the function, $sn^{-1}(\tilde{z}_4,k^2)$ is given by 
\be
sn^{-1}(\tilde{z}_4,k^2) = i \int ^{e^{\f{-\pi}{2\epsilon}(t_1+(P_2-X_2))}}_{0}\f{dq}{\sqrt{(1+q^2)(1+k^2 q^2)}}\stackrel{\tilde{q}=kq}{=}i\int^{e^{\f{-\pi}{2\epsilon}(P_1-X_2+t_1)}}_0 \f{d\tilde{q}}{\sqrt{(k^2+\tilde{q}^2)(1+\tilde{q}^2)}}.
\ee
\subsubsection*{In the early time regime, $X_2-P_1>t_1>0$}
Here, we may approximate $sn^{-1}(\tilde{z}_4,k^2)$ in the early time region, $X_2-P_1-t_1>0$, as 
\be
\begin{split}
    sn^{-1}(\tilde{z}_4,k^2) &=i\left[\int^{\infty}_{0}\f{d\tilde{q}}{\sqrt{(k^2+\tilde{q}^2)(1+\tilde{q}^2)}}-\int^{\infty}_{e^{\f{-\pi(P_1-X_2+t_1)}{2\epsilon}}} \f{d\tilde{q}}{\sqrt{(k^2+\tilde{q}^2)(1+\tilde{q}^2)}}\right]\\
  &  \approx i\left[\int^{\infty}_{0}\f{d\tilde{q}}{\sqrt{(k^2+\tilde{q}^2)(1+\tilde{q}^2)}}-\int^{\infty}_{e^{\f{-\pi(P_1-X_2+t_1)}{2\epsilon}}} \f{d\tilde{q}}{\tilde{q}^2}\right]\\
    &i \left[K(1-k^2)-e^{\f{\pi}{2\epsilon}(P_1-X_2+t_1)}\right],
\end{split}
\ee
where we assumed that $e^{\f{\pi}{2\epsilon}(P_1-X_2+t_1)}\ll 1$.
\subsubsection*{In the late time regime, $t_1>X_2-P_1$}
Here, we may approximate $sn^{-1}(\tilde{z}_4,k^2)$ in the early time region, $t_1>X_2-P_1$, as 
\be
\begin{split}
    sn^{-1}(\tilde{z}_3,k^2) \approx \f{i}{k} \int^{e^{\f{-\pi}{2\epsilon}(P_1-X_2+t_1)}}_{0}d\tilde{y}=i e^{\f{-\pi}{2\epsilon}(P_2-X_2+t_1)},
\end{split}
\ee
where we assumed that $e^{\f{-\pi}{2\epsilon}(P_2-X_2+t_1)} \ll 1$.

\subsection*{$sn^{-1}(\overline{\tilde{z}}_1,k^2)$}
After the analytic continuation to real time, the function $sn^{-1}(\overline{\tilde{z}}_1,k^2)$ is given by 
\be
sn^{-1}(\tilde{z}_1,k^2) = i \int ^{e^{\f{\pi}{2\epsilon}(X_1-P_2-t_2)}}_{0}\f{dq}{\sqrt{(1+q^2)(1+k^2 q^2)}}\stackrel{\tilde{q}=kq}{=}i\int^{e^{\f{\pi}{2\epsilon}(X_1-P_1-t_2)}}_0 \f{d\tilde{q}}{\sqrt{(k^2+\tilde{q}^2)(1+\tilde{q}^2)}}.
\ee
\subsubsection*{In the early time-regime, $X_1-P_1>t_2>0$}
Here, we may approximate $sn^{-1}(\overline{\tilde{z}}_1,k^2)$ in the early time, $X_1-P_1>t_2>0$, as 
\be
\begin{split}
    sn^{-1}(\overline{\tilde{z}}_1,k^2) &=i\left[\int^{\infty}_{0}\f{d\tilde{q}}{\sqrt{(k^2+\tilde{q}^2)(1+\tilde{q}^2)}}-\int^{\infty}_{e^{\f{\pi(X_1-P_1-t_2)}{2\epsilon}}} \f{d\tilde{q}}{\sqrt{(k^2+\tilde{q}^2)(1+\tilde{q}^2)}}\right]\\
  &  \approx i\left[\int^{\infty}_{0}\f{d\tilde{q}}{\sqrt{(k^2+\tilde{q}^2)(1+\tilde{q}^2)}}-\int^{\infty}_{e^{\f{\pi(X_1-P_1-t_2)}{2\epsilon}}} \f{d\tilde{q}}{\tilde{q}^2}\right]\\
    &i \left[K(1-k^2)-e^{-\f{\pi}{2\epsilon}(X_1-P_1-t_2)}\right],
\end{split}
\ee
where we assumed that $e^{\f{-\pi}{2\epsilon}(X_1-P_1-t_2)}\ll 1$.
\subsubsection*{In the late time regime, $t_2>X_1-P_1$}
Here, we may approximate $sn^{-1}(\overline{\tilde{z}}_1,k^2)$ in the late time, $t_2>X_1-P_1$, as 
\be
\begin{split}
    sn^{-1}(\overline{\tilde{z}}_1,k^2) \approx \f{i}{k} \int^{e^{\f{\pi}{2\epsilon}(X_1-P_1-t_2)}}_{0}d\tilde{y}=i e^{\f{\pi}{2\epsilon}(X_1-P_1-t_2)},
\end{split}
\ee
where we assumed that $e^{\f{\pi}{2\epsilon}(X_1-P_2-t_2)} \ll 1$. 
\subsection*{$sn^{-1}(\overline{\tilde{z}}_2,k^2)$}
After the analytic continuation to real time, the function $sn^{-1}(\overline{\tilde{z}}_2,k^2)$ is given by 
\be
sn^{-1}(\tilde{z}_2,k^2) = i \int ^{e^{\f{\pi}{2\epsilon}(X_2-P_2-t_2)}}_{0}\f{dq}{\sqrt{(1+q^2)(1+k^2 q^2)}}\stackrel{\tilde{q}=kq}{=}i\int^{e^{-\f{\pi}{2\epsilon}(t_2+P_1-X_2)}}_0 \f{d\tilde{q}}{\sqrt{(k^2+\tilde{q}^2)(1+\tilde{q}^2)}}.
\ee
\subsubsection*{In the early time regime, $X_2-P_1>t_2>0$}
Here, we may approximate $sn^{-1}(\overline{\tilde{z}}_1,k^2)$ in the early time, $X_1-P_1>t_2>0$, as 
\be
\begin{split}
    sn^{-1}(\overline{\tilde{z}}_2,k^2) &=i\left[\int^{\infty}_{0}\f{d\tilde{q}}{\sqrt{(k^2+\tilde{q}^2)(1+\tilde{q}^2)}}-\int^{\infty}_{e^{\f{-\pi(t_2+P_1-X_2)}{2\epsilon}}} \f{d\tilde{q}}{\sqrt{(k^2+\tilde{q}^2)(1+\tilde{q}^2)}}\right]\\
  &  \approx i\left[\int^{\infty}_{0}\f{d\tilde{q}}{\sqrt{(k^2+\tilde{q}^2)(1+\tilde{q}^2)}}-\int^{\infty}_{e^{\f{-\pi(t_2+P_1-X_2)}{2\epsilon}}} \f{d\tilde{q}}{\tilde{q}^2}\right]\\
    &i \left[K(1-k^2)-e^{\f{\pi}{2\epsilon}(P_1-X_2+t_2)}\right],
\end{split}
\ee
where we assumed that $e^{\f{\pi}{2\epsilon}(P_1-X_2+t_2)}\ll 1$.
\subsubsection*{In the late time regime, $t_2>X_2-P_1$}
Here, we may approximate $sn^{-1}(\overline{\tilde{z}}_2,k^2)$ in the late time, $X_1-P_1>t_2>0$, as 
\be
\begin{split}
    sn^{-1}(\overline{\tilde{z}}_2,k^2) \approx \f{i}{k} \int^{e^{\f{-\pi}{2\epsilon}(P_1-X_2+t_2)}}_{0}d\tilde{y}=i e^{\f{\pi}{2\epsilon}(X_2-P_2-t_2)},
\end{split}
\ee
where we assumed that  $e^{\f{\pi}{2\epsilon}(X_2-P_2-t_2)} \ll 1$.
\subsection*{$sn^{-1}(\overline{\tilde{z}}_3,k^2)$}
After the analytic continuation to real time, the function $sn^{-1}(\overline{\tilde{z}}_3,k^2)$ is given by 
\be
sn^{-1}(\tilde{z}_2,k^2) = -i \int ^{e^{\f{\pi}{2\epsilon}(Y_1-P_2+t_1)}}_{0}\f{dq}{\sqrt{(1+q^2)(1+k^2 q^2)}}\stackrel{\tilde{q}=kq}{=}-i\int^{e^{\f{\pi}{2\epsilon}(Y_1-P_1+t_1)}}_0 \f{d\tilde{q}}{\sqrt{(k^2+\tilde{q}^2)(1+\tilde{q}^2)}}.
\ee
\subsubsection*{In the early time regime, $P_1-Y_1>t_1$}
Here, we may approximate $sn^{-1}(\overline{\tilde{z}}_3,k^2)$ in the late time, $P_1-Y_1>t_1$, as 
\be
\begin{split}
    sn^{-1}(\overline{\tilde{z}}_3,k^2) \approx \f{-i}{k} \int^{e^{\f{\pi}{2\epsilon}(Y_1-P_1+t_1)}}_{0}d\tilde{y}=-i e^{\f{\pi(Y_1-P_2+t_1)}{2\epsilon}},
\end{split}
\ee
where we assumed that $ e^{\f{\pi(Y_1-P_2+t_1)}{2\epsilon}} \ll 1$.
\subsubsection*{In the late time-regime, $t_1>P_1-Y_1$}
Here, we may approximate $sn^{-1}(\overline{\tilde{z}}_3,k^2)$ in the late time, $t_1>P_1-Y_1$, as
\be
\begin{split}
    sn^{-1}(\overline{\tilde{z}}_3,k^2) &=-i\left[\int^{\infty}_{0}\f{d\tilde{q}}{\sqrt{(k^2+\tilde{q}^2)(1+\tilde{q}^2)}}-\int^{\infty}_{e^{\f{\pi(Y_1-P_1+t_1)}{2\epsilon}}} \f{d\tilde{q}}{\sqrt{(k^2+\tilde{q}^2)(1+\tilde{q}^2)}}\right]\\
  &  \approx -i\left[\int^{\infty}_{0}\f{d\tilde{q}}{\sqrt{(k^2+\tilde{q}^2)(1+\tilde{q}^2)}}-\int^{\infty}_{e^{\f{\pi(Y_1-P_1+t_1)}{2\epsilon}}} \f{d\tilde{q}}{\tilde{q}^2}\right]\\
    &=-i \left[K(1-k^2)-e^{-\f{\pi}{2\epsilon}(Y_1-P_1+t_1)}\right],
\end{split}
\ee
where we assumed that $e^{-\f{\pi}{2\epsilon}(X_2-P_1-t_2)}\ll 1$.
\subsection*{$sn^{-1}(\overline{\tilde{z}}_4,k^2)$}
After the analytic continuation to real time, the function $sn^{-1}(\overline{\tilde{z}}_3,k^2)$ is given by 
\be
sn^{-1}(\tilde{z}_2,k^2) = -i \int ^{e^{\f{\pi}{2\epsilon}(Y_2-P_2+t_1)}}_{0}\f{dq}{\sqrt{(1+q^2)(1+k^2 q^2)}}\stackrel{\tilde{q}=kq}{=}-i\int^{e^{\f{\pi}{2\epsilon}(Y_2-P_1+t_1)}}_0 \f{d\tilde{q}}{\sqrt{(k^2+\tilde{q}^2)(1+\tilde{q}^2)}}.
\ee
\subsubsection*{In the early time regime, $P_1-Y_2>t_1$}
Here, we may approximate $sn^{-1}(\overline{\tilde{z}}_4,k^2)$ in the late time, $P_1-Y_2>t_1$, as 
\be
\begin{split}
    sn^{-1}(\overline{\tilde{z}}_4,k^2) \approx \f{-i}{k} \int^{e^{\f{\pi}{2\epsilon}(Y_2-P_1+t_1)}}_{0}d\tilde{y}=-i e^{\f{\pi(Y_2-P_2+t_1)}{2\epsilon}},
\end{split}
\ee
where we assumed that $ e^{\f{\pi(Y_2-P_2+t_1)}{2\epsilon}} \ll 1$.
\subsubsection*{In the late time regime, $t_1>P_1-Y_2$}
Here, we may approximate $sn^{-1}(\overline{\tilde{z}}_4,k^2)$ in the late time, $t_1>P_1-Y_2$, as 
\be
\begin{split}
    sn^{-1}(\overline{\tilde{z}}_4,k^2) &=-i\left[\int^{\infty}_{0}\f{d\tilde{q}}{\sqrt{(k^2+\tilde{q}^2)(1+\tilde{q}^2)}}-\int^{\infty}_{e^{\f{\pi(Y_2-P_1+t_1)}{2\epsilon}}} \f{d\tilde{q}}{\sqrt{(k^2+\tilde{q}^2)(1+\tilde{q}^2)}}\right]\\
  &  \approx -i\left[\int^{\infty}_{0}\f{d\tilde{q}}{\sqrt{(k^2+\tilde{q}^2)(1+\tilde{q}^2)}}-\int^{\infty}_{e^{\f{\pi(Y_2-P_1+t_1)}{2\epsilon}}} \f{d\tilde{q}}{\tilde{q}^2}\right]\\
    &=-i \left[K(1-k^2)-e^{-\f{\pi}{2\epsilon}(Y_2-P_1+t_1)}\right],
\end{split}
\ee
where we assumed that  $e^{-\f{\pi}{2\epsilon}(Y_2-P_1+t_1)}\ll 1$.
)\fi
\bibliographystyle{ieeetr}
\bibliography{reference}
\end{document}